\documentclass[twocolumn, a4paper, 11pt,book,fleqn,usenatbib]{archive}

\usepackage{graphicx}
\usepackage{txfonts}
\usepackage{color}
\usepackage[dvipsnames]{xcolor}

\usepackage{txfonts}
\usepackage{natbib}
\usepackage{amssymb}
\usepackage{tikz} 
\usetikzlibrary{calc}
\usepackage{newtxtext,newtxmath}
\usepackage[T1]{fontenc}
\usepackage{graphicx}	
\usepackage{amsmath}	
\usepackage{amssymb}	

\usepackage{color}
\usepackage{multirow,bigdelim}

\usepackage{multirow,bigdelim}
\usepackage{xcolor}
\usepackage{colortbl}
\usepackage{array}
\usepackage{arydshln}
\def\Xint#1{\mathchoice{\XXint\displaystyle\textstyle{#1}}
{\XXint\textstyle\scriptstyle{#1}}
{\XXint\scriptstyle\scriptscriptstyle{#1}}
{\XXint\scriptscriptstyle\scriptscriptstyle{#1}}
\!\int}\def\XXint#1#2#3{{\setbox0=\hbox{$#1{#2#3}{\int}$}
\vcenter{\hbox{$#2#3$}}\kern-.5\wd0}}

\def\dashint{\Xint-}
\begin{document} 
\twocolumn[{%
 \centering
%
{\center \bf \huge Influence of grain growth on CO$_2$ ice spectroscopic profiles}\\
{\center \bf \Large Modelling for dense cores and disks}\\
\vspace*{0.25cm}

{\Large 
E. Dartois \inst{1},
J. A. Noble  \inst{2},
N. Ysard \inst{3},
K. Demyk \inst{4},
M. Chabot \inst{5}}\\
\vspace*{0.25cm}

$^1$      Institut des Sciences Mol\'eculaires d'Orsay, CNRS, Universit\'e Paris-Saclay, 
B\^at 520, Rue Andr\'e Rivi\`ere, 91405 Orsay, France\\
              \email{emmanuel.dartois@universite-paris-saclay.fr}\\
$^2$      	CNRS, Aix-Marseille Universit\'e, Laboratoire PIIM, Marseille, France\\
$^3$      	Institut d'Astrophysique Spatiale, CNRS, Universit\'e Paris-Saclay, B\^at. 121, 91405 Orsay cedex, France\\
$^4$      	IRAP, Universit\'e de Toulouse, CNRS, UPS, IRAP, 9 Av. colonel Roche, BP 44346, 31028, Toulouse Cedex 4, France\\
$^5$               Laboratoire de physique des deux infinis Ir\`ene Joliot-Curie,CNRS-IN2P3, Universit\'e Paris-Saclay, 91405 Orsay, France\\

 \vspace*{0.5cm}
{keywords: ISM: lines and bands, Radiative transfer,  ISM: dust, extinction,  Protoplanetary disks, ISM: clouds, Infrared: ISM}\\
 \vspace*{0.5cm}
{\it \large Submitted to Astronomy \& Astrophysics}\\
 \vspace*{0.5cm}
  }]
  \section*{Abstract}
Interstellar dust grain growth in dense clouds and protoplanetary disks, even moderate, affects the observed interstellar ice profiles as soon as a significant fraction of dust grains is in the size range close to the wave vector at the considered wavelength. The continuum baseline correction made prior to analysing ice profiles influences the subsequent analysis and hence the estimated ice composition, typically obtained by band fitting using thin film ice mixture spectra.
   {We explore the effect of grain growth on the spectroscopic profiles of ice mantle constituents, focusing particularly on carbon dioxide, with the aim of understanding how it can affect interstellar ice mantle spectral analysis and interpretation.}
   {Using the Discrete Dipole Approximation for Scattering and Absorption of Light, the mass absorption coefficients of several distributions of grains -- composed of ellipsoidal silicate cores with water and carbon dioxide ice mantles -- are calculated. A few models also include amorphous carbon in the core and pure carbon monoxide in the ice mantle. We explore the evolution of the size distribution starting in the dense core phase in order to simulate the first steps of grain growth up to three microns in size. The resulting mass absorption coefficients are injected into RADMC-3D radiative transfer models of spherical dense core and protoplanetary disk templates to retrieve the observable spectral energy distributions. Calculations are performed using the full scattering capabilities of the radiative transfer code. We then focus on the particularly relevant calculated profile of the carbon dioxide ice band at 4.27~$\mu$m.}
   {The carbon dioxide antisymmetric stretching mode profile is a meaningful indicator of grain growth. The observed profile toward dense cores with the Infrared space observatory and Akari satellites already showed profiles possibly indicative of moderate grain growth.}
   {The observation of true protoplanetary disks at high inclination with the JWST should present distorted profiles that will allow constraints to be placed on the extent of dust growth. The more evolved the dust size distribution, the more the extraction of the ice mantle composition will require both understanding and taking into account grain growth.
%

\section{Introduction}
%

\begin{figure*}
  \centering
\includegraphics[width=\columnwidth]{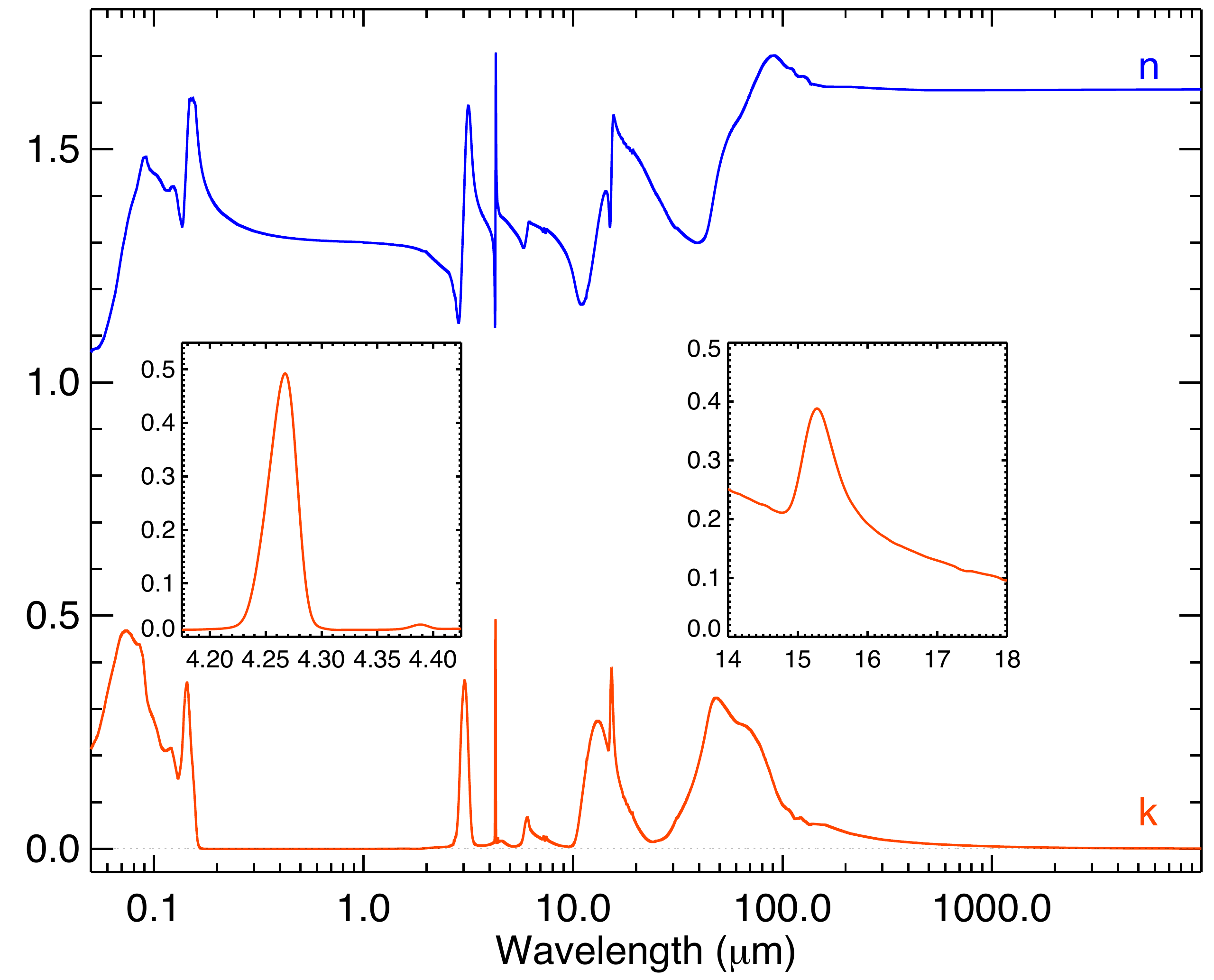}
\includegraphics[width=\columnwidth]{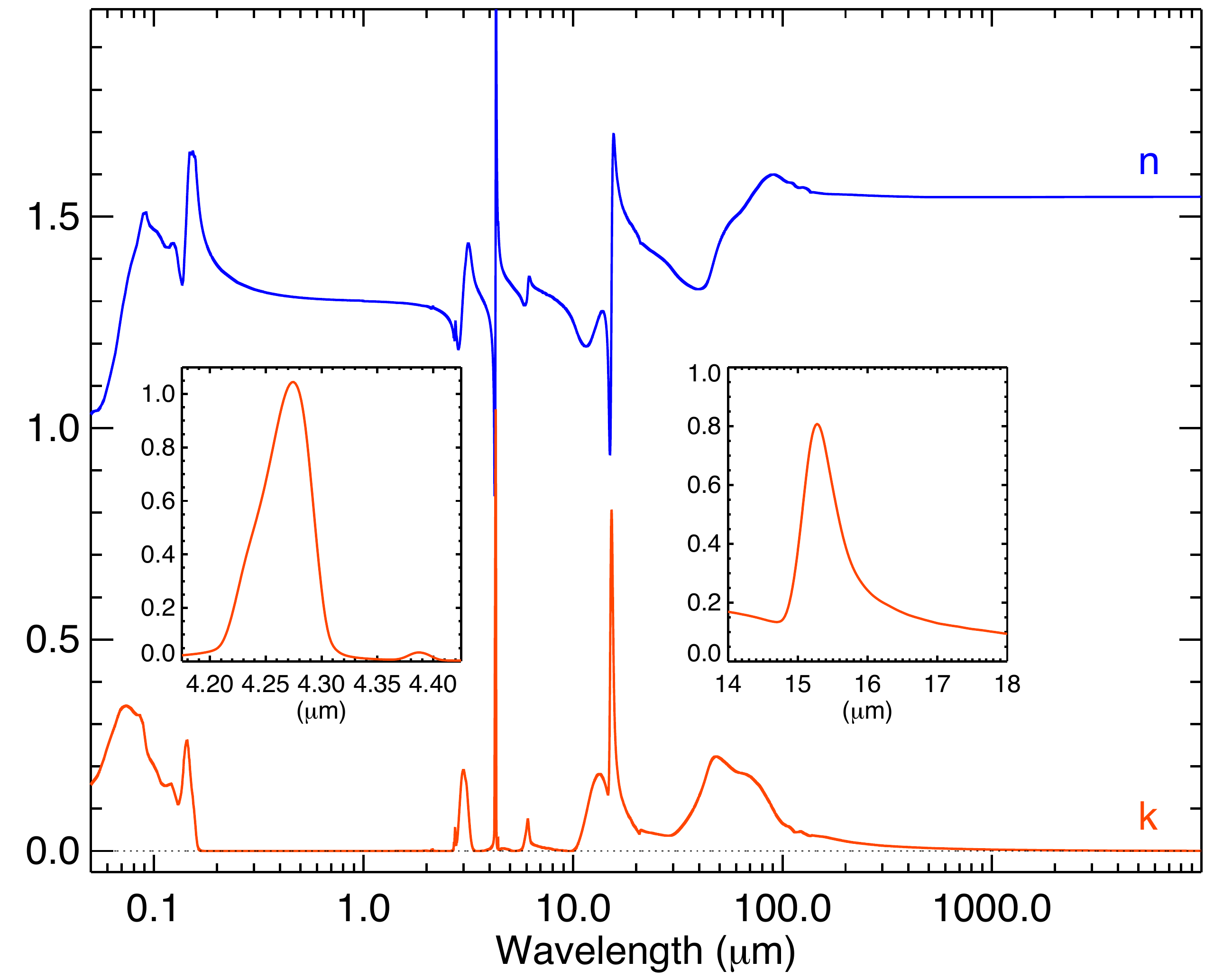}
\includegraphics[width=\columnwidth]{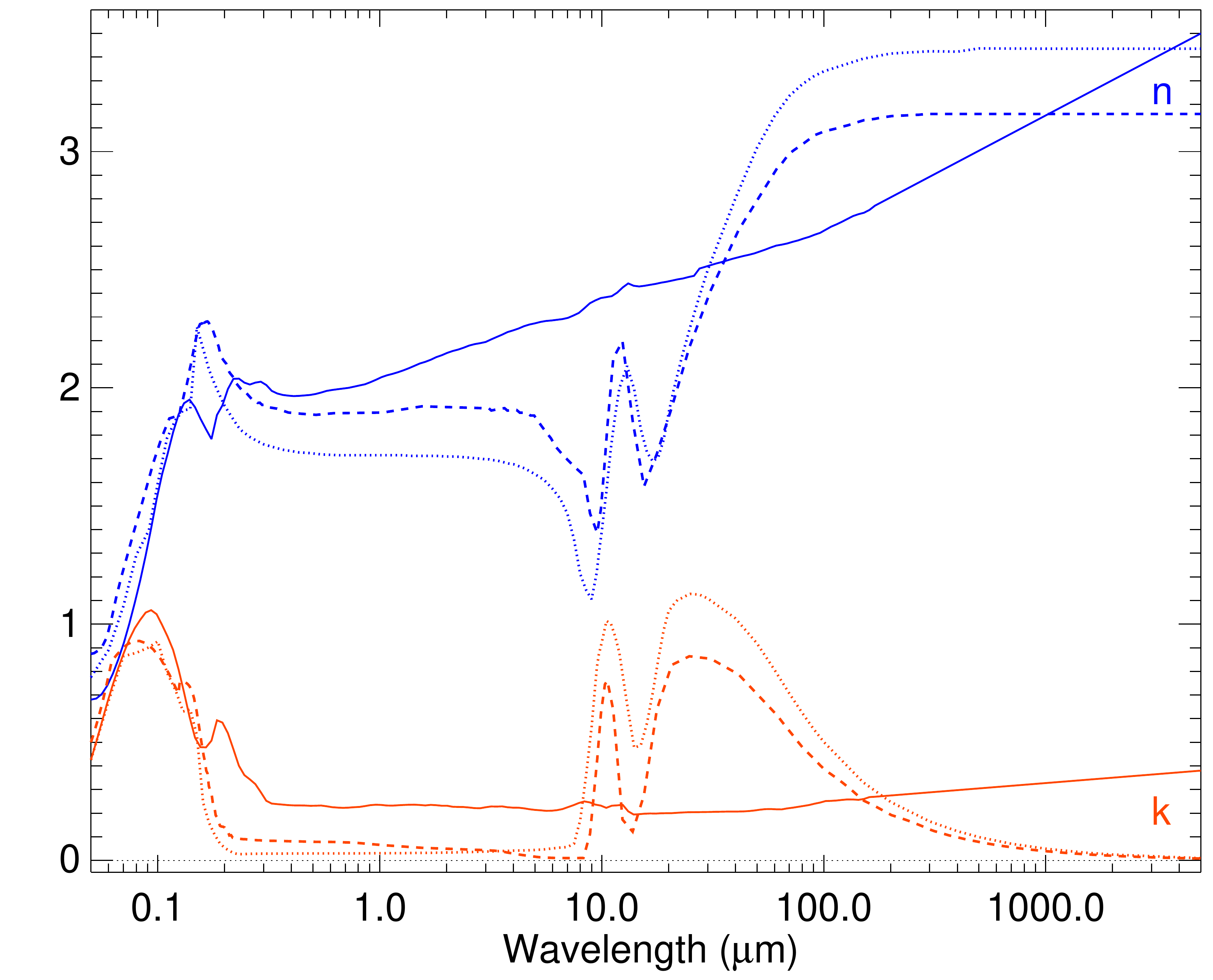}
\includegraphics[width=\columnwidth]{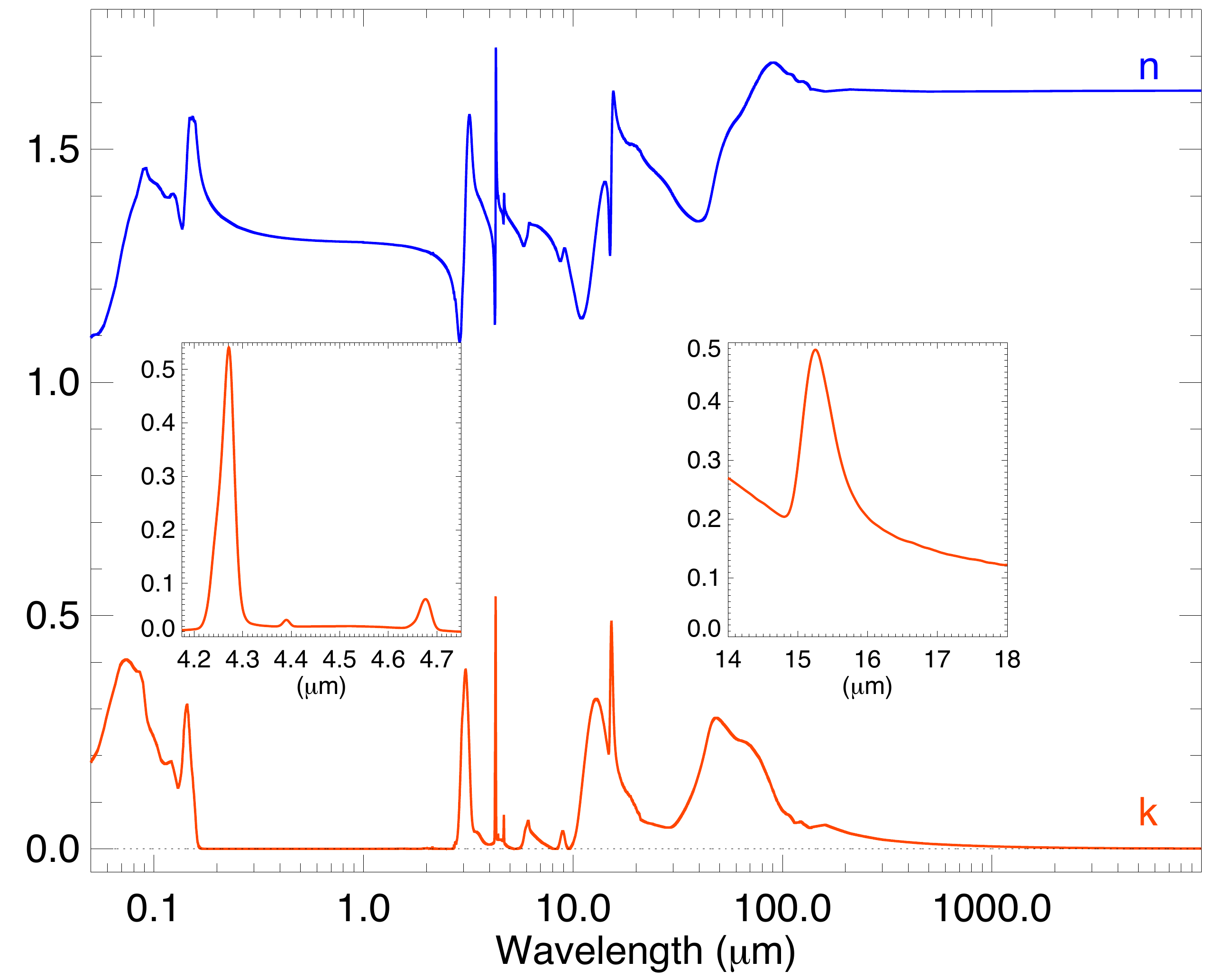}
  \caption{Upper left: Complex optical constants for the CO$_2$:H$_2$O 15:100 ice mixture (M15). Close-ups are shown for the imaginary part of the constant for the CO$_2$ stretching and bending modes. Upper right: Complex optical constants for the CO$_2$:H$_2$O 50:100 ice mixture (M50). Close-ups are shown for the imaginary part of the constant for the CO$_2$ stretching and bending modes. Lower left: complex optical constants for the "astrosilicates" \citep{Draine1984} and amorphous carbon AC1 from \cite{Rouleau1991}. Lower right: Complex optical constants for the H$_2$O:CO$_2$:CO:NH3 100:16:8:8 ice mixture. Close-ups are shown for the imaginary part of the constant for the CO$_2$ and CO stretching modes and CO$_2$ bending mode.
  See text for details.}
  \label{Fig_co}%
\end{figure*}
%

In the first stages of star formation, protostars are still embedded
in their parental cloud, where an active gas-grain chemistry is at
work. Either using background stars for dense clouds, a nascent protostellar object once it is able to emit sufficient
light flux in the vibrational infrared wavelength range, or in a few protoplanetary disks well inclined toward the observer,
the infrared pencil beam allows the probing of the composition of the cloud or circumstellar dust.
The low temperature ice mantles formed on top of or
mixed with refractory dust (silicates, organics) can be retrieved.  
A harvest of
astronomical observations from ground based (e.g. UKIRT, IRTF, CFHT, VLT)
or satellites (e.g. IRAS, ISO, Akari, Spitzer) of such lines of sight has led,
since the late seventies, to the deciphering of the chemical compositions, column
densities, and variations associated with these ice mantles \citep[e.g.][and references therein]{Boogert2015, Oberg2011, Boogert2008, Dartois2005, vanDishoeck2004, Gibb2004, Keane2001, Dartois1999, Brooke1999}. 
The interpretation of these observed spectra is
mainly based on their comparison with the infrared spectra of laboratory produced ice films of well
 controlled composition and cryogenic
temperatures \citep[e.g.][]{Hudson2021,Palumbo2020, Rachid2020, Terwisscha2018, Hudson2014, Oberg2007, Dartois2003, Dartois1999b, Dartois1999, Moore1998, Ehrenfreund1997, Gerakines1995, Hudgins1993}. The routes investigated are the influence of ice mixture
on the line width and position, temperature modifications, segregation
(phase separation), and/or intermolecular interactions (polar/apolar
ices, molecular complexes). Sometimes, the impact of a distribution of grain
shapes, mainly in the Rayleigh regime, is also explored. The literature is dominated by analyses based on the decomposition of the observed astronomical profiles into principal components from different ice mixtures.
When dust grains evolve from the diffuse interstellar medium to the dense phase and the protoplanetary phases, grains grow. It will affect the observed profiles.
It is expected to be, at least partly, responsible for enhanced scattering effects in dense cloud evolution, often referred to as cloudshine/coreshine effects \citep{Ysard2018, Saajasto2018,Ysard2016, Jones2016, Steinacker2015, Lefevre2014}.
{ The growth can also be guessed by the evolution of the silicate-to-K band ratio ($\tau_{9.7}/A_K$,  e.g. \cite{Madden2022, VanBreemen2011, Chiar2007}.}

It is already evidenced from direct spectroscopic profile evolution for silicates observed in emission coming from the surface of some disks \citep[e.g.][]{vanBoekel2005, Meeus2011}. Not only grain growth is important but also depletion of the smallest grains in the distribution.
Grain growth is a parameter that will take on an increasing importance for the interpretation of observed ice mantle
band profiles, especially in the study of protoplanetary disks \citep{Tazaki2021a,Tazaki2021b,Terada2017,Terada2007,Honda2009}.
The inventory of solid-state material sometimes combines ice feature investigation with radiative transfer codes to simulate the observed spectral energy distribution and/or chemical models \citep[e.g.][]{Pontoppidan2005, Ballering2021}.
The ice band spectroscopic profiles observed at medium to high spectral resolution intrinsically contain information to constrain the extent of, and are affected by, grain growth.
Carbon dioxide display several characteristics that are particularly interesting for probing grain growth. It is one of the main and ubiquitous ice mantle constituents, along with water and, depending on the line of sight, carbon monoxide.
The carbon dioxide stretching mode around 4.27 microns possesses a fairly narrow absorption band with a typical full width at half maximum (FWHM) of tens of cm$^{-1}$, depending on the exact ice mixture environment, \cite[e.g.][]{Ehrenfreund1996,Ehrenfreund1999}, whereas the water ice FWHM is several hundreds of cm$^{-1}$.
In addition to the relatively high contrast expected in the CO$_2$ ice profile due its narrowness, carbon dioxide absorbs in a relatively clean region of the infrared spectrum. 
For the absorption band of water ice centred at 3.1~$\mu$m, the red wing of the profile is modified not only by grain growth but also by additional absorption from e.g. methanol and the 3.47$\mu$m band assigned to the presence of ammonia in the water mantle.
The carbon monoxide stretching mode, lying at slightly higher wavelength than that of carbon dioxide, and also in a relatively clean region. { Some sources show a significant absorption at 4.62 $\mu$m, attributed to the presence of OCN$^-$, that can affect mainly the blue side of the CO absorption profile}. It has been investigated { and discussed} in \cite{Dartois2006}, where it was shown that grain growth to micron sizes can { still} produce an observed large red component in its absorption profiles toward some lines of sight. 
{ Some YSO spectra can also harbour hydrogen lines in emission, such as Pfund $\beta$ (4.654~$\mu$m) and Brackett $\alpha$ (4.051~$\mu$m), that have to be taken into account. Their contribution can either be estimated from the set of observed hydrogen lines and/or taken out of the profile analysis if significant for spectra with high enough spectral resolution given their small profile widths}.
The integrated absorption cross-section of the carbon dioxide band is relatively high, higher than carbon monoxide, an additional reason to make it a good target to look at how grain growth affects spectroscopic band profiles.\\
This article is dedicated to the prediction of the CO$_2$ ice mantle spectral profile behaviour expected for grain size distributions that have evolved, starting from the diffuse interstellar medium. We describe in \S~2 the ice mixtures and optical constant calculations used to build the ice mantle models. We discuss in \S~3 the dust grain shapes adopted to represent the diversity of shapes in the distribution, and in \S~4 the discrete dipole approximation method to evaluate the  absorption and scattering matrices for these grains. 
We apply the method to evolved dust grain size distributions resulting from previous literature models in \S~5.  In \S~6 the RADMC3D Monte Carlo radiative transfer model is used to calculate the emerging spectra from fiducial spherical dust clouds and a protoplanetary disk observed at various inclination angles along the line of sight, with a particular focus on the evolution of the CO$_2$ ice stretching band. Finally, in \S~7 we draw conclusions on the interpretation of principal component analysis of ice profiles and make predictions on grain growth constraints in the perspective of JWST observations.
\section{Experiments and methods}

In order to build spectroscopic profiles of ice mantles, the first modelling ingredients to define are interstellar 
relevant ice mixtures, as recorded in the laboratory, and deriving their optical constants. Then appropriate dust grains shape and size distributions are adopted and their absorption and scattering properties calculated. These steps are described below.

\subsection{Ice mixtures and optical constants calculations}

We use two binary water and carbon dioxide ice mixtures to explore the effect of a moderate to high CO$_2$ ice proportion.
These mixtures, called M15 and M50 are CO$_2$/H$_2$O low temperature amorphous mixed ices, with a carbon dioxide to water content of 15\% and 50\%, respectively. These values cover the CO$_2$ range observed towards most lines of sight, with 15\% being the closer to Massive Young Stellar Objects (MYSOs) or comets \citep[e.g., Fig.8][]{Boogert2015}, { whereas a higher CO$_2$ fraction can be observed towards low mass young stellar objects (LYSOs)} and 50\% represents a possible, while unusually CO$_2$-rich, mixture.

The set of ice optical constants used is built from ice film
laboratory experiments; a co-deposited CO$_2$/H$_2$O mixture measured
in the near to mid-infrared in our laboratory for the high CO$_2$ mixture 50\% (M50), while
the CO$_2$/H$_2$O 15\% ice mixture (M15) is from \cite{Ehrenfreund1996}.
Far-infrared water ice optical constants adopted are from \cite{Trotta1996}.
The millimeter and UV to visible optical constants are interpolated from pure
H$_2$O ice literature data \citep{Warren1984}. 
Real measurements for the same CO$_2$/H$_2$O mixtures over the full range would be, of course,
better but are not available. Such an extension will however have little influence on the
calculated profiles in the near to mid-infrared range, where the correct
optical constants for the mixed ices are used. They are extrapolated outside this range
in order to implement them into the radiative transfer model used in the 
final step of the analysis.
The scale of the imaginary part (k) of the complex refractive index is validated using 
\begin{equation} 
\rm A = \int_{band} 4 \pi k(\bar{\nu}) d\bar{\nu} M / (N_A \rho)
\end{equation} 
where A is the integrated band strength (cm.molec$^{-1}$), M the molar mass, $\rm N_A$ the Avogadro number and $\rm\rho$ the density of the ice, by checking that it falls within the range of expected band strengths. In the M15 mixture, the estimated water ice stretching mode band strength is about 1.9$\times$10$^{-16}$cm.molec$^{-1}$, whereas for the M50 mixture it is about 1.6$\times$10$^{-16}$cm.molec$^{-1}$, in agreement with what is expected \citep[e.g. Fig.4 of][]{Oberg2007}.
Self consistency for the real and imaginary components of the optical constants is ensured by calculating
the refractive index from a Kramers-Kronig transformation of the
imaginary part of the complex index, rescaled to the visible real
part of the index assumed to be 1.3, a typical value for H$_2$O ice, and also close to that of many ices of
astrophysical interest \citep{Trotta1996, Satorre2008}. 

The real part of the complex refractive index, n, is calculated using the Kramers-Kronig integral dispersion relation, related to the imaginary part k by
\begin{equation}
\rm n(\bar{\nu}) = n(\infty) + \frac{2}{\pi} \; \dashint_0^{\infty} 
\frac{\bar{x}\;k(\bar{x})} {\bar{x}^2 - \bar{\nu}^2} d\bar{x}
\end{equation}
To calculate numerically n over a finite frequency interval $\rm\{\bar{\nu}_{min};\bar{\nu}_{max}\}$, the subtractive Kramers-Kronig relation is preferred
\begin{equation}
\rm n(\bar{\nu}) = n(\bar{\nu}_0) 
+ \frac{2}{\pi} \; \dashint_{\bar{\nu}_{min}}^{\bar{\nu}_{max}}
\left( \frac{\bar{x} k(\bar{x}) - \bar{\nu} k(\bar{\nu})} { \bar{x}^2-\bar{\nu}^2 } -
 \frac{\bar{x} k(\bar{x}) - \bar{\nu}_0 k(\bar{\nu}_0)} { \bar{x}^2-\bar{\nu}_0^2} \right)
d\bar{x} 
\end{equation}
using the anchor point at frequency $\bar{\nu}_0$, away from strong absorptions, with a known value $\rm n(\bar{\nu}_0)$.
For ices, generally, as stated above, a value in the visible domain where most of them do not absorb significantly, is adopted. { Our spectra used in the optical constant derivation were recorded in transmittance at normal or close to normal incidence. In the absence of a dedicated experiment to record simultaneously the refractive index, the infrared spectrum and the ice density, and an experiment designed specifically for a complete optical constants inversion process, the absolute values of k will vary slightly with the refractive index and scale with the density values. Uncertainties for the derived optical constants in the main mid-infrared bands, the core of the analysis in this article,} are conservatively estimated to be below 20\%.\\
The refractory material is assumed to be represented by the so-called ''astronomical silicates'' optical constants from \citep{Draine1984}.
The adopted pure silicate cores and one ice mixture for each model is a simplification over all the possible ice mixtures and core compositions. This consideration was made based on main components, and to avoid mixing the effect of too many parameters in the resulting comparisons. 
In order to compensate for this potential oversimplification, we expand our test set with two additional models.
One includes a possible additional pure carbon monoxide component in the ice mantle, as, towards some lines of sight, at high visual extinction, the condensation of pure CO onto the mantle has been observed \citep[e.g.][]{Pontoppidan2006}, i.e. CO not mixed with the other components in the ice mantle.
For this model we take the optical constants from \cite{Palumbo2006}.
The interstellar dust distribution in the ISM is comprised of siliceous and carbonaceous components. We thus also explore a model with a possible refractory core mixture including amorphous carbon with optical constants taken from \cite{Rouleau1991}.
%

\begin{figure}
  \centering
\includegraphics[width=\columnwidth]{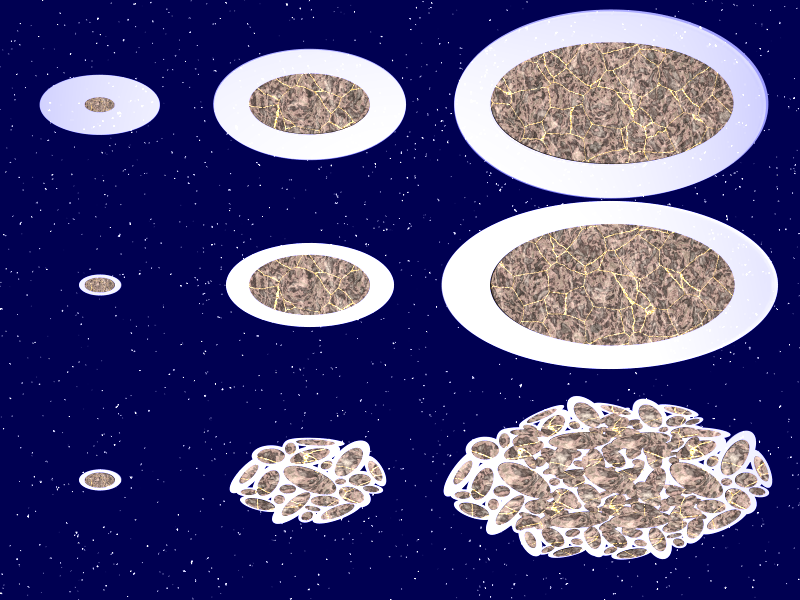}
  \caption{Sketch of the different ice distributions with respect to refractory grains during grain growth addressed in the modelling: (i) constant ice thickness, (ii) constant ice/core volume ratio (QCDE), (iii) stochastic mixing (CSDE).}
  \label{schema}%
\end{figure}
\begin{figure*}[h]
  \centering
\includegraphics[width=2\columnwidth]{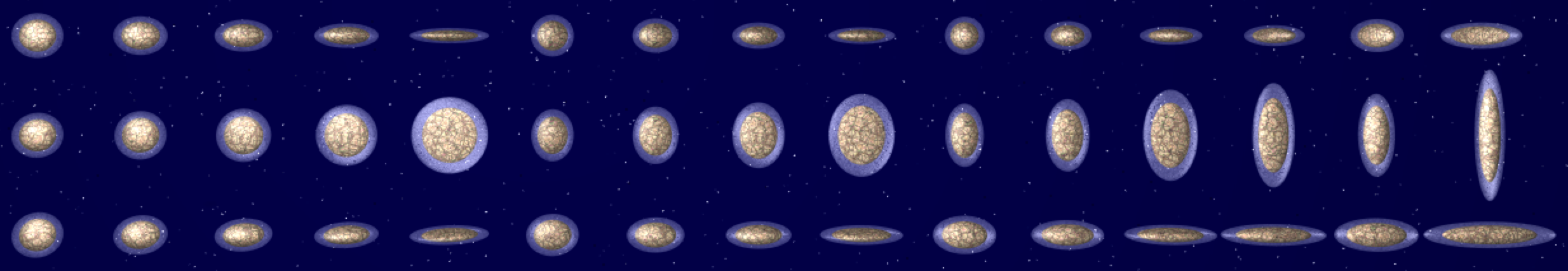}
  \caption{3D view of the fifteen ellipsoid shapes used in the calculation of the adopted distribution of ice mantle coated ellipsoids, with a quadratic weighting scheme, as used in \cite{Dartois2006}, also called CDE2 \citep[e.g.][]{Ossenkopf1992} and presented along their three main axes, as in Fig. 3 of \cite{Draine2017}. The probability of their occurrence (weighting) in the distribution is the same as in \cite{Dartois2006}.}
  \label{Fig_ellipsoides}%
\end{figure*}

\subsection{Dust grain shape distribution}
The exact shape distribution of interstellar grains is not known, but from polarisation considerations is known not to be represented adequately by pure spheres. To sample the expected diversity, several shape distributions may be adopted \citep[e.g.][]{Fabian2001,Min2003}.
Among the most convenient shapes are ellipsoids for their mathematical properties, as discussed in, e.g., \cite{Draine2017}. In the possible continuous distributions of ellipsoidal shapes, one of the most popular is the uniform weighting (named CDE or CDE1), where any shape has an equal probability of occurring. With such a weighting scheme, extreme shapes that have a nonphysical probability of being present in space, such as when the ellipsoid tends to be like an infinite rod or to be planar, have the same weight as more compact or spheroidal shapes. This is unrealistic and has led to the adoption of a quadratic weighting scheme, used in \citep[e.g.][]{Draine2017, Dartois2006, Fabian2001, Ossenkopf1992} under the name of QCDE or CDE2. This distribution explores a variety of ellipsoidal shapes, but with such a weighting scheme, due to the interdependence of geometrical factors, that if one axis of the ellipsoid is far from the others its occurrence, and thus contribution to the distribution, drops. An ellipsoid with axis ratios commensurate to 5:1:1 will have a probability of about 0.33 with respect to a sphere, and an ellipsoid commensurate to 100:1:1, i.e. a needle, or a 100:100:1, i.e a disc, would contribute as little as 0.0028 and 0.0011, respectively.
%
\subsection{Ice distributions}
We can consider three different scenarios of growth and acquirement of an ice mantle, that will define the distribution of ices with respect to the refractory material in the grains (the representations of such models are shown in Fig.\ref{schema}):\\
(i) the simplest model assumes the impact of molecules, atoms and/or radicals on the initial ISM grain distribution builds the ice mantle from the gas phase.
Therefore the grain volume growth due to the ice mantle is given by:
\begin{equation}
\rm{4}\pi\rm{a}^2\dot{a}=\pi\rm{a}^2\sum_{i=1}^{N}\rm{n}_{i}\;\frac{\rm{m}_{i}}{\rho}\;\rm{v}_{i}\;s_i
\label{eq_growth}
\end{equation}
where a is the grain radius, $\rm{n}_{i}$ and $\rm{m}_{i}$ are the
density and mass of impinging molecules or nuclei with velocity
$\rm{v}_{i}$, $\rho$ the mass volume density of the accreting mantle,
$s_i$ is an ``efficiency factor'' which includes the sticking
coefficient of the impactor, its reaction rate, etc, (the ratio
$\rm s_i\,\rm{m}_{i}$/$\rho$ is the volume increase per impact).  
Equation \ref{eq_growth} shows that the grain radius growth $\rm\dot{a}$ is independent of the grain radius. As a consequence, the acquired mantle thickness is constant on each grain size, and the ice to core volume ratio will be very high for the small grains and very low for the big grains in the distribution. Extending this to bigger grains when the upper size end of the distribution increases due to grain aggregation is equivalent to assuming that ice growth proceeds only after refractory core aggregation is fully completed. We do not model this distribution here, as it is probably the least physical among the growth properties. In addition we expect only mild spectral changes for the ice features with respect to an MRN distribution, as most of the ice volume is carried, within such an assumption, by the smallest grains (see, e.g. Fig. 2 from \cite{Dartois2006}).\\ 
(ii) the second model assumes an ellipsoid refractory core coated by the ice mantle, ascribing each grain an individual constant Vice/Vcore ratio equivalent to the mean value observed. Such a model 
can be interpreted as the continuous aggregation of ice mantle/refractory grains accompanied by, if ices are mobile enough e.g. upon energetic events, a progressive settling of the refractory core mainly inside the grain, migrating partly to the surface. This distribution is called the quadratic continuous distribution of ellipsoids, hereafter QCDE.\\
(iii) the third scenario is a stochastic sticking of the initial ISM dust grain distribution, each coated with an ice mantle, aggregating into a bigger grain. This distribution is called hereafter the compact stochastic distribution of ellipsoids (CSDE).
These differences in ice distributions will have consequences on the resulting ice spectral features. We consider only QCDE and CSDE in the following models.
%
\begin{table*}[htp]
\caption{Summary of model parameters in this work}
\begin{center}
\begin{tabular}{llcccccccccccc}
\hline
Set    &Model	&\multicolumn{2}{c}{Ice mantle mixture} 	&\multicolumn{4}{c}{Ice mantle mixture}		&\multicolumn{4}{c}{Distribution}			&\multicolumn{2}{c}{Aggregation type}	\\
    &\#		&Silicates 		&Amorphous carbon		&M15		&M50	&pure CO &MX					&MRN		&$\tau_1$		&$\tau_3$   &$\tau_5$		&QCDE		&CSDE\\
\hline
\arrayrulecolor{lightgray}
\ldelim\{{3}{5pt}[{\it A$\;\;$}] &1		&\checkmark	&	&\checkmark	&			& & &\checkmark	&			&			&   &\checkmark	&\\ \hdashline
    &2		&\checkmark	&   					&\checkmark	&	&           & &			&\checkmark	&			& &\checkmark	&\\ \hdashline
    &3		&\checkmark	&					&\checkmark	&			& &&			&			&\checkmark	& &\checkmark	&\\ \hline
\ldelim\{{4}{5pt}[{\it B$\;\;$}]     &4		&\checkmark	&					&\checkmark	&&			&&\checkmark	&			&			&			&   &\checkmark	\\ \hdashline
    &5		&\checkmark	&					&\checkmark	&&			&&			&\checkmark	&			&			&   &\checkmark	\\ \hdashline
    &6		&\checkmark	&					&\checkmark	&&			&&			&			&\checkmark	&			&   &\checkmark	\\ \hdashline
    &7		&\checkmark	&					&\checkmark	&&			&&			&			&   &\checkmark	&			  &\checkmark	\\ \hline
\ldelim\{{3}{5pt}[{\it C$\;\;$}]     &8		&\checkmark	&					&			&\checkmark	&&&\checkmark	&			&			& &\checkmark	&\\ \hdashline
    &9		&\checkmark	&					&			&\checkmark	&&&			&\checkmark	&			& &\checkmark	&\\ \hdashline
    &10		&\checkmark	&					&			&\checkmark	&&&			&			&\checkmark	& &\checkmark	&\\ \hline
\ldelim\{{3}{5pt}[{\it D$\;\;$}]     &11		&\checkmark	&					&			&\checkmark	&&&\checkmark	&			&			&			&   &\checkmark	\\ \hdashline
    &12		&\checkmark	&					&			&\checkmark	&&&			&\checkmark	&			&			&   &\checkmark	\\ \hdashline
    &13		&\checkmark	&					&			&\checkmark	&&&			&			&\checkmark	&			&   &\checkmark	\\ \hline
\ldelim\{{3}{5pt}[{\it E$\;\;$}]     &14		&\checkmark	&\checkmark			&\checkmark	&&			&&\checkmark	&			&			&			&   &\checkmark	\\ \hdashline
    &14		&\checkmark	&\checkmark			&\checkmark	&&			&&			&\checkmark	&			&			&   &\checkmark	\\ \hdashline
    &16		&\checkmark	&\checkmark			&\checkmark	&&			&&			&			&\checkmark	&			&   &\checkmark	\\ \hline
\ldelim\{{3}{5pt}[{\it F$\;\;$}]     &17		&\checkmark	&					&\checkmark	&&\checkmark			&	&\checkmark	&			&			&			&   &\checkmark	\\ \hdashline
    &18		&\checkmark	&					&\checkmark	&&\checkmark			&	&			&\checkmark	&			&			&   &\checkmark	\\ \hdashline
    &19		&\checkmark	&					&\checkmark	&&\checkmark			&	&			&			&\checkmark	&			&   &\checkmark	\\ \hline
    \ldelim\{{3}{5pt}[{\it G$\;\;$}]     &20		&\checkmark	&					& &	&			&\checkmark	&\checkmark	&			&			&			&   &\checkmark	\\ \hdashline
    &21		&\checkmark	&					& &	&			&\checkmark	&			&\checkmark	&			&			&   &\checkmark	\\ \hdashline
    &22		&\checkmark	&					& &	&			&\checkmark	&			&			&\checkmark	&			&   &\checkmark	\\
\arrayrulecolor{black}
\hline
\end{tabular}
\end{center}
\label{summary_models}
\end{table*}%
%

\begin{figure}
  \centering
\includegraphics[width=\columnwidth]{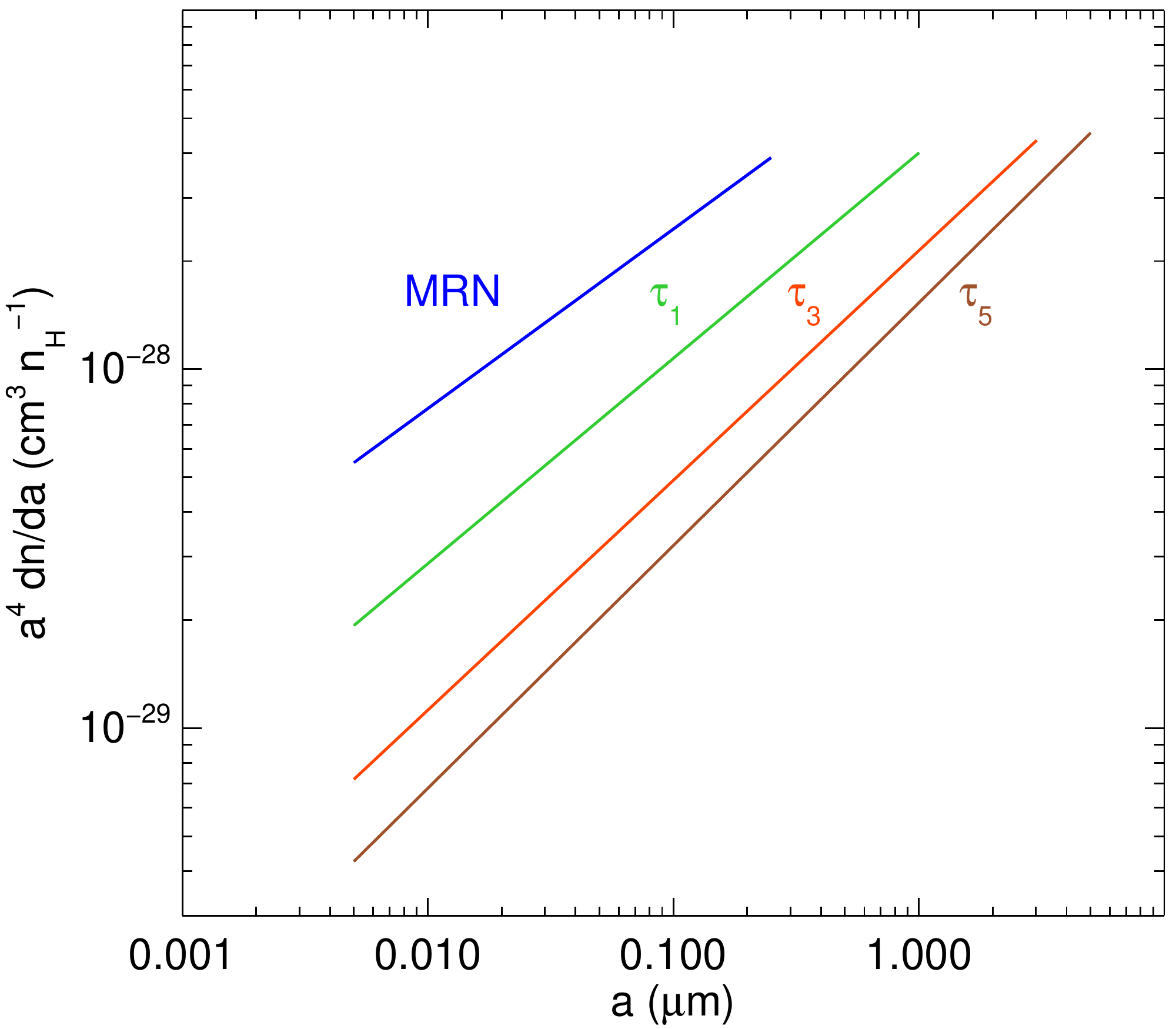}
  \caption{Size distributions reproducing qualitatively the expected size evolution behaviour for dust grain growth in aggregation models. The lower bound of the grain sizes is kept constant at 0.005$\mu$m. The upper bound of the distribution is varied from 0.25$\mu$m (MRN distribution) to 1$\mu$m ($\tau_1$) and 3$\mu$m ($\tau_3$), and the slope defined to maintain the same total mass. See text for details.}
  \label{Fig_size_evolution}%
\end{figure}

\begin{figure*}
  \centering
\includegraphics[width=0.66\columnwidth]{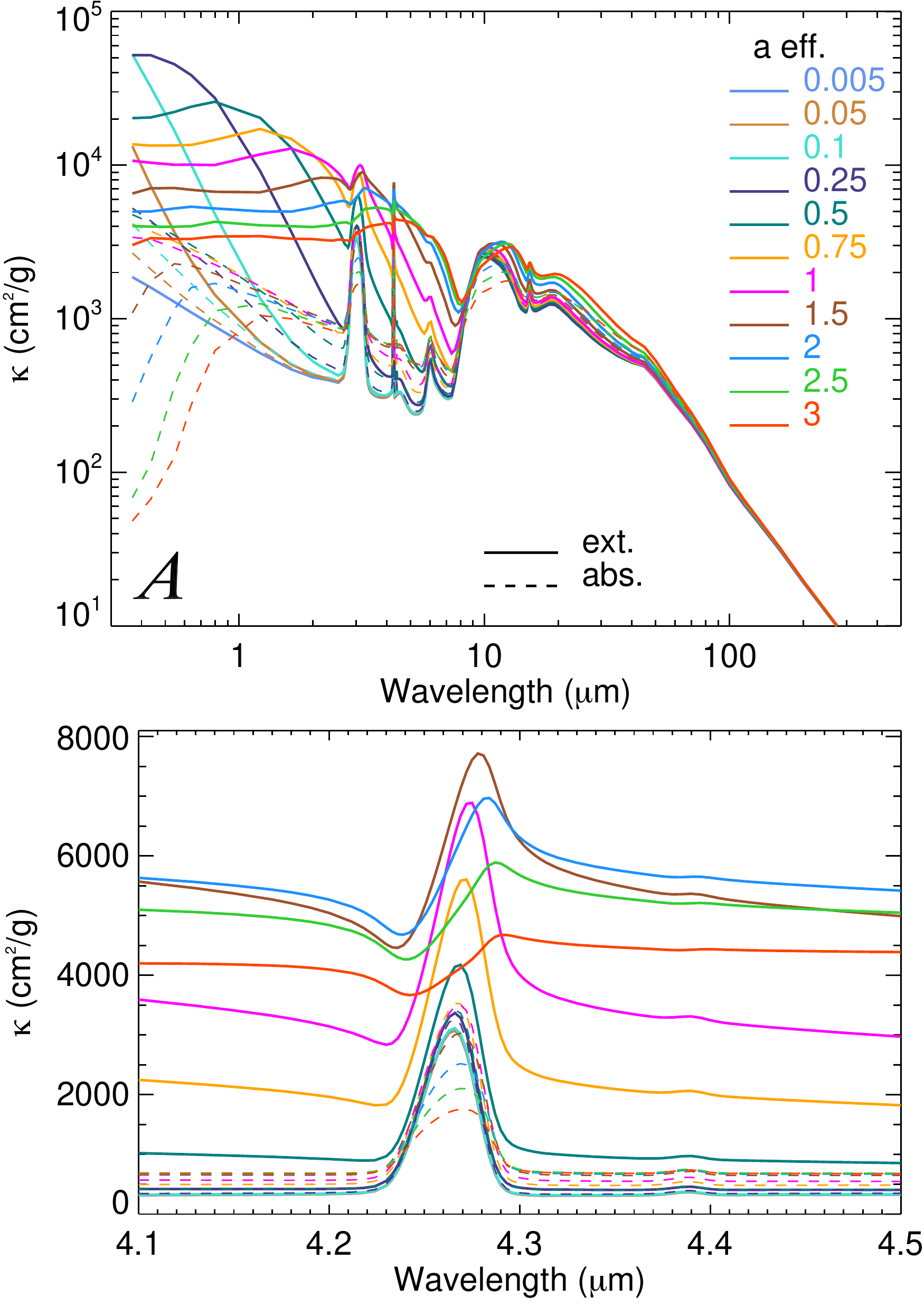}
\includegraphics[width=0.66\columnwidth]{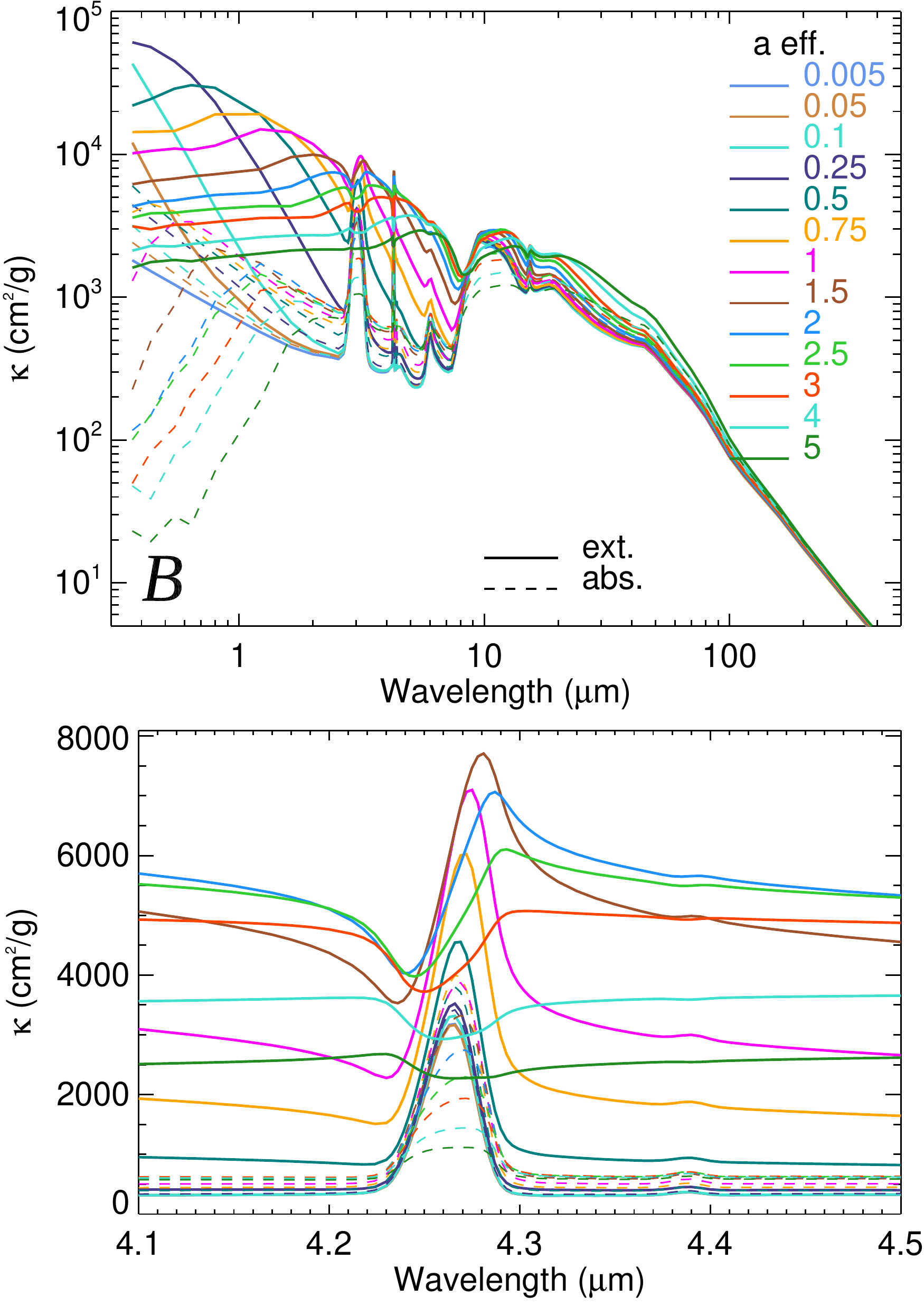}
\includegraphics[width=0.66\columnwidth]{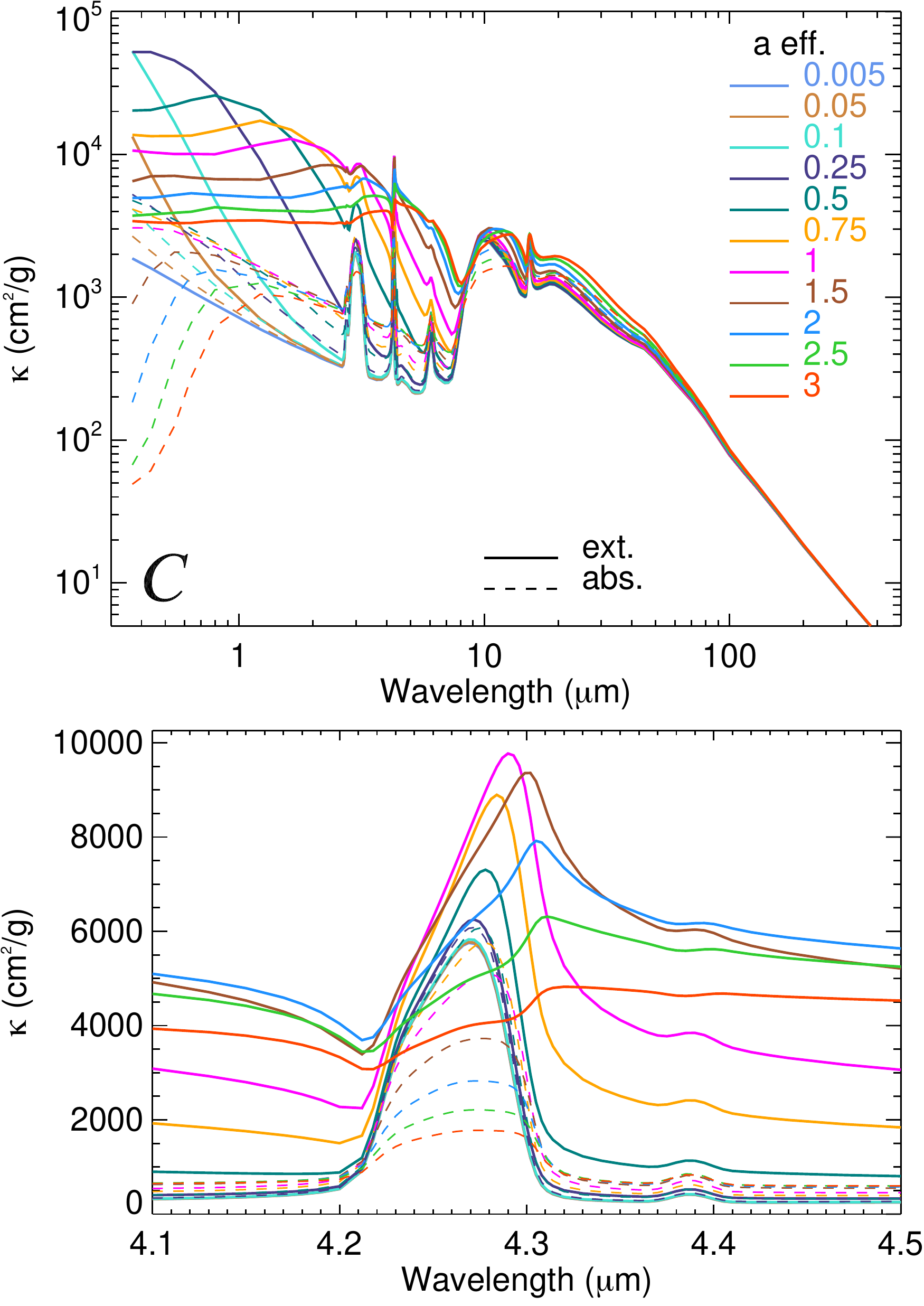}
\includegraphics[width=0.66\columnwidth]{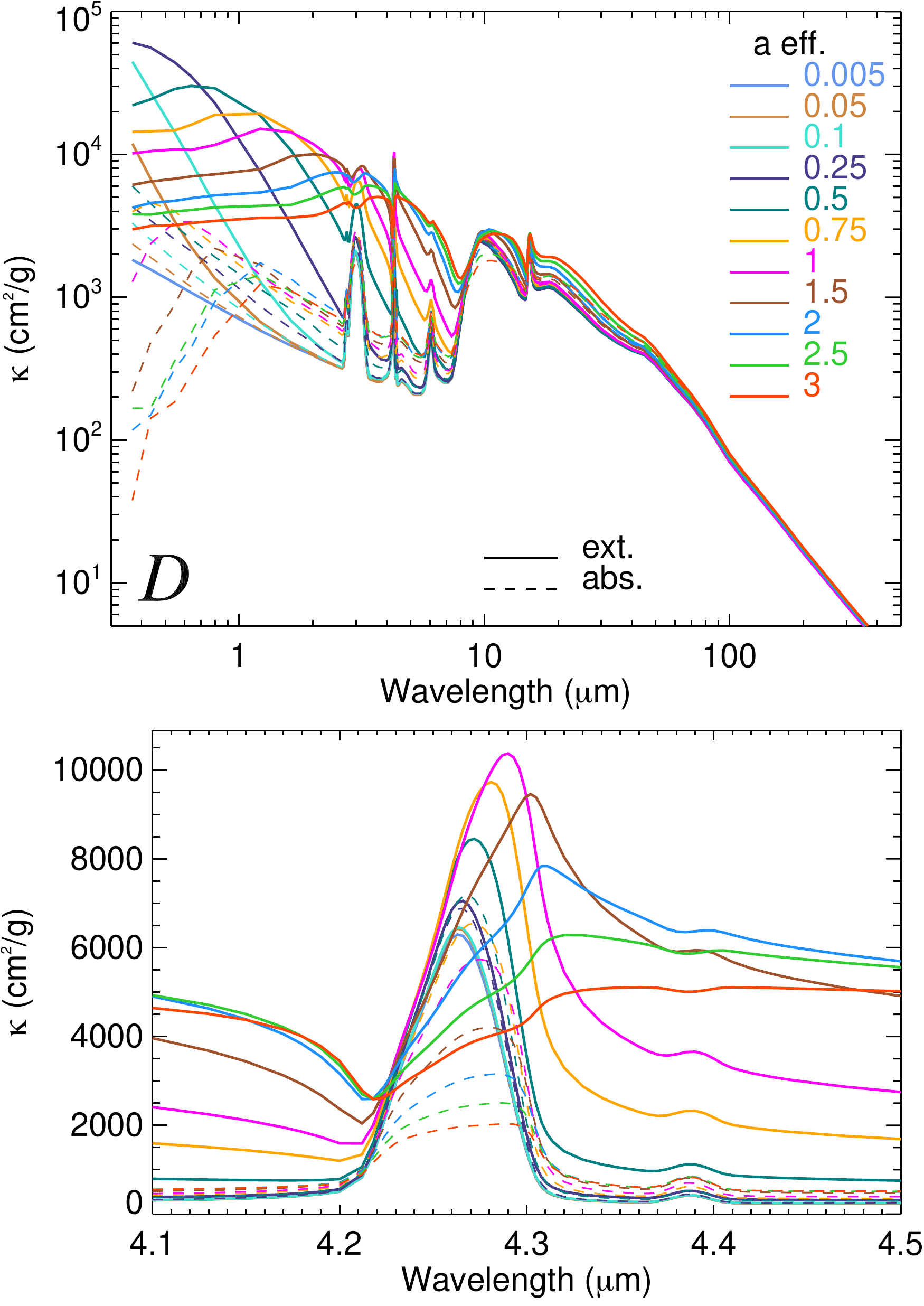}
\includegraphics[width=0.66\columnwidth]{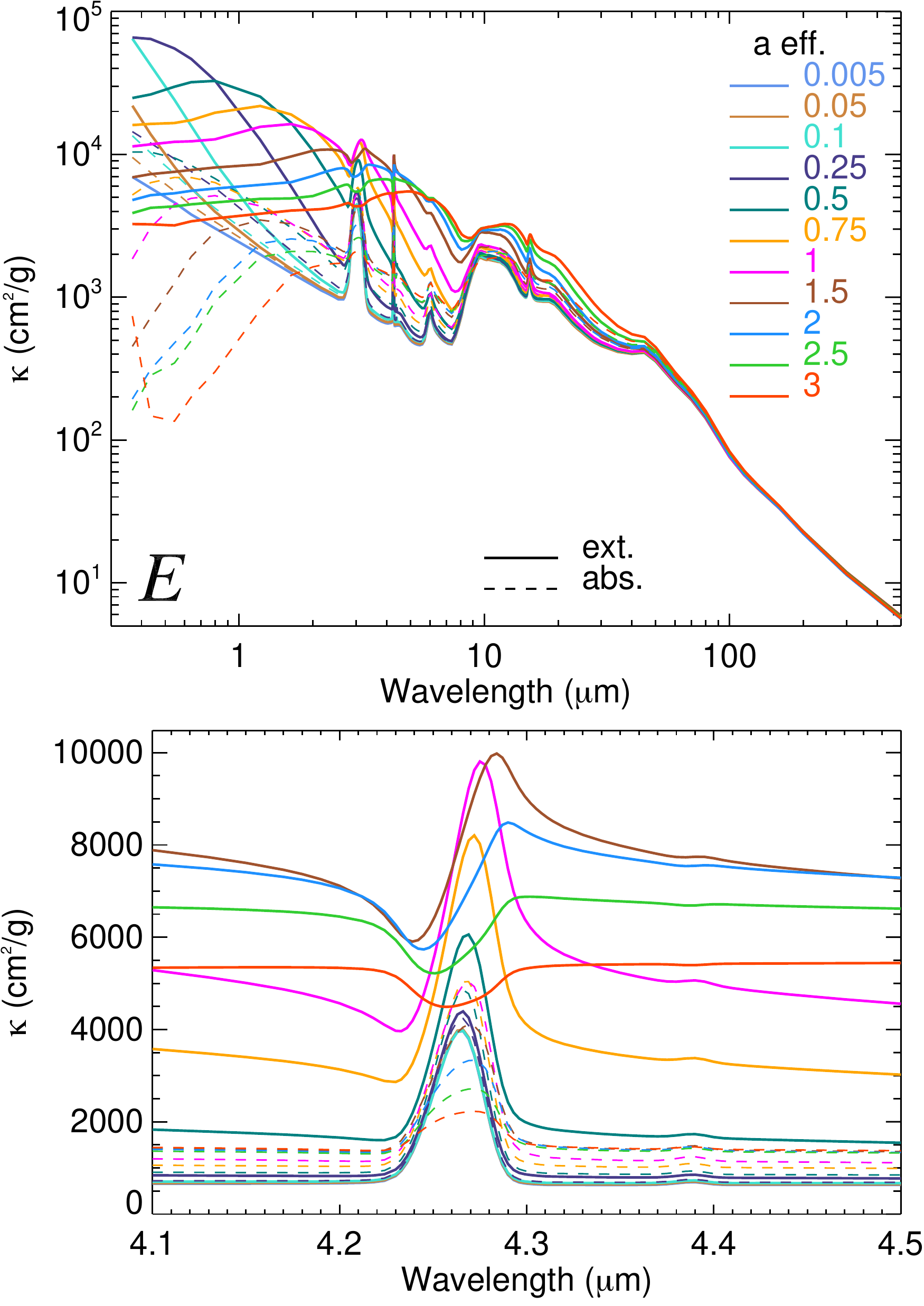}
\includegraphics[width=0.66\columnwidth]{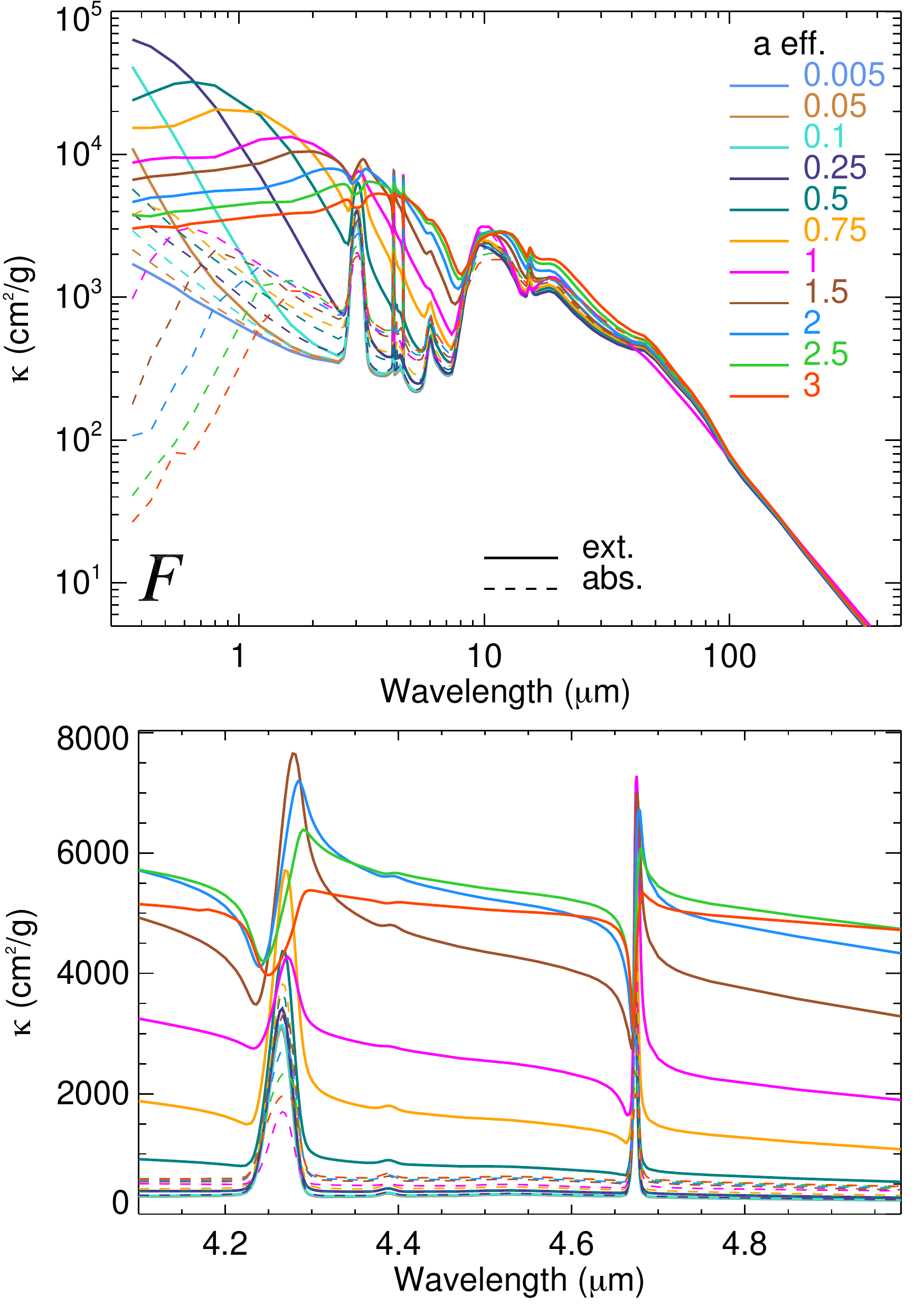}
\begin{minipage}{0.66\columnwidth}
\includegraphics[width=\columnwidth]{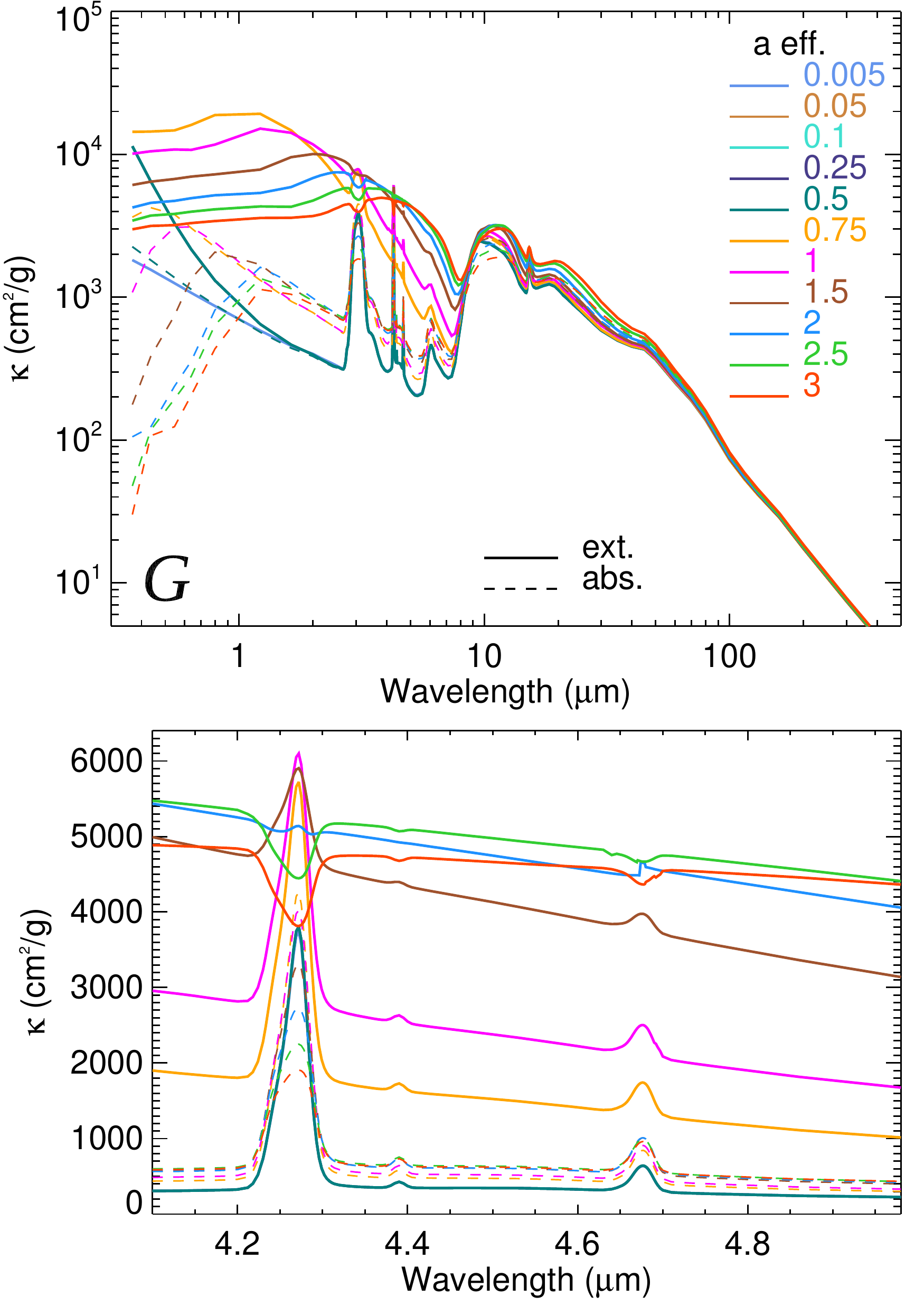}
\end{minipage}
\begin{minipage}{0.66\columnwidth}
$\;$
\end{minipage}
\begin{minipage}{0.66\columnwidth}
  \caption{Grain size dependent mass extinction coefficients (solid lines), absorption (dashed line) coefficients, from upper left to lower right for models composition corresponding to model sets {\it A, B, C, D, E, F, G}, respectively. { The mass extinction coefficient for increasing effective grain radius a eff are shown. See text for details.
  For each set, a close-up of the CO$_2$ stretching mode (as well as including CO for the last two panels) is shown below.}}
\end{minipage}
\begin{minipage}{0.165\columnwidth}
$\;$
\end{minipage}
  \label{Fig_MAC_individuels}%
\end{figure*}

\subsection{Discrete Dipole Approximation calculations}
To model an individual ice mantle coated refractory ellipsoid grain, or mixed ice/refractory grains, we make use of the Discrete Dipole Approximation (DDA) program DDSCAT \citep[version 7.3,][]{Draine2013,Draine2008,Draine2000}.
The QCDE model calculations are performed with a similar scheme as presented in the appendix of \cite{Dartois2006}, i.e. over all possible ellipsoids with relative non-degenerated integer aspect ratios of their three axes, between 1 and 5, having fixed the first axis to the highest integer. This provides 15 possibilities in total, that are shown in Fig.\ref{Fig_ellipsoides}. The orientation average of the cross section for each ellipsoid, is performed over 16 angles equally sampling the relative orientation of individual ellipsoids with respect to the incoming plane wave.
This calculation is performed over two incident orthogonal polarisations, then averaged over angles orientations, for grain sizes covering the range of our distributions (described in the following), sampled from 0.005~$\mu$m up to 3~$\mu$m (at eleven sizes). The number of dipoles used in the calculations is dependent on two parameters: the first one is the overall grain size, in order to set constraints on the code precision, and the second one is to satisfy enough accuracy of the modeled ice mantle to silicates core volume ratio.
The number of dipoles for  0.005~$\mu$m grains is as low as 15000, whereas above 0.5~$\mu$m it is up to 64000.
We calculate the absorption and scattering properties, as well as the Mueller scattering matrices for each ellipsoidal ice coated grain, with an equivalent effective radius { a eff, the radius of a sphere corresponding to the same volume} corresponding to each grain size.
Once these calculations have been made, they are used to build the mass absorption and scattering coefficient for a prescribed dust grain size distribution, that we discuss in the grain growth modelling section.\\
For the CSDE model we stochastically generate dipoles with the refractory or ice composition attribute in the volume of the desired shape, with relative proportions corresponding to their volume ratios, and proceed as for the QCDE model for the remaining calculations. 
This CSDE stochastic mixing aims at simulating the aggregation of many small ice coated grains. This is formally not the exact situation, as describing the aggregation of thousands of small ice coated grains with a fixed mantle to core volume would require a dramatically higher number of DDA dipoles. The mixing described by the model is only an approximation to this. Neither is this a classical effective medium theory approach when one phase is diluted in another matrix phase. Given the high ice to core volume ratio adopted, we are definitely not in a regime of domination of one volume component over the other and we are above the percolation threshold, there are very few, if any, embedded isolated dipoles with one component properties. This model thus represents a fair compromise to simulate the aggregation of many small ice coated grains.
For CSDE, the number of dipoles can be as low as 2000 because of the relaxed condition on the core-mantle shape.
This saves calculation time without compromising the required accuracy. The error tolerance is set to better than 10$^{-5}$, except for the first wavelength points in the UV-visible where it is better than 10$^{-2}$, again to speed up the calculations. The accuracy in the UV-visible is thus lower, but high enough to serve for the purpose of radiative transfer calculations.
Each data set required about a week (CSDE) to several weeks (QCDE) of calculation time.

%

\section{Models}
\subsection{Ice mantle to core volume ratio}

The volume ratio between ice mantles and the refractory dust grain core \cite[e.g.,][]{Dartois2006} is given by:
\begin{equation}
\frac{\rm{V}(\rm{ice})}{\rm{V}(\rm{sil})}
=
\left( 
\frac {\rm{M}_{\rm{ice}}} {\rho_{\rm{ice}}}
{\rm{N}_{\rm{H}_2\rm{O}}}
\right)/
\left(
\frac {\rm{M}_{\rm{sil}}} {\rho_{\rm{sil}}}
\frac {\rm{N}_{\rm{Si}}} {\rm{n}_{\rm{Si}}}
\right)
\end{equation}
$$
\rm{with}\;\;
{\rm{N}_{\rm{H}_2\rm{O}}} \approx
\frac{\tau_{\rm{ice}}\Delta\nu_{\rm{ice}}}{\rm{A}_{\rm{ice}}}
\;\;\;\rm{and}\;\;\;
{\rm{N}_{\rm{Si}}} \approx
\frac{\tau_{\rm{sil}}\Delta\nu_{\rm{sil}}}{\rm{A}_{\rm{sil}}}
$$ 
where $\rm{M}_{\rm{ice}}$, $\rho_{\rm{ice}}$, $\rm{M}_{\rm{sil}}$, and
$\rho_{\rm{sil}}$ are the molar mass (g/mol) and density (g/cm$^3$)
for water ice and silicates, respectively.  $\rm{N}_{\rm{H}_2\rm{O}}$
and $\rm{N}_{\rm{Si}}$ are the water molecules and silicon atoms in
the observed silicates column density,
respectively. $\rm{n}_{\rm{Si}}$ is the number of silicon atoms in a
mole of a given silicate. $\tau_{\rm{ice}}$, $\Delta\nu_{\rm{ice}}$,
$\rm{A}_{\rm{ice}}$, $\tau_{\rm{sil}}$, $\Delta\nu_{\rm{sil}}$, $\rm{A}_{\rm{sil}}$, are the line
center optical depth, line full width at half maximum (cm$^{-1}$) and
integrated absorption cross section (cm/molecule) of ice and silicates.
$\rm{M}_{\rm{sil}}/(\rho_{\rm{sil}}\rm{n}_{\rm{Si}})$ is
between about 30 and 40~cm$^3$/mol/Si for magnesium-rich silicates
\citep{Reddy2005} and $\rm{M}_{\rm{H}_2\rm{O}}/\rho_{\rm{H}_2\rm{O}}$
is about 19.6~cm$^3$/mol for non-porous ice.  Using
$\Delta\nu_{\rm{H}_2\rm{O}}(\rm{3}\mu\rm{m})\approx$300~cm$^{-1}$,
$\rm{A}_{\rm{H}_2\rm{O}}(\rm{3}\mu\rm{m})=$2.10$^{-16}$cm/molecule
\citep{Oberg2007,Gerakines1995,ldh1986,Hagen1981} and
$\Delta\nu_{\rm{sil}}(\rm{10}\mu\rm{m})\approx$300~cm$^{-1}$,
$\rm{A}_{\rm{sil}}(\rm{10}\mu\rm{m})\approx$1.6-2.10$^{-16}$cm/molecule.

\begin{equation}
f_{\rm{vol}} = 
\frac{\rm{V}(\rm{ice})}{\rm{V}(\rm{sil})}
\approx (0.4\sim0.66) \frac{ \tau_{\rm{ice}} } {\tau_{\rm{sil}} }
\end{equation}

The water ice and silicates optical depths are evaluated with respect to A$\rm_V$ towards many lines of sights, or directly as a ratio in some cases. \cite{Bowey2004} and \cite{Rieke1985} evaluate $\rm{A}_{V}/\tau(\rm{silicates})\approx19.2$ and $\rm{A}_{V}/\tau(\rm{silicates})\approx16.6$, respectively. \cite{Murakawa2000} gives $\tau(\rm{H}_2\rm{O}\;\rm{ice})\approx0.067\;\rm (\rm{A}_{V}-\rm{A}_{V\;threshold})$, whereas \cite{Whittet1988} gives $\tau(\rm{H}_2\rm{O}\;\rm{ice})\approx0.093\;\rm (\rm{A}_{V}-\rm{A}_{V\;threshold})$ and $\rm{A}_{V}/\tau(\rm{silicates})\approx12-25$.
\cite{Brooke1999, Brooke1996}.
 \cite{Boogert2011} evaluates $\tau(\rm{silicates})=0.36\pm0.09+0.36\pm0.06 \;\tau(\rm{H}_2\rm{O})$.
Combining the various possibilities, we get ${ \tau_{\rm{ice}} }/ {\tau_{\rm{sil}} }\approx 0.8-2.5$, and deduce
$$
f_{\rm{vol}}\approx 1\pm0.5
$$
This value establishes a lower limit to the
ice-to-refractory mantle volume ratio, as one should include all ices
and recall that one integrates the silicate optical depth along lines
of sight where some of the grains are uncoated (in particular where
the grain temperatures are above mantle evaporation limits, near
protostellar objects). We adopt an ice-to-core volume ratio of 1 in the calculations for the main models.
In the additional models including additional pure CO in the mantle, we adopt an additional 0.25 volume ratio of CO with respect to the H$_2$O:CO$_2$ ice mantle.
In the mixed amorphous carbon/silicates core model, we adopt an amorphous carbon to silicates ratio for the core of 2/3, a ratio close to the one adopted in several interstellar dust models \citep[e.g.][]{Hensley2021, Jones2017, Zubko2004}.
\subsection{Dust grains growth and size distribution}

A simple analytical dust size distribution to describe the diffuse interstellar medium was given early by \cite{Mathis1977}, the so-called MRN distribution, with a power law describing the number density of grains as a function of grain radius, with a minimum radius $\rm a_{min} \sim 0.005 \;\mu m$, and an upper bound $\rm a_{max} \sim 0.25 \;\mu m$. The power law follows $\rm dn(a) \propto a^{\beta}da$, with $\beta=-3.5$.
In the dense phases of the Galaxy,
dust grains will grow in size: moderately by accreting ice mantles in the cold dense molecular regions, or more significantly by aggregation. Large size increase due to grain growth is expected to
be primarily due to aggregation rather than gas phase freeze out, the latter
being unable to provide sufficient material to significantly increase
the larger grain sizes (see \cite{Dartois2006} for a discussion).
This dense phase is the very first step initiating the subsequent evolution, that will eventually build large planetesimals in protoplanetary disks.

The aggregation, clustering, and assembling of dust particles leading to grain growth is investigated in literature models. The result of the time dependent evolution of the dust size distribution is shown in e.g. \cite{Silsbee2020,Paruta2016,Ormel2011,Ormel2009,Weingartner2001}. 
In these models, at the very first stage of aggregation, the size boundaries of the distribution do not move significantly, as there is mostly a redistribution of the most numerous small grains aggregating to other grains of the distribution, and the parameter affected is the slope of the distribution. At larger dynamical times, corresponding to a large fraction of the cloud lifetime, the size distribution shifts toward bigger clustered grains. Starting from the MRN size distribution, we fix a new slope for the power law implying a decrease of the amount of small grains, and calculate the distribution that satisfies two essential criteria: the total mass and the lower size limit in the distribution are conserved. With such calculations, we reproduce qualitatively the range of evolution of the model results presented in e.g. Fig.2-3 of \cite{Paruta2016} or Fig.4-6 of \cite{Silsbee2020}.

Dust grain mass absorption and scattering coefficients are calculated for the three size distributions presented in Fig.~\ref{Fig_size_evolution}, for the quadratic ellipsoid weighting scheme (QCDE) and compact stochastic distribution of ellipsoid (CSDE), and with an ice mantle to core volume ratio of $f_{\rm{vol}}=1$. We apply the model to the M15 (mixture with low CO$_2$) and M50 (mixture with high CO$_2$) ice mixtures.
We also explore a model with a possible core mixture including amorphous carbon, and a model including a possible pure CO component in the ice mantle. A summary of the models calculated is given in Table~\ref{summary_models}.

The mass extinction and absorption coefficients are presented in Fig.~\ref{Fig_MAC_individuels} for different grain sizes for the different sets of models. As expected, the shape of the bands changes when the grain size increases: the bands broaden, become asymmetric and the peak/continuum ratio decreases. The mass coefficients for the different size distribution are presented in Fig.~\ref{Fig_MAC}.
The models show that when grains grow, leading to the $\rm \tau_1$ and $\rm \tau_3$ like size distributions, the antisymmetric CO$_2$ ice vibrational mode extinction profile is affected. Averaged over all orientations of the ellipsoids, the mass absorption coefficients behave, to the first order, similarly for the QCDE and CSDE models.

\begin{figure*}
  \centering
\includegraphics[width=0.66\columnwidth]{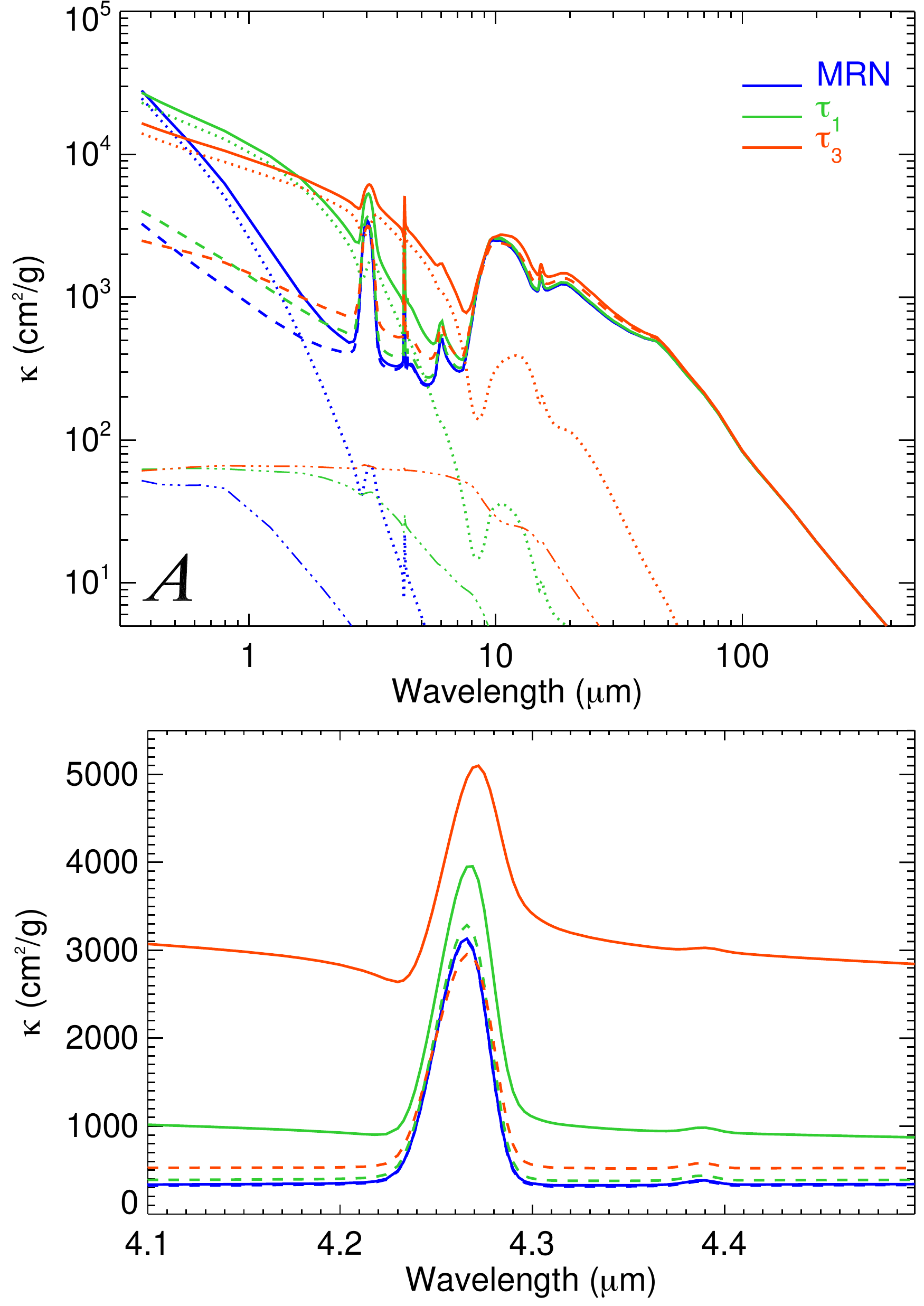}
\includegraphics[width=0.66\columnwidth]{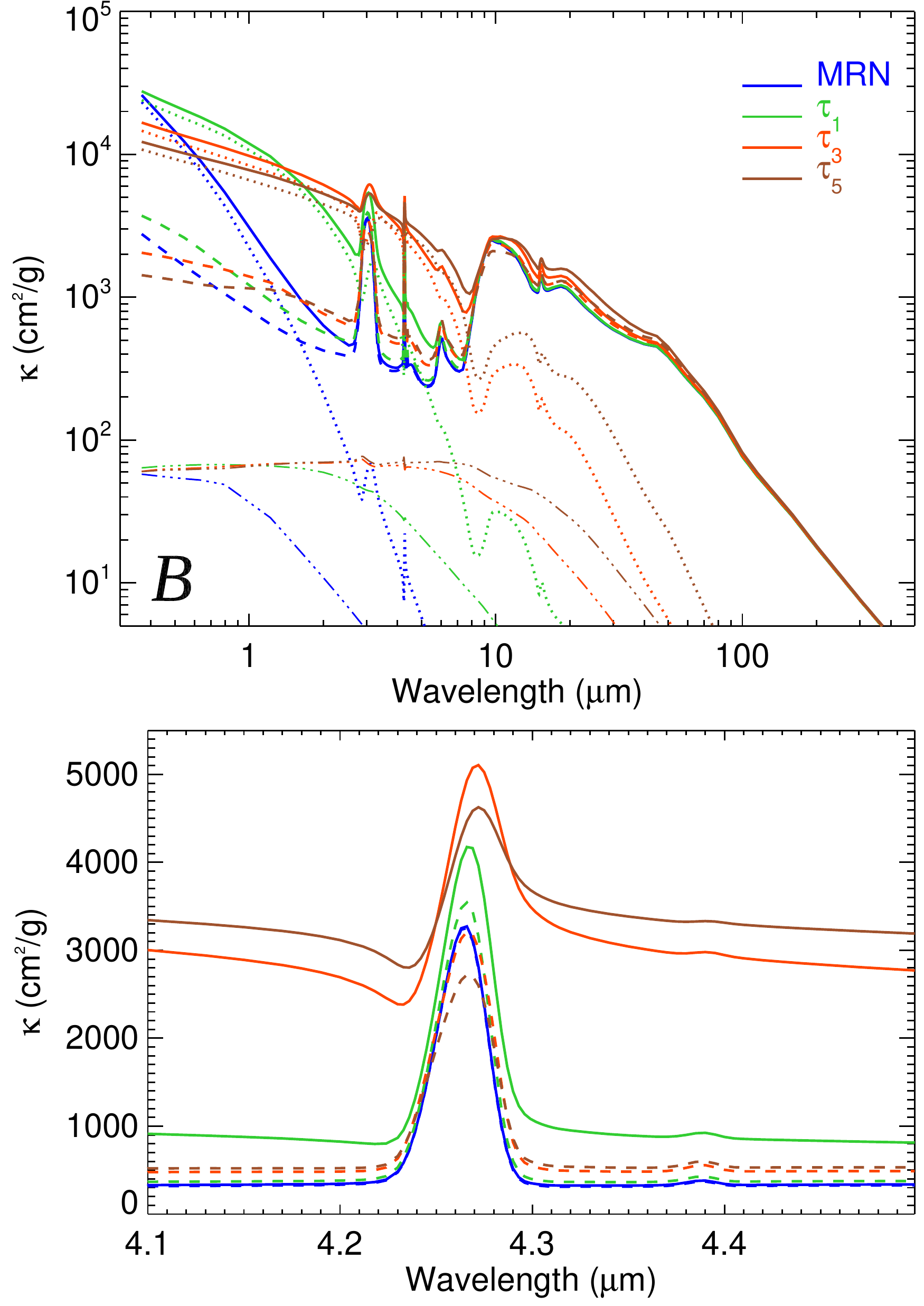}
\includegraphics[width=0.66\columnwidth]{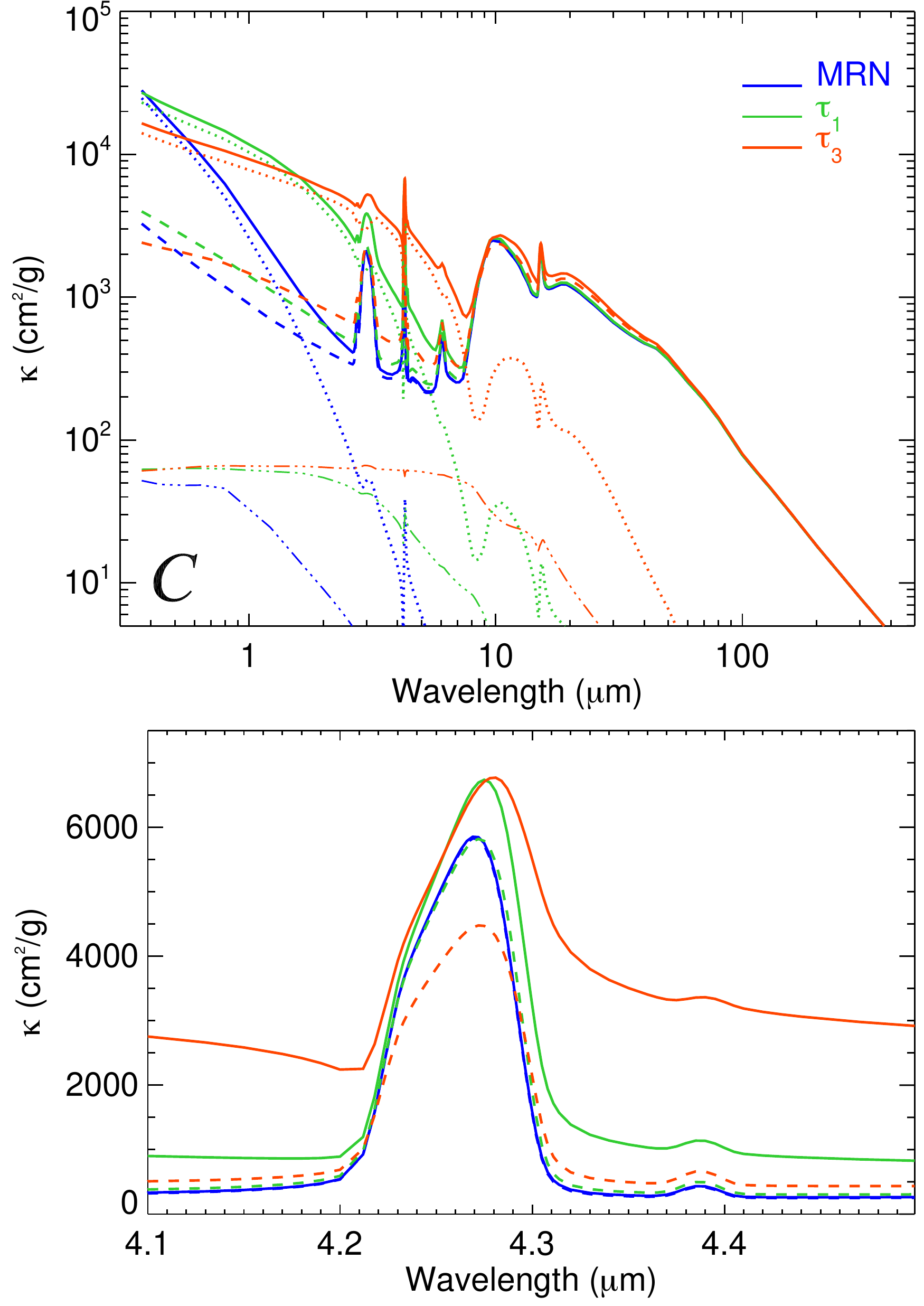}
\includegraphics[width=0.66\columnwidth]{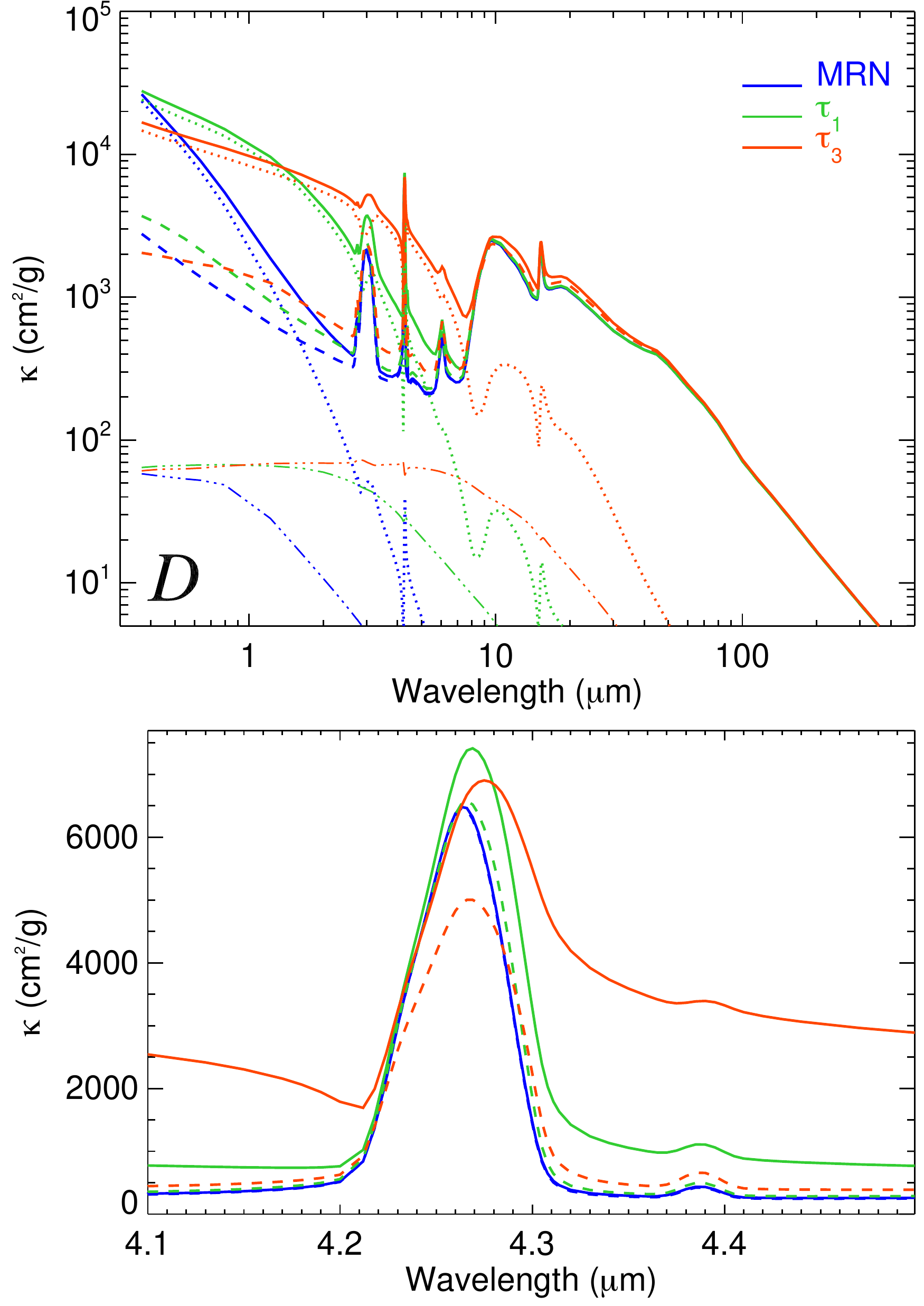}
\includegraphics[width=0.66\columnwidth]{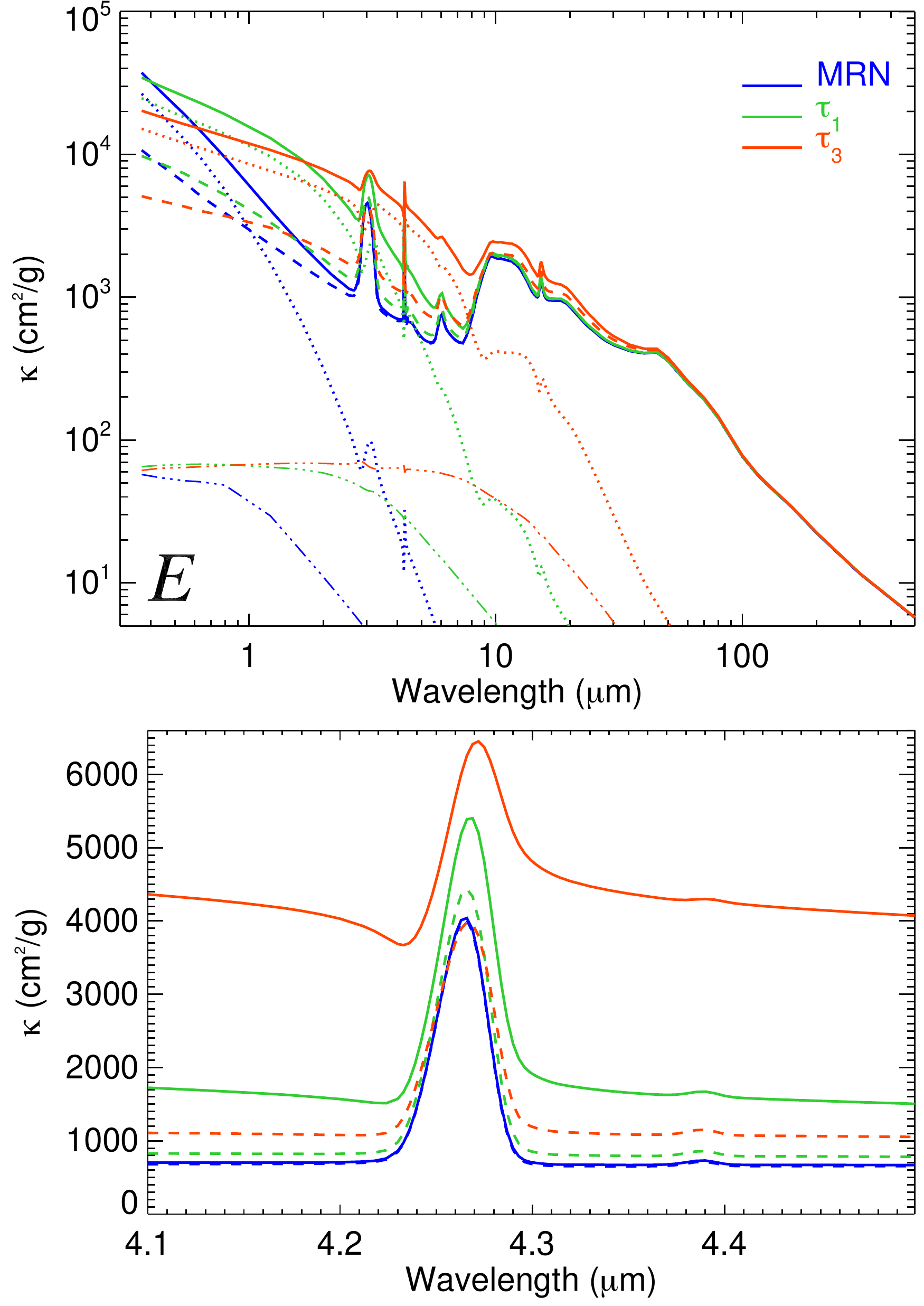}
\includegraphics[width=0.66\columnwidth]{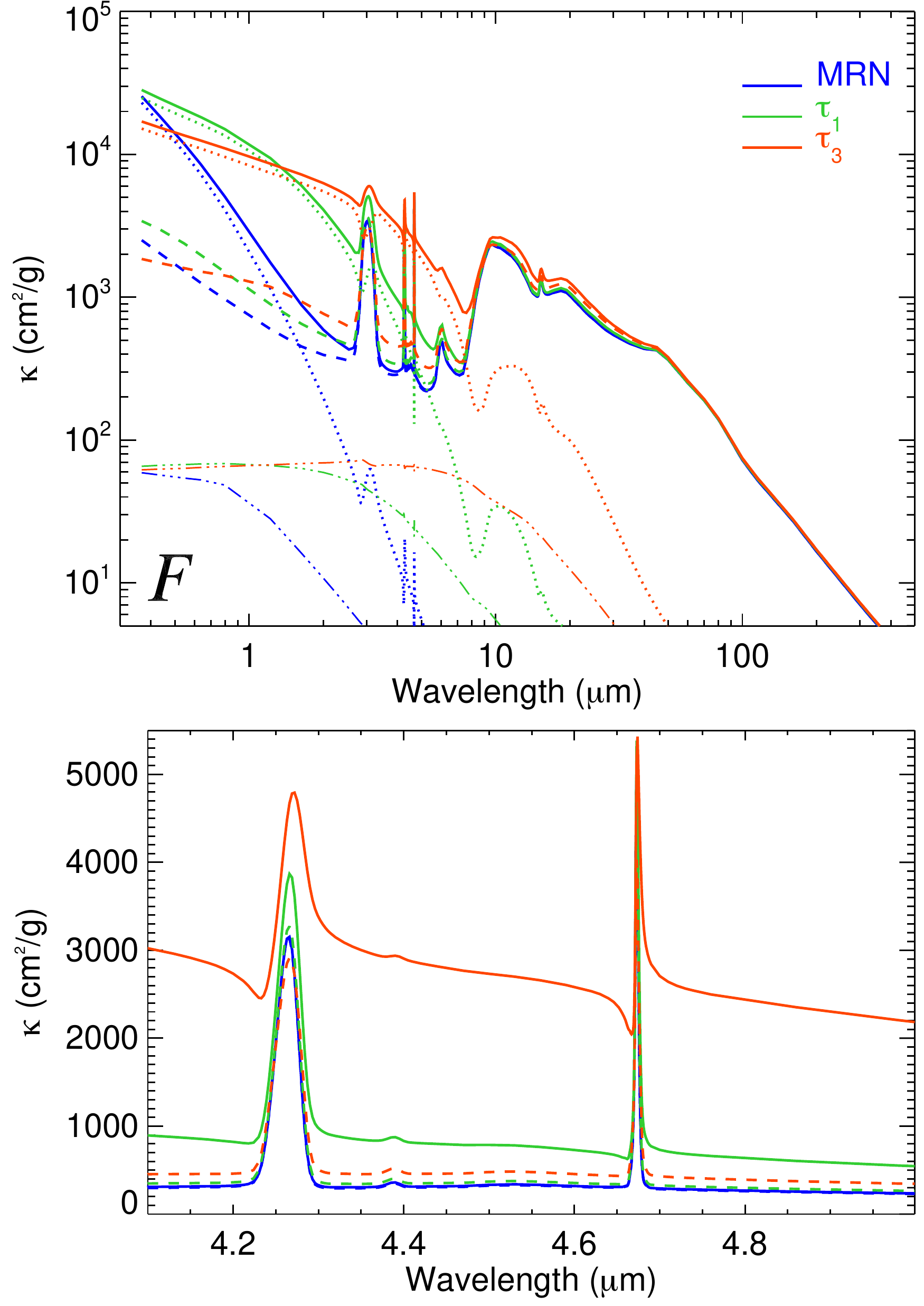}
\begin{minipage}{0.66\columnwidth}
\includegraphics[width=\columnwidth]{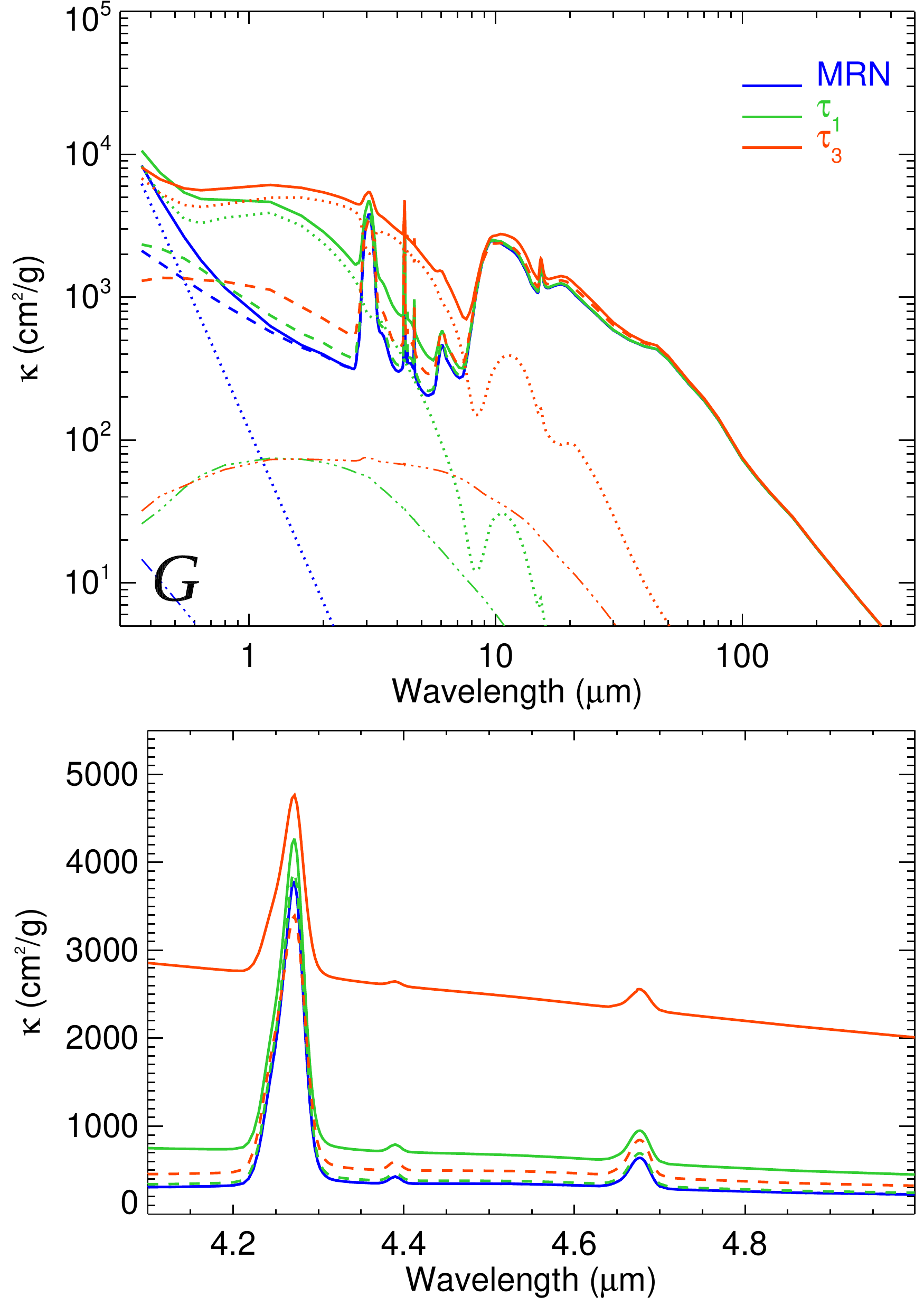}
\end{minipage}
\begin{minipage}{0.165\columnwidth}
$\;$
\end{minipage}
\begin{minipage}{\columnwidth}
  \caption{Mass extinction coefficients (solid lines), absorption coefficients (dashed line), Henyey-Greenstein phase function g (dot-dashed line, x100), for model sets {\it A, B, C, D, E, F, G} (3 models per panel corresponding to the MRN ($\rm a_{max}~=~0.25\mu m$), $\rm \tau_1$ ($\rm a_{max}~=~1\mu m$) and $\rm \tau_3$ ($\rm a_{max}~=~3\mu m$) size distributions), respectively. { Model set B includes a size distribution extending to $\rm \tau_5$ ($\rm a_{max}~=~5\mu m$).}
  Below each panel is shown a close-up of the CO$_2$ stretching mode (as well as including CO for the last two sets).
}
\end{minipage}
\begin{minipage}{0.165\columnwidth}
$\;$
\end{minipage}
  \label{Fig_MAC}
\end{figure*}

 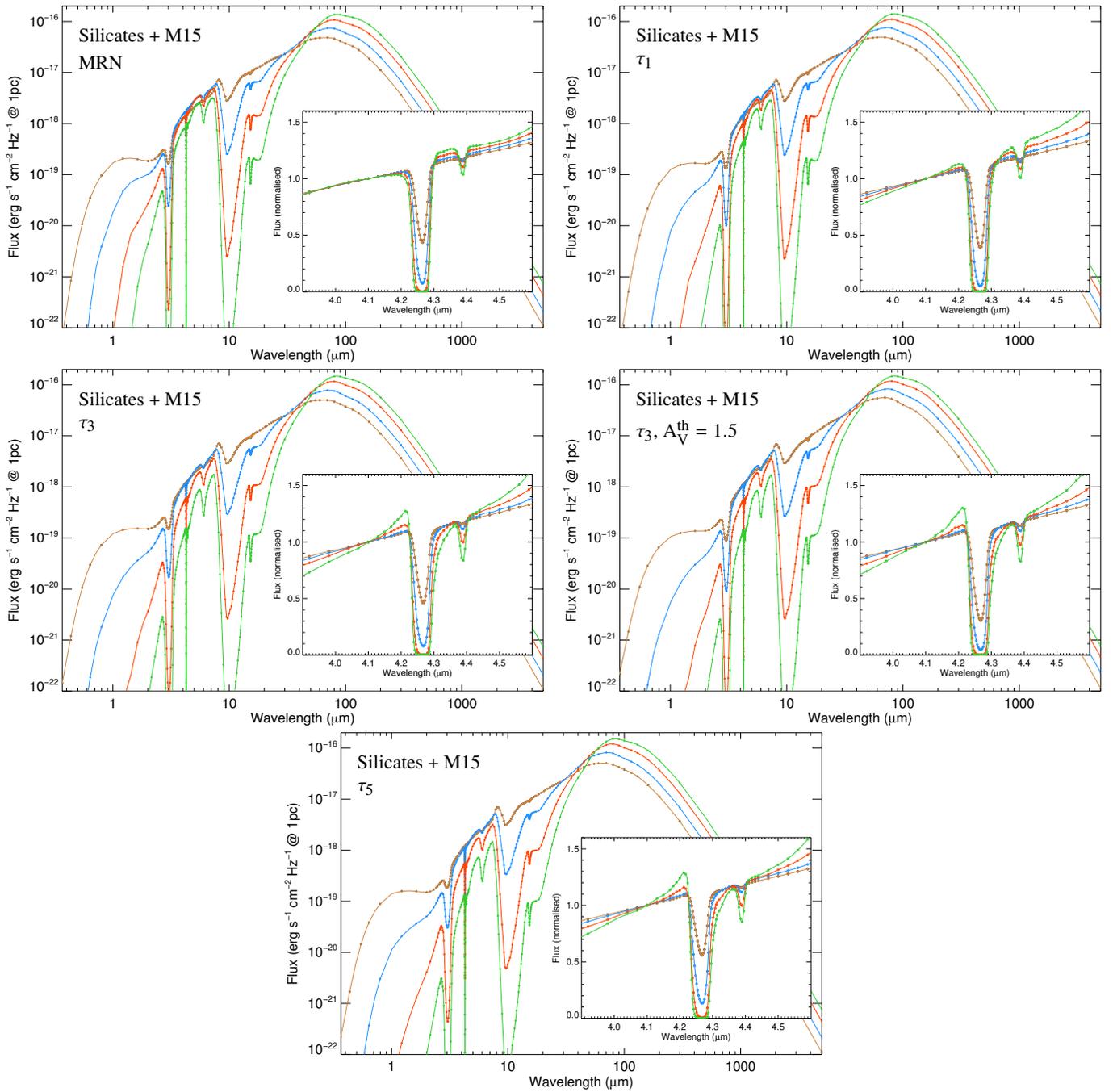
\begin{figure*}
      \centering
\begin{minipage}{\columnwidth}
\begin{tikzpicture}
\node [
    above right,
    inner sep=0] (image) at (0,0) {  \includegraphics[width=.99\columnwidth]{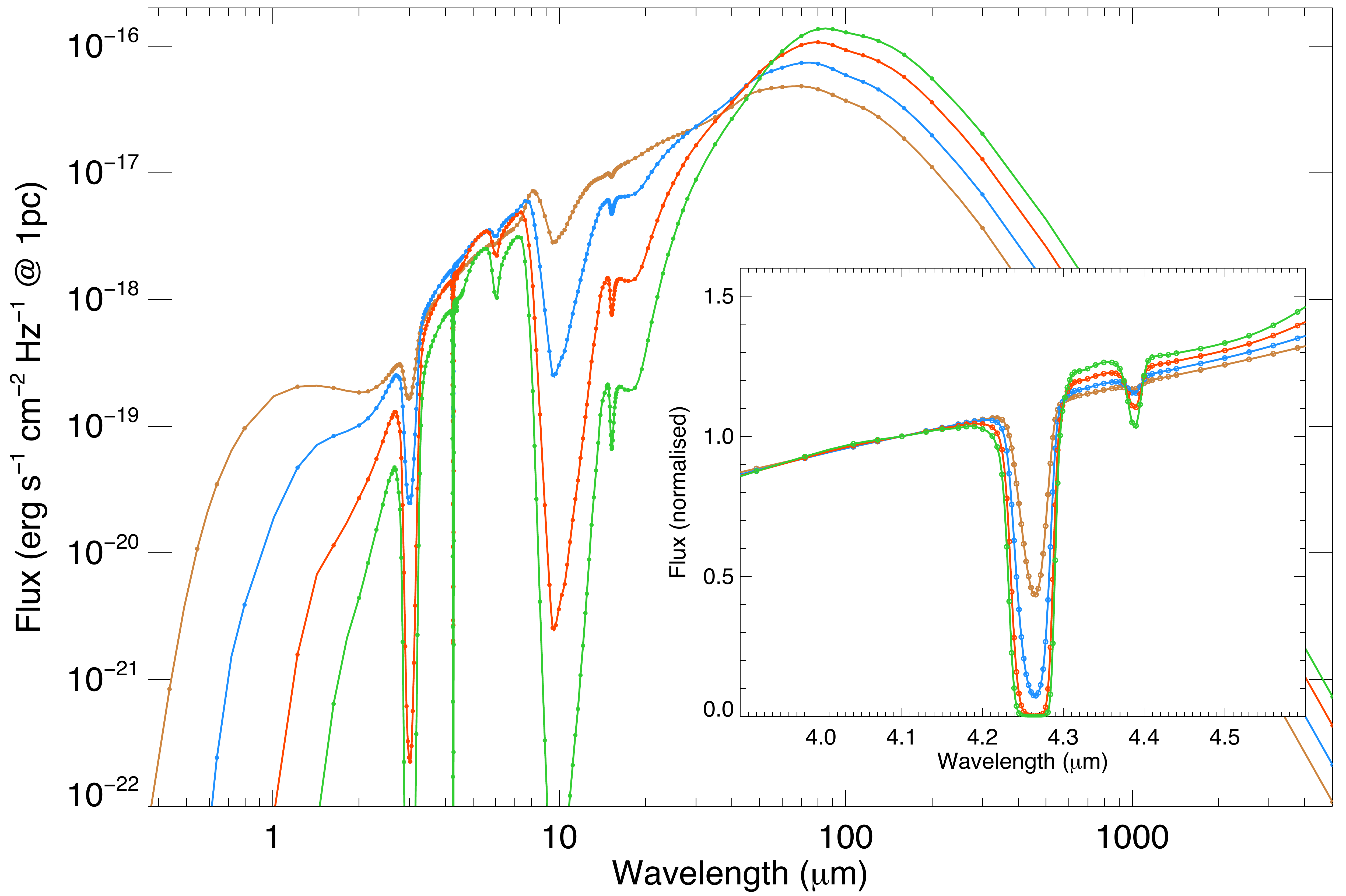}};
\begin{scope}[
x={($0.1*(image.south east)$)},
y={($0.1*(image.north west)$)}]
    \draw[very thick,green] (1.25,9.5) 
       node[below right,black]{\small Silicates + M15};
    \draw[very thick,green] (1.25,8.75) 
       node[below right,black]{\small MRN};
\end{scope}
\end{tikzpicture}
\end{minipage}
  \begin{minipage}{\columnwidth}
\begin{tikzpicture}
\node [
    above right,
    inner sep=0] (image) at (0,0) {  \includegraphics[width=.99\columnwidth]{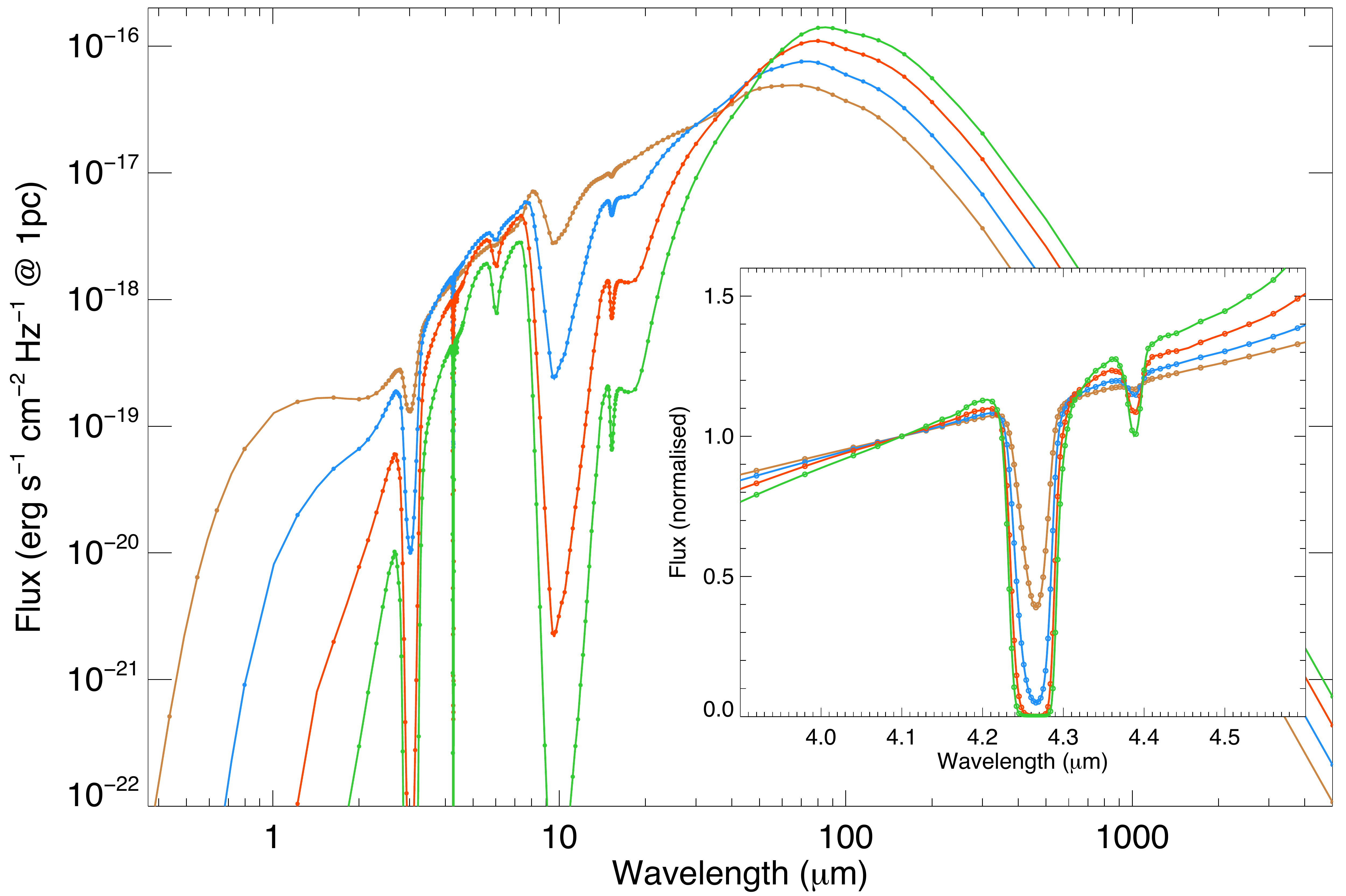}};
\begin{scope}[
x={($0.1*(image.south east)$)},
y={($0.1*(image.north west)$)}]
     \draw[very thick,green] (1.25,9.5) 
       node[below right,black]{\small Silicates + M15};
    \draw[very thick,green] (1.25,8.75) 
       node[below right,black]{\small $\tau_1$};
\end{scope}
\end{tikzpicture}
\end{minipage}
  \begin{minipage}{\columnwidth}
\begin{tikzpicture}
\node [
    above right,
    inner sep=0] (image) at (0,0) {  \includegraphics[width=.99\columnwidth]{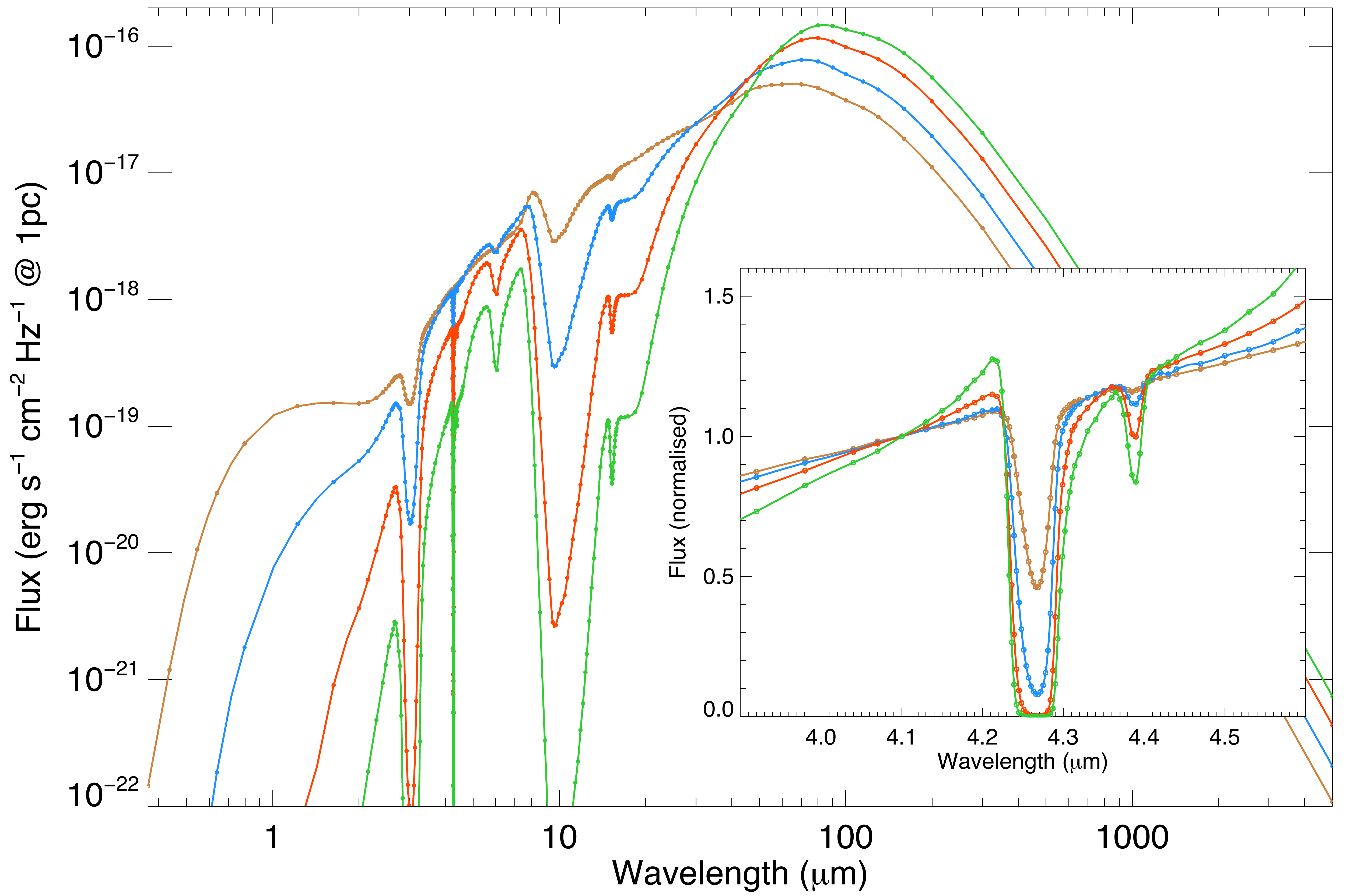}};
\begin{scope}[
x={($0.1*(image.south east)$)},
y={($0.1*(image.north west)$)}]
     \draw[very thick,green] (1.25,9.5) 
       node[below right,black]{\small Silicates + M15};
    \draw[very thick,green] (1.25,8.75) 
       node[below right,black]{\small $\tau_3$};
\end{scope}
\end{tikzpicture}
\end{minipage}
  \begin{minipage}{\columnwidth}
\begin{tikzpicture}
\node [
    above right,
    inner sep=0] (image) at (0,0) {  \includegraphics[width=.99\columnwidth]{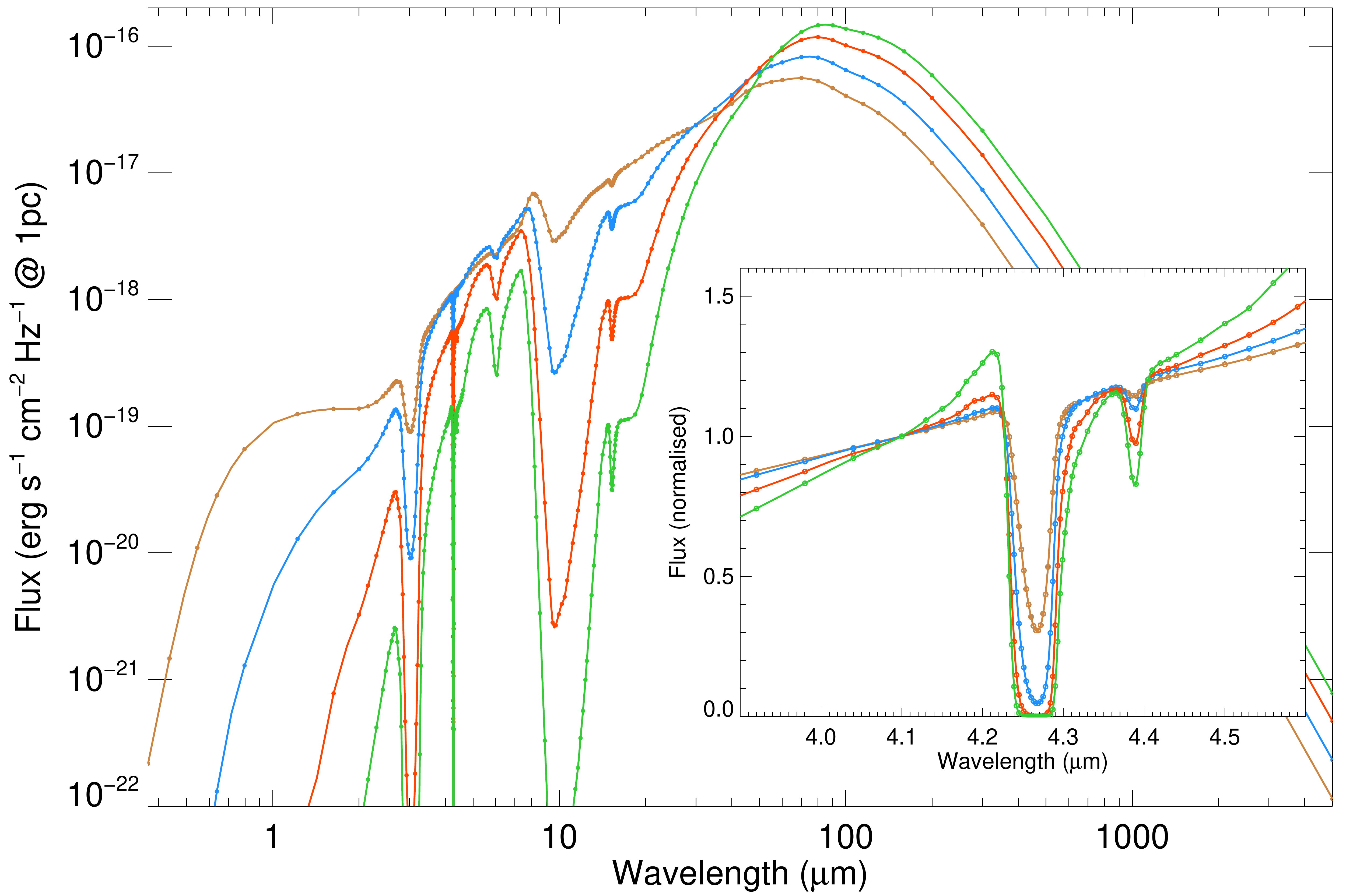}};
\begin{scope}[
x={($0.1*(image.south east)$)},
y={($0.1*(image.north west)$)}]
     \draw[very thick,green] (1.25,9.5) 
       node[below right,black]{\small Silicates + M15};
    \draw[very thick,green] (1.25,8.75) 
       node[below right,black]{\small $\tau_3$, \small $\rm A_V^{th}=1.5$};
\end{scope}
\end{tikzpicture}
\end{minipage}
  \begin{minipage}{\columnwidth}
\begin{tikzpicture}
\node [
    above right,
    inner sep=0] (image) at (0,0) {  \includegraphics[width=.99\columnwidth]{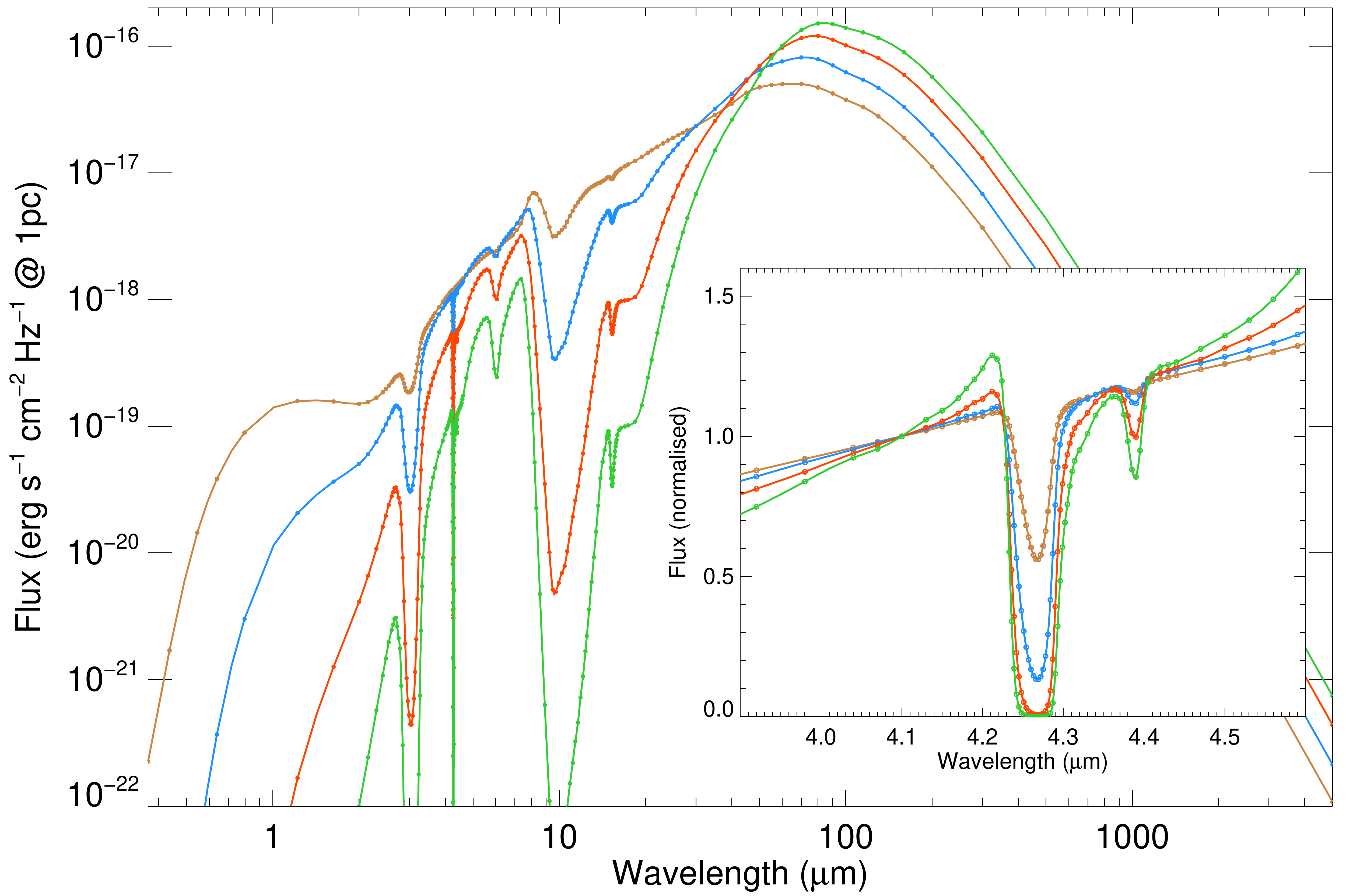}};
\begin{scope}[
x={($0.1*(image.south east)$)},
y={($0.1*(image.north west)$)}]
     \draw[very thick,green] (1.25,9.5) 
       node[below right,black]{\small Silicates + M15};
    \draw[very thick,green] (1.25,8.75) 
       node[below right,black]{\small $\tau_5$};
\end{scope}
\end{tikzpicture}
\end{minipage}

  \caption{{ Spectral energy distributions for the spherical radiative transfer for the M15 ice and core composition, and varying the dust size distribution, for different visual extinctions (A$\rm_V$ = 15, brown; 30, blue; 60, red; 100, green). Distributions corresponding to, from top left to lower right: silicates core, M15 ice mantle composition and size distribution
  MRN, $\rm a_{max}~=~0.25\mu m$ (model \#4);
 $\rm \tau_1$, $\rm a_{max}~=~1\mu m$, (model \#5);
  $\rm \tau_3$, $\rm a_{max}~=~3\mu m$, (model \#6);
  $\rm \tau_3$, $\rm a_{max}~=~3\mu m$, (model \#6) with ice threshold is set at A$\rm_V^{th}(ice)$ = 1.5 instead of 3;
  $\rm \tau_5$, $\rm a_{max}~=~5\mu m$, (model \#7). 
The inserts display close-ups of the CO$_2$ ice band, with spectra normalised at 4.1~$\mu$m.}
}
  \label{Fig_modeles_sphere_1}%
\end{figure*}

 \begin{figure*}
      \centering
  \begin{minipage}{\columnwidth}
\begin{tikzpicture}
\node [
    above right,
    inner sep=0] (image) at (0,0) {  \includegraphics[width=.99\columnwidth]{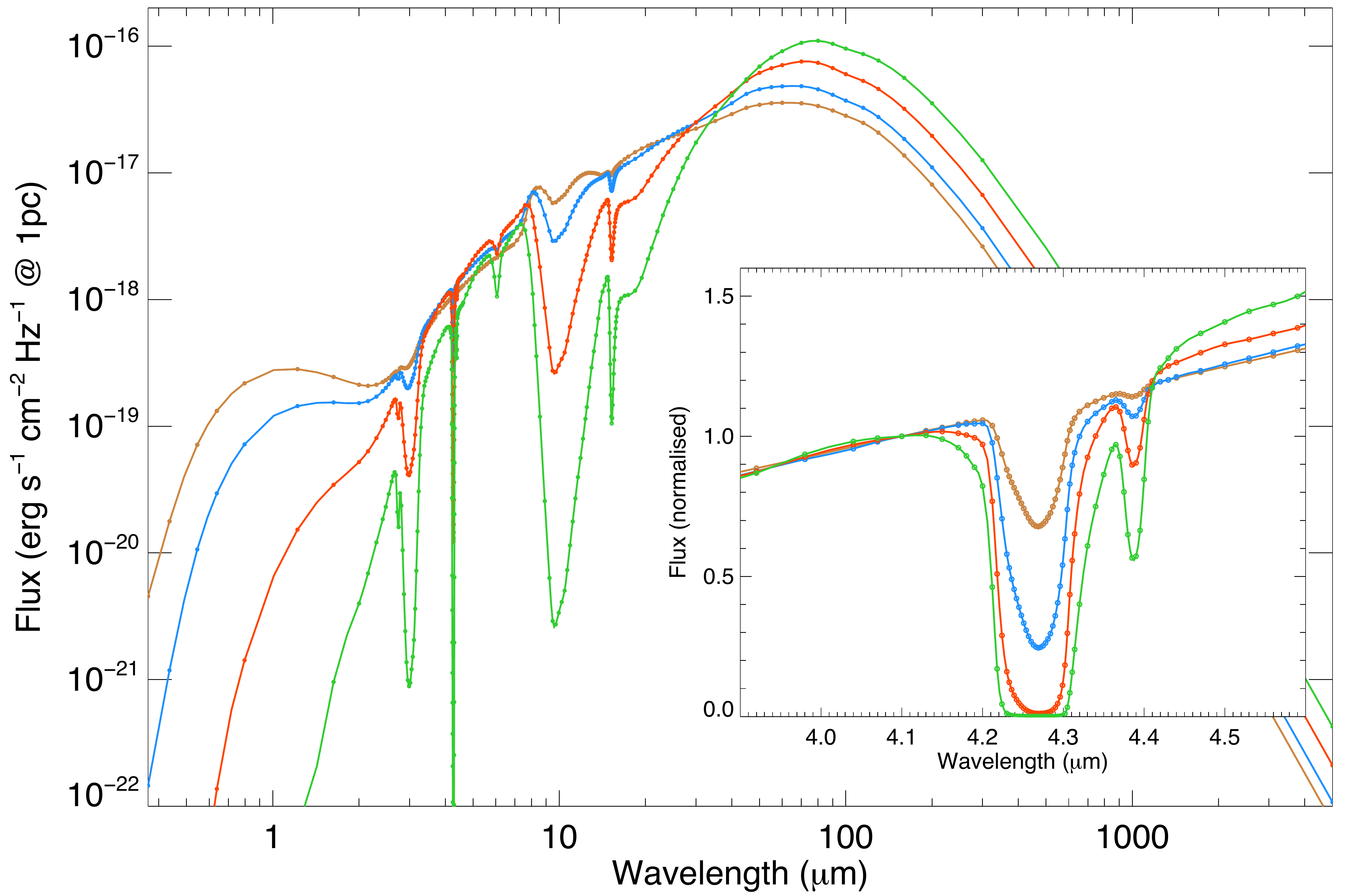}};
\begin{scope}[
x={($0.1*(image.south east)$)},
y={($0.1*(image.north west)$)}]
     \draw[very thick,green] (1.25,9.5) 
       node[below right,black]{\small Silicates + M50};
    \draw[very thick,green] (1.25,8.75) 
       node[below right,black]{\small $\tau_3$};
\end{scope}
\end{tikzpicture}
\end{minipage}
  \begin{minipage}{\columnwidth}
\begin{tikzpicture}
\node [
    above right,
    inner sep=0] (image) at (0,0) {  \includegraphics[width=.99\columnwidth]{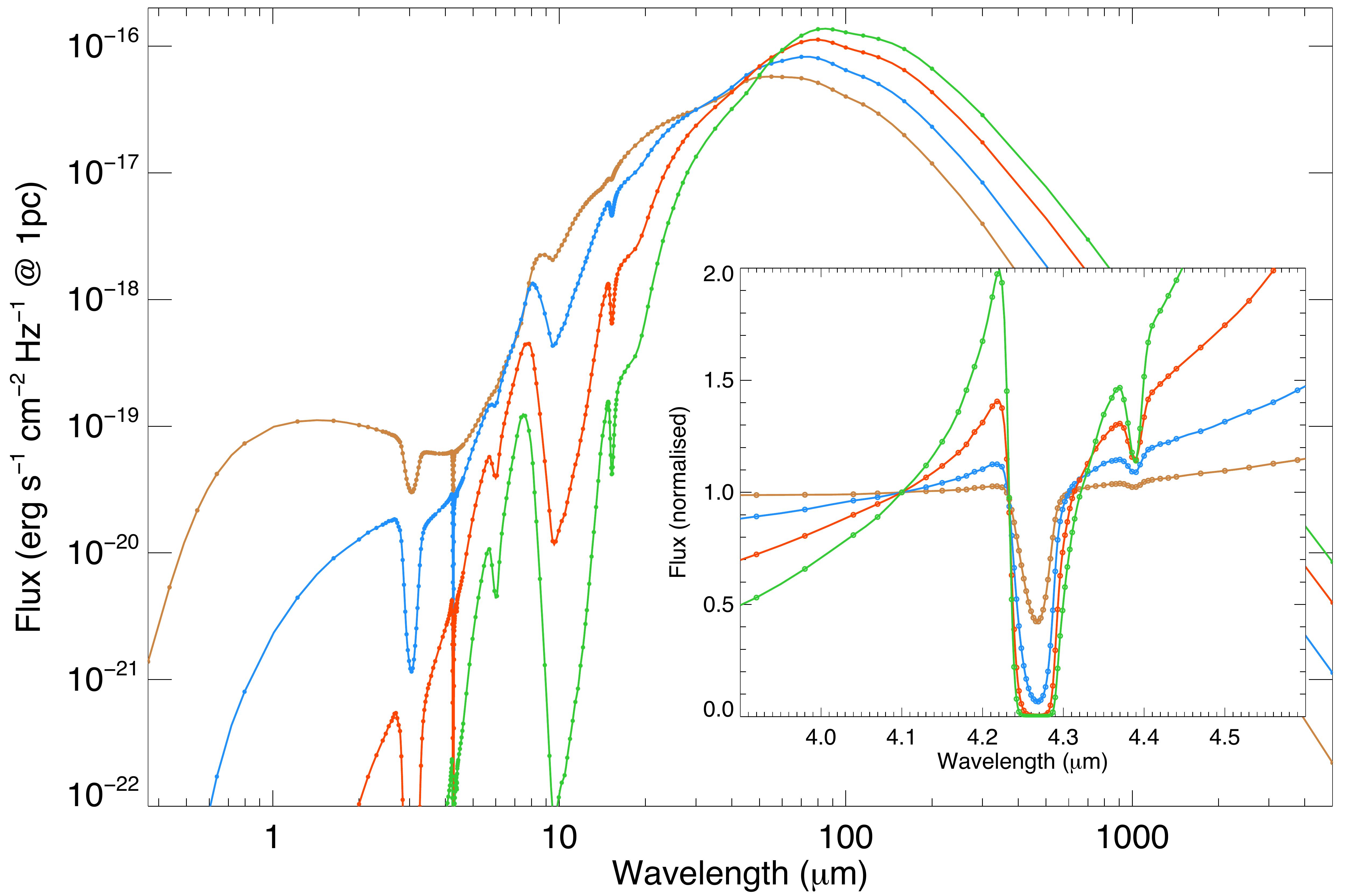}};
\begin{scope}[
x={($0.1*(image.south east)$)},
y={($0.1*(image.north west)$)}]
     \draw[very thick,green] (1.25,9.5) 
       node[below right,black]{\small Silicates + am. carbon + M15};
    \draw[very thick,green] (1.25,8.75) 
       node[below right,black]{\small $\tau_3$};
\end{scope}
\end{tikzpicture}
\end{minipage}
  \begin{minipage}{\columnwidth}
\begin{tikzpicture}
\node [
    above right,
    inner sep=0] (image) at (0,0) {  \includegraphics[width=.99\columnwidth]{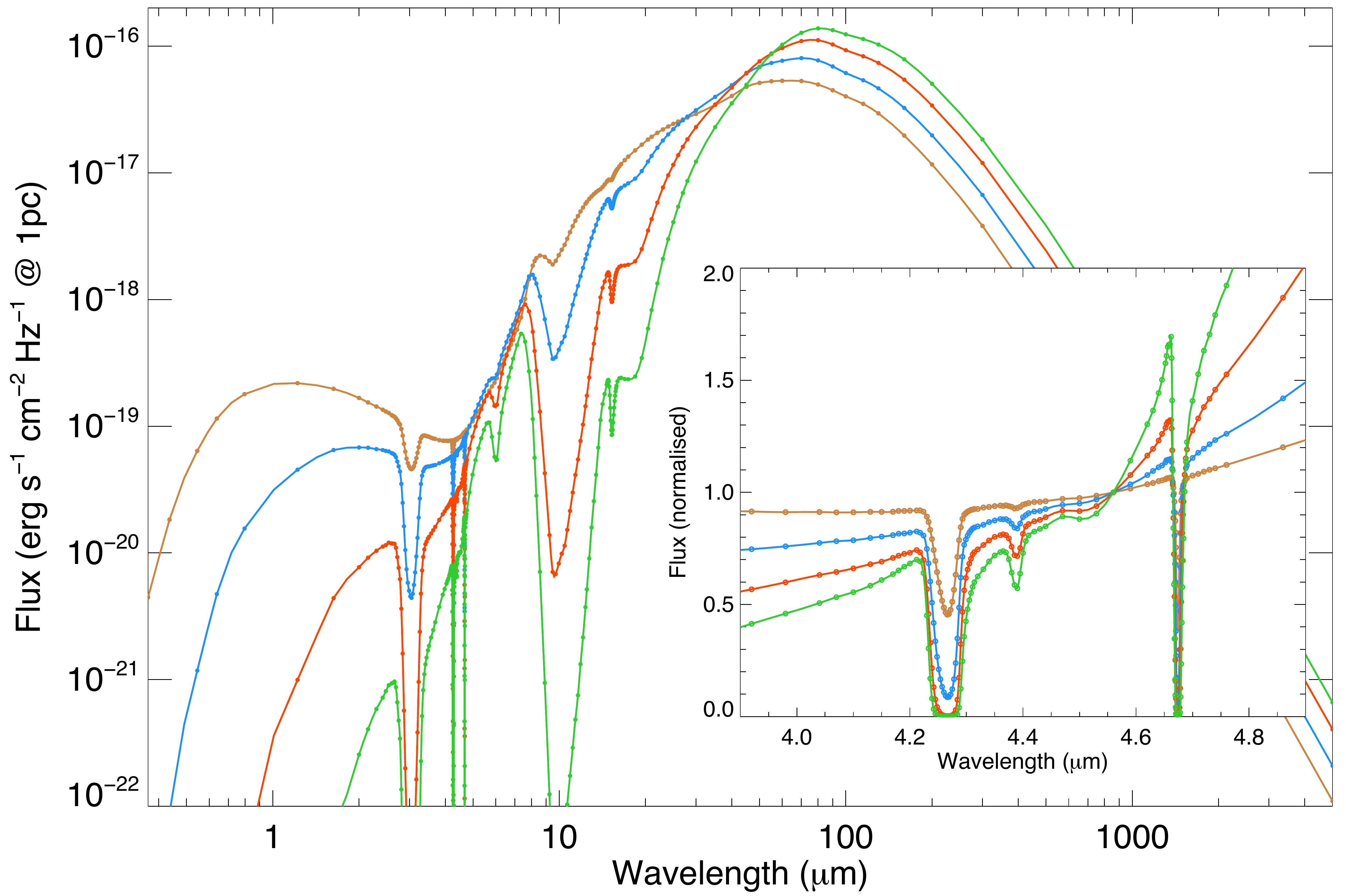}};
\begin{scope}[
x={($0.1*(image.south east)$)},
y={($0.1*(image.north west)$)}]
     \draw[very thick,green] (1.25,9.5) 
       node[below right,black]{\small Silicates + M15 + CO};
    \draw[very thick,green] (1.25,8.75) 
       node[below right,black]{\small $\tau_3$};
\end{scope}
\end{tikzpicture}
\end{minipage}
  \begin{minipage}{\columnwidth}
\begin{tikzpicture}
\node [
    above right,
    inner sep=0] (image) at (0,0) {  \includegraphics[width=.99\columnwidth]{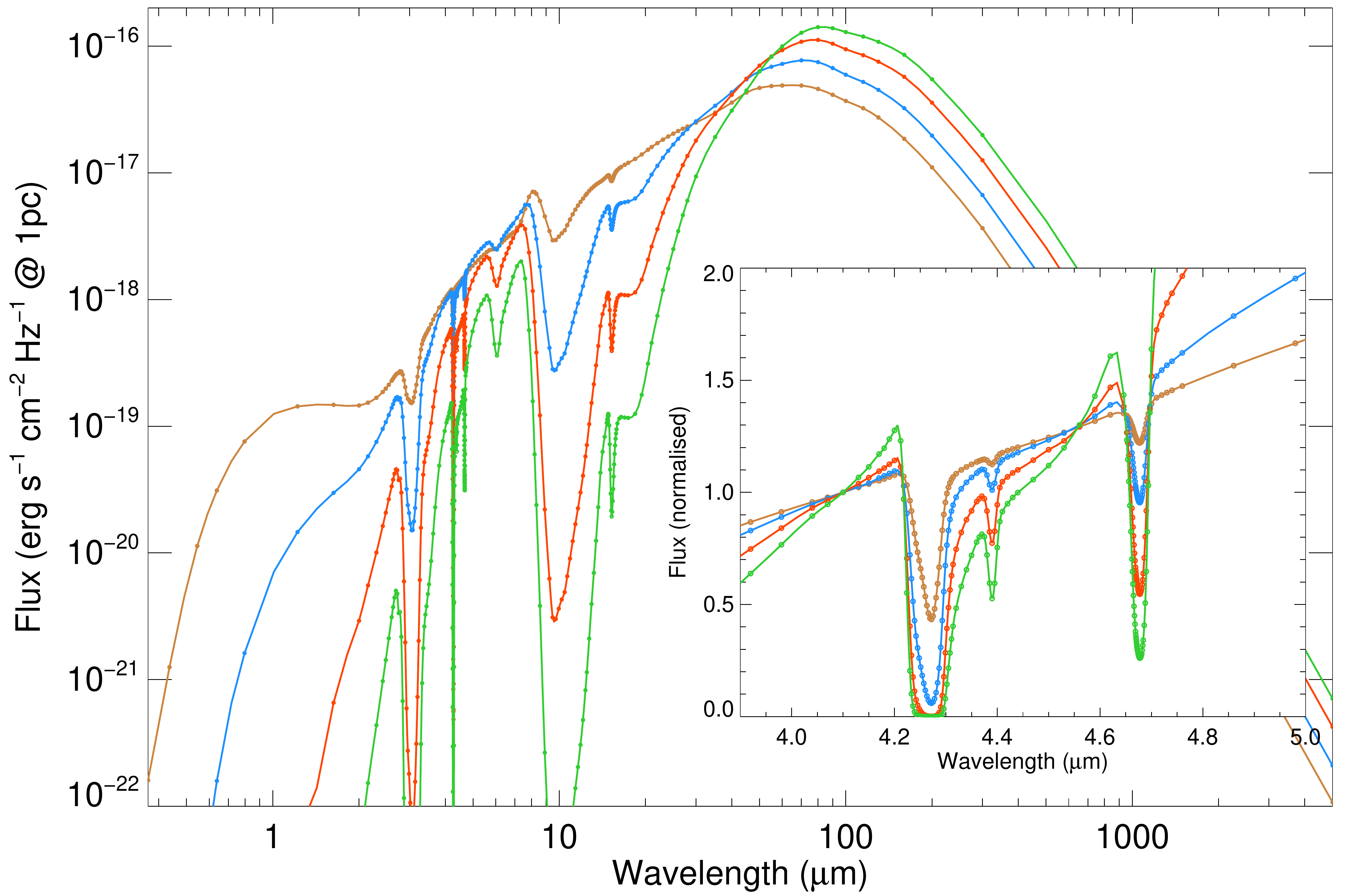}};
\begin{scope}[
x={($0.1*(image.south east)$)},
y={($0.1*(image.north west)$)}]
     \draw[very thick,green] (1.25,9.5) 
       node[below right,black]{\small Silicates + MX};
    \draw[very thick,green] (1.25,8.75) 
       node[below right,black]{\small $\tau_3$};
\end{scope}
\end{tikzpicture}
\end{minipage}
  \caption{{ Same as Fig.~\ref{Fig_modeles_sphere_1} but including the other models considered for the ice and core composition. From top left to lower right: 
silicates core, M50 ice mantle composition and size distribution $\rm \tau_3$, $\rm a_{max}~=~3\mu m$, (model \#13);
silicates and amorphous carbon core, M15 ice composition and size distribution $\rm \tau_3$, $\rm a_{max}~=~3\mu m$, (model \#16);
silicates core, M15 ice mantle composition, including pure CO and size distribution $\rm \tau_3$, $\rm a_{max}~=~3\mu m$, (model \#19);
silicates core, MX ice mantle composition, including H2O, CO2, CO and NH3 and size distribution $\rm \tau_3$, $\rm a_{max}~=~3\mu m$, (model \#22).
The inserts display close-ups of the CO$_2$ (and CO for the last one) ice band, with spectra normalised at 4.1~$\mu$m (4.55~$\mu$m in the last one).}
}
  \label{Fig_modeles_sphere_2}%
\end{figure*}

%
\section{Radiative transfer modelling}

In order to make relevant comparisons with the expected global
mid-infrared spectrum of embedded infrared sources, the UV to
millimeter mass extinction coefficient calculated above is used as an
input to modelling.

Using the calculated dust distributions, we perform radiative transfer calculations to model the expected spectra from fiducial spherical dust clouds and disks, with a higher spectral resolution set on the CO$_2$ ice profile. The calculations are performed using the Monte Carlo RADMC-3D\footnote{https://www.ita.uni-heidelberg.de/$\sim$dullemond/software/radmc-3d/} software \citep{Dullemond2012}, in the full anisotropic scattering mode for the dust radiative transfer. 

\subsection{Spherical clouds}

We model a fiducial spherical cloud with an adopted density 
\begin{equation}
\rm \rho(r) = \rho_0 (r/r_{in})^{-p}  \; , \; r \geq r_{in} \; , \; r \leq r_{out}
\end{equation}
$\rm r_{in}$ is the inner boundary of the cloud, $\rm r_{out}$ the outer one, r is the radial distance to the star.
$\rm \rho(r)$ is the dust density ($\rm g/cm^3$).

DDA calculations for bare silicates were also performed to add them
into the radiative transfer code for regions where the temperature, {determined self consistently during the radiative transfer calculation}, is
above the ice sublimation temperature (we adopt $\rm T_{subl}=$~100K), i.e. towards the
inner core and also for the regions where the visual extinction is below a
given $\rm A_V$ threshold.

We adopt as a model template a prescription close to the one described by \cite{Siebenmorgen1997} in the case of HH100 IRS, with p=1, $\rm r_{in}=0.3 au$, $\rm r_{out}=3000 au$. We adjust the value of $\rho_0$ to obtain an ice absorption with our conditions close to those observed.
Our goal is not to vary all the possible parameters, but to show the effect of the distributions with different levels of grain growth on the CO$_2$ ice profile.

{ The ice onset threshold is distinct for different clouds and different ices
\citep[e.g.][and references therein]{Whittet1988, Murakawa2000, Boogert2015}.
The minimum onset value for water ice and quiescent clouds is about 3 (the measured value includes back and front of the cloud and thus correspond to about $\rm A_V^{th} \sim 1.5$ on one side). This threshold value can be higher in other clouds (e.g. Serpens, rho Ophiuchus) with higher thresholds up to above 10 \citep{Tanaka1990, Eiroa1989} related to some star formation activity or higher local external UV field.
For disks, the typical threshold value is not well known yet, and, if the underlying physics is the same, the equilibrium threshold value might be higher. We thus adopt a visual extinction threshold value $\rm A_V^{th} = 3$ that will be used for both spherical clouds and disk models to keep this parameter fixed for comparison, and in a few models only we explored a threshold of $\rm A_V^{th} = 1.5$.}
Results of the calculations are shown in Fig.\ref{Fig_modeles_sphere_1}-\ref{Fig_modeles_sphere_2} for MRN to $\tau_5$ icy dust size distributions with an M15 mantle, $\tau_3$ with an M50 mantle, a model including amorphous carbon in the refractory core, and a model including pure CO in the mantle, for different total column densities corresponding to visual extinctions $\rm A_{V}^{total}\approx 15,30,60,100$.

\subsection{Disks}

We model fiducial protoplanetary disks at various inclination angles along the line of sight. 
The axisymmetric disk model parametrisation is given by:
\begin{equation}
\rm n(r,z) = n(r) e^{-(\frac{z}{H(r)})^2} \; , \; r \geq r_{in} \; , \; r \leq r_{out}
\end{equation}
\begin{equation}
\rm n(r) = n_0 (r/r_0)^{-s} \; , \; 	  H(r)= H_0 (r/r_0)^{h} 
\end{equation}
where r is the radial distance to the star and z the height from the disk midplane, $\rm n(r,z)$ the gas density and $\rm H(r)$ is the hydrostatic scale height.
We assume a gas-to-dust ratio of 100. 
We adopt typical values for the parameters of the disk being considered \citep[e.g.][]{Dartois2003, Pietu2007, Pinte2008, Ansdell2016, Simon2000}: $\rm r_0$ is set at 100 au, $\rm r_{in}=0.1$~au, $\rm r_{out}=100$~au,  a value of 1.2 for the hydrostatic scale height exponent, $\rm H_0=20$~au, defining the flaring of the disk, and $s=3$ for the radial density exponent. The density of the disk at 100 au in the midplane, $n_0$,  is of the order of $5\times10^{8}$~cm$^{-3}$, corresponding to a total mass for the disk of about 0.02 $\rm M\odot$.
In the radiative transfer code the region where ices are present is defined with two simultaneous constraints:
T must be below the ice sublimation temperature (set to 100~K) and the visual extinction must be above a
given $\rm A_V$ threshold (following the
fact that thresholds are observed in dark clouds for ice appearance, as mentioned previously).
For the region where the visual extinction is below the
given $\rm A_V$ threshold or the temperature above ice sublimation, we use bare silicate grains.
We use a 3D cartesian grid with 128$^3$ points. To provide a better sampling of the inner disk region, the grid is refined first with an inner grid of 32$^3$ points (a linear factor of 2, factor 8 in volume). This refined inner grid is refined again in the 8$^3$ points of this subgrid (linear factor of 2, factor 8 in volume).

A sketch of the disk model is shown in Fig.\ref{Fig_modele_disk_1}.
\begin{figure}
  \centering
\includegraphics[width=\columnwidth]{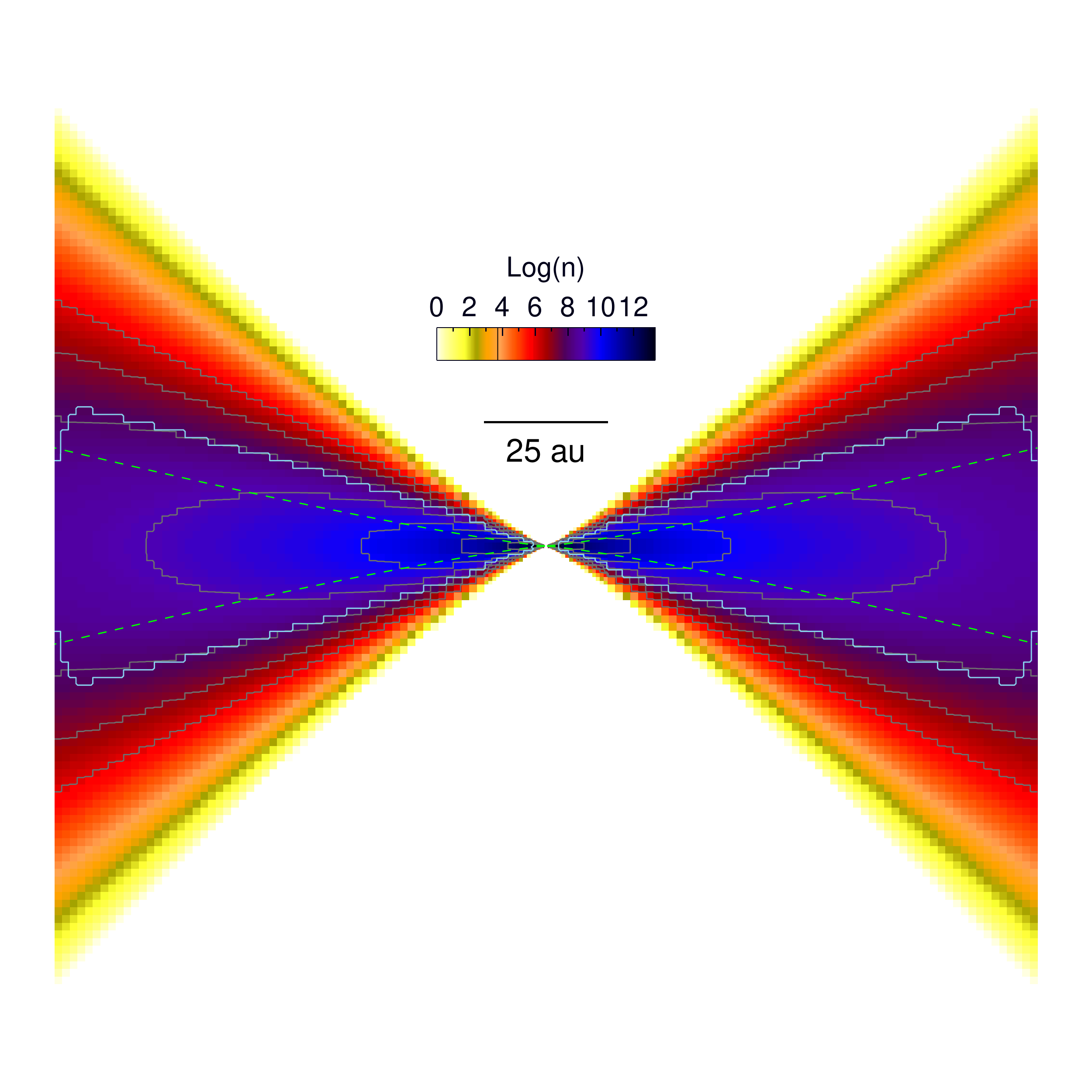}
  \caption{Adopted disk model density distribution. The dashed lines show the hydrostatic scale height as a function of distance to the star, responsible for the disk flaring. The cyan line defines the boundary of the region where ice mantles are present because of the existence of the visual extinction threshold.}
  \label{Fig_modele_disk_1}
\end{figure}

Results of the calculations as observed for different disk inclinations above the onset of ice absorption, for the different models of the ice and core composition, and dust size distribution are shown in Fig.\ref{Fig_modeles_disk_1}-\ref{Fig_modeles_disk_2}. A high diversity of CO$_2$ ice absorption profiles can be observed for different disk inclinations, column density and extent of grain growth, sometimes showing a marked asymmetry.

\begin{figure*}
  \centering
\begin{minipage}{\columnwidth}
\begin{tikzpicture}
\node [
    above right,
    inner sep=0] (image) at (0,0) {  \includegraphics[width=.99\columnwidth]{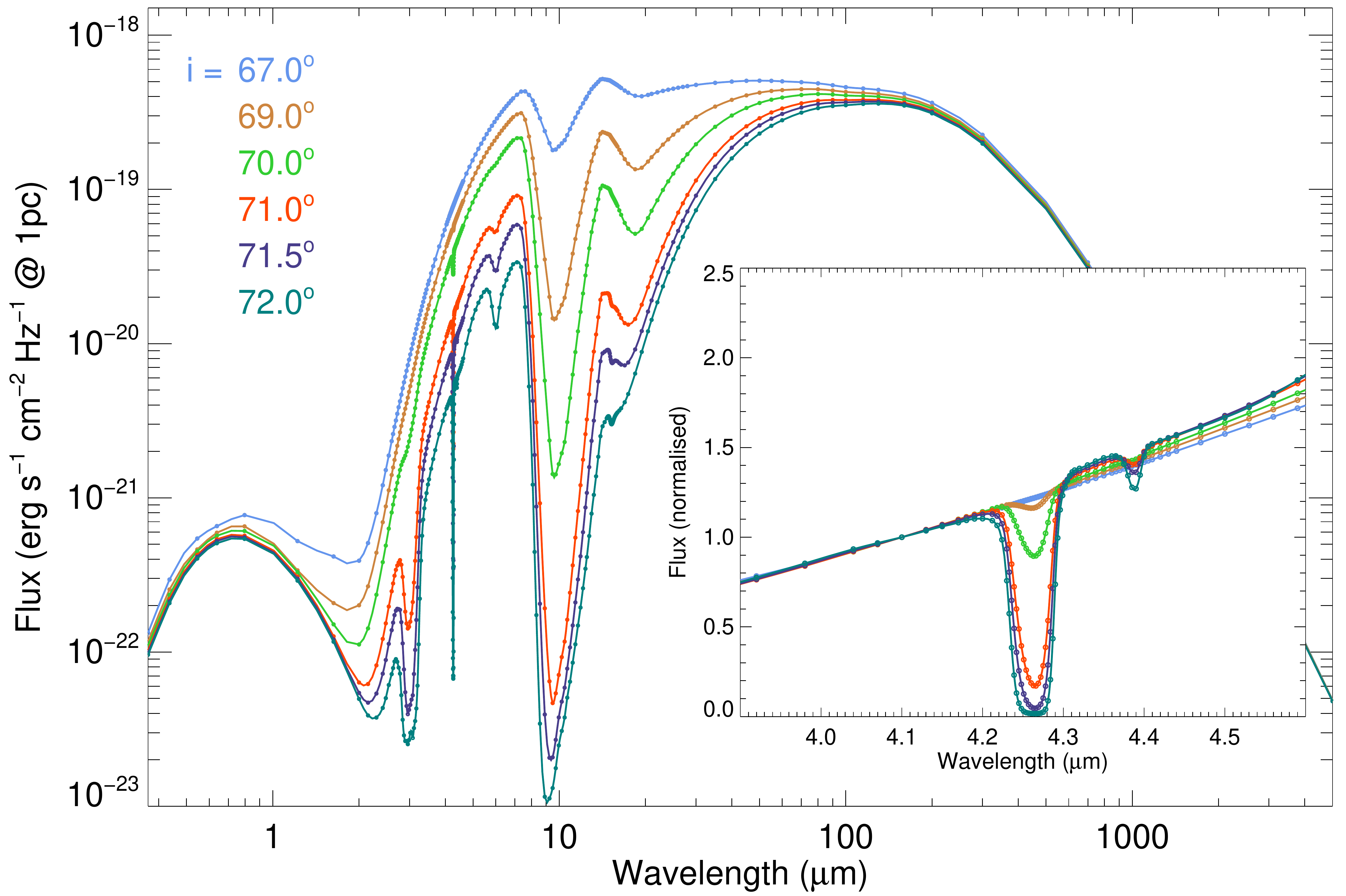}};
\begin{scope}[
x={($0.1*(image.south east)$)},
y={($0.1*(image.north west)$)}]
    \draw[very thick,green] (9.75,9.7) 
       node[below left,black]{\small Silicates + M15};
    \draw[very thick,green] (9.75,8.75) 
       node[below left,black]{\small MRN};
\end{scope}
\end{tikzpicture}
\end{minipage}
\begin{minipage}{\columnwidth}
\begin{tikzpicture}
\node [
    above right,
    inner sep=0] (image) at (0,0) {  \includegraphics[width=.99\columnwidth]{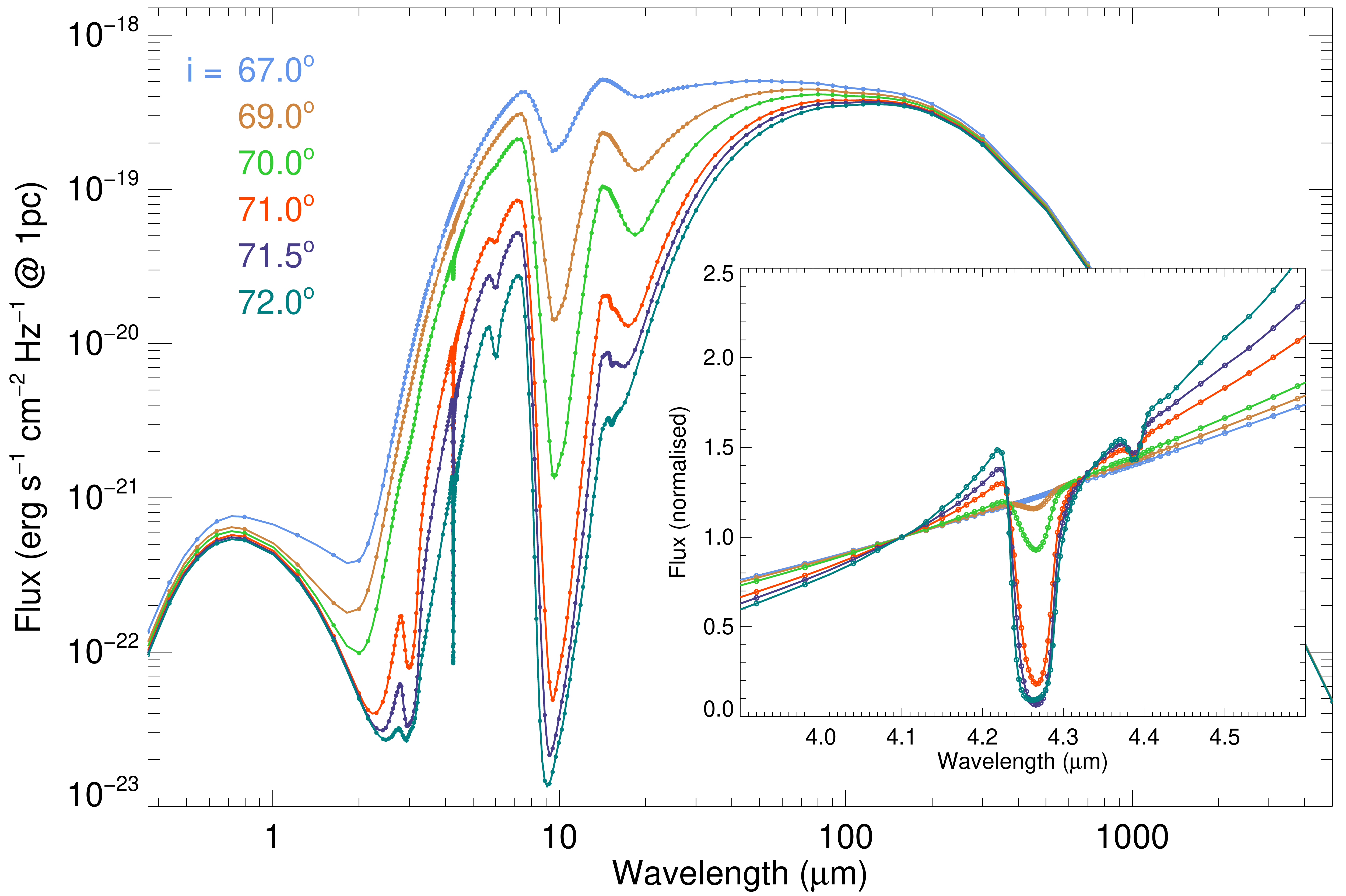}};
\begin{scope}[
x={($0.1*(image.south east)$)},
y={($0.1*(image.north west)$)}]
    \draw[very thick,green] (9.75,9.7) 
       node[below left,black]{\small Silicates + M15};
    \draw[very thick,green] (9.75,8.75) 
       node[below left,black]{\small $\tau_1$};
\end{scope}
\end{tikzpicture}
\end{minipage}
\begin{minipage}{\columnwidth}
\begin{tikzpicture}
\node [
    above right,
    inner sep=0] (image) at (0,0) { \includegraphics[width=.99\columnwidth]{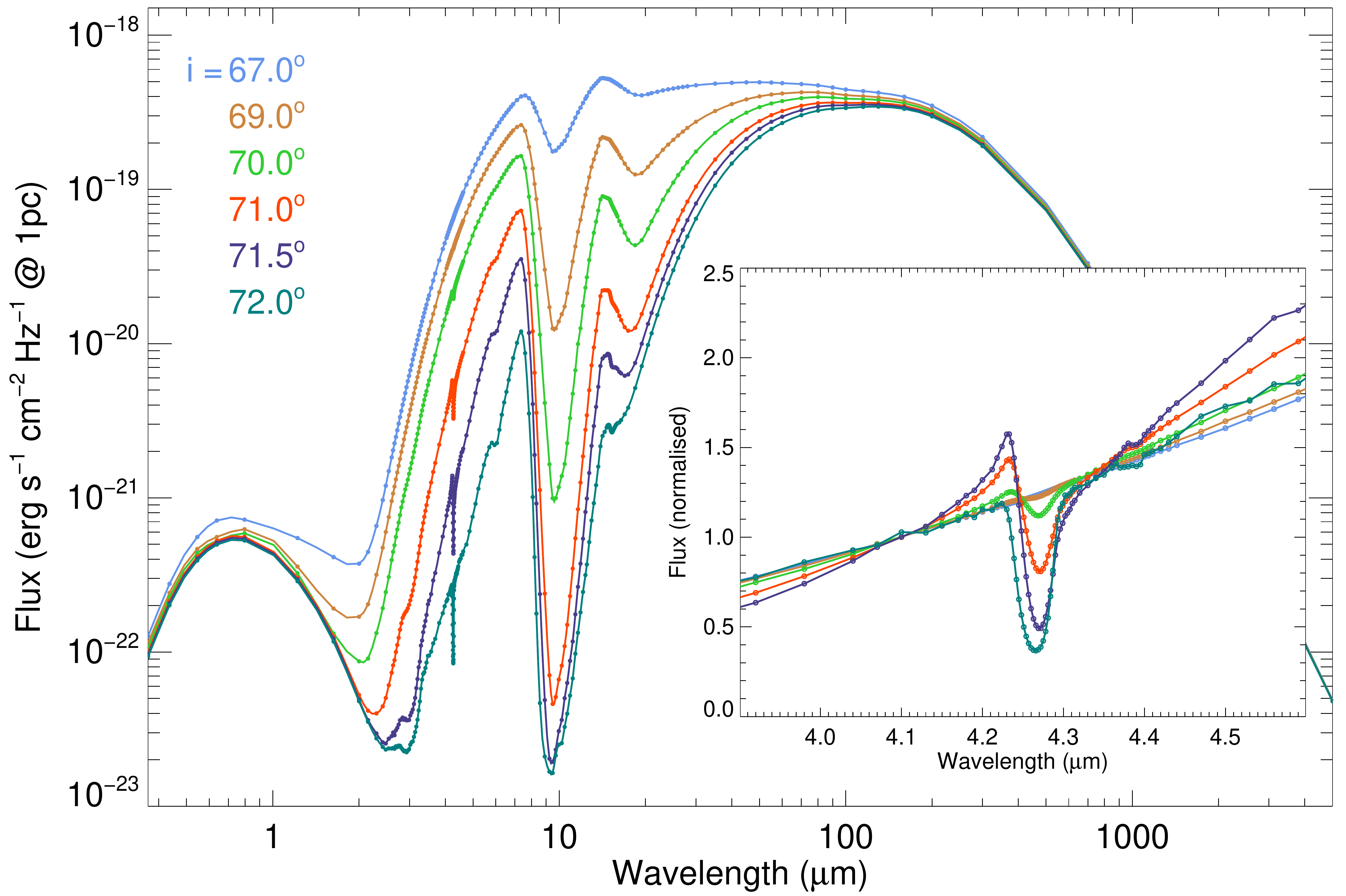}};
\begin{scope}[
x={($0.1*(image.south east)$)},
y={($0.1*(image.north west)$)}]
    \draw[very thick,green] (9.75,9.7) 
       node[below left,black]{\small Silicates + M15};
    \draw[very thick,green] (9.75,8.75) 
       node[below left,black]{\small $\tau_3$};
\end{scope}
\end{tikzpicture}
\end{minipage}
\begin{minipage}{\columnwidth}
\begin{tikzpicture}
\node [
    above right,
    inner sep=0] (image) at (0,0) {  \includegraphics[width=.99\columnwidth]{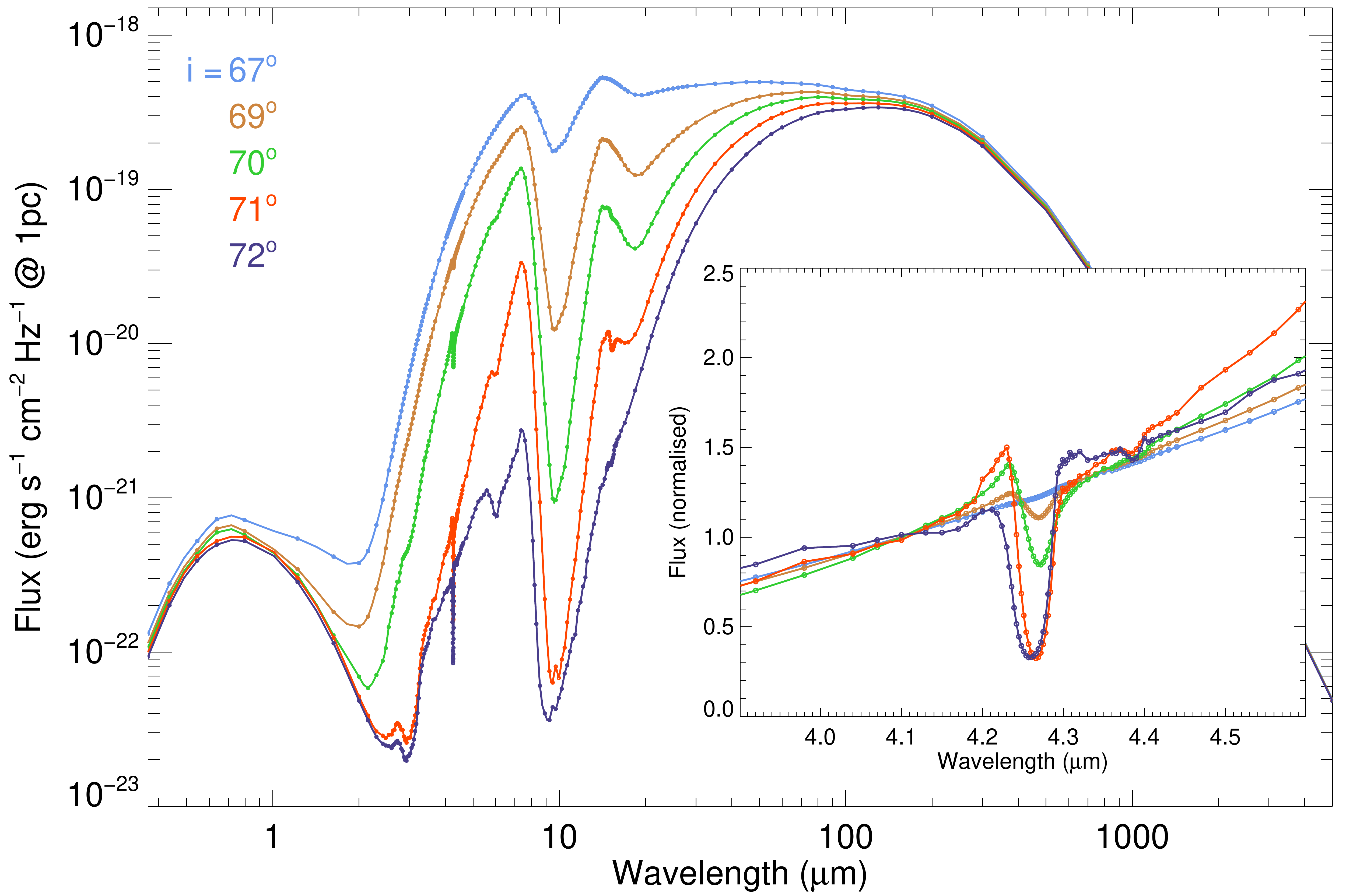}};
\begin{scope}[
x={($0.1*(image.south east)$)},
y={($0.1*(image.north west)$)}]
    \draw[very thick,green] (9.75,9.7) 
       node[below left,black]{\small Silicates + M15};
    \draw[very thick,green] (9.75,8.75) 
       node[below left,black]{\small $\tau_3$, $\rm A_V^{th}=1.5$};
\end{scope}
\end{tikzpicture}
\end{minipage}
\begin{minipage}{\columnwidth}
\begin{tikzpicture}
\node [
    above right,
    inner sep=0] (image) at (0,0) {  \includegraphics[width=.99\columnwidth]{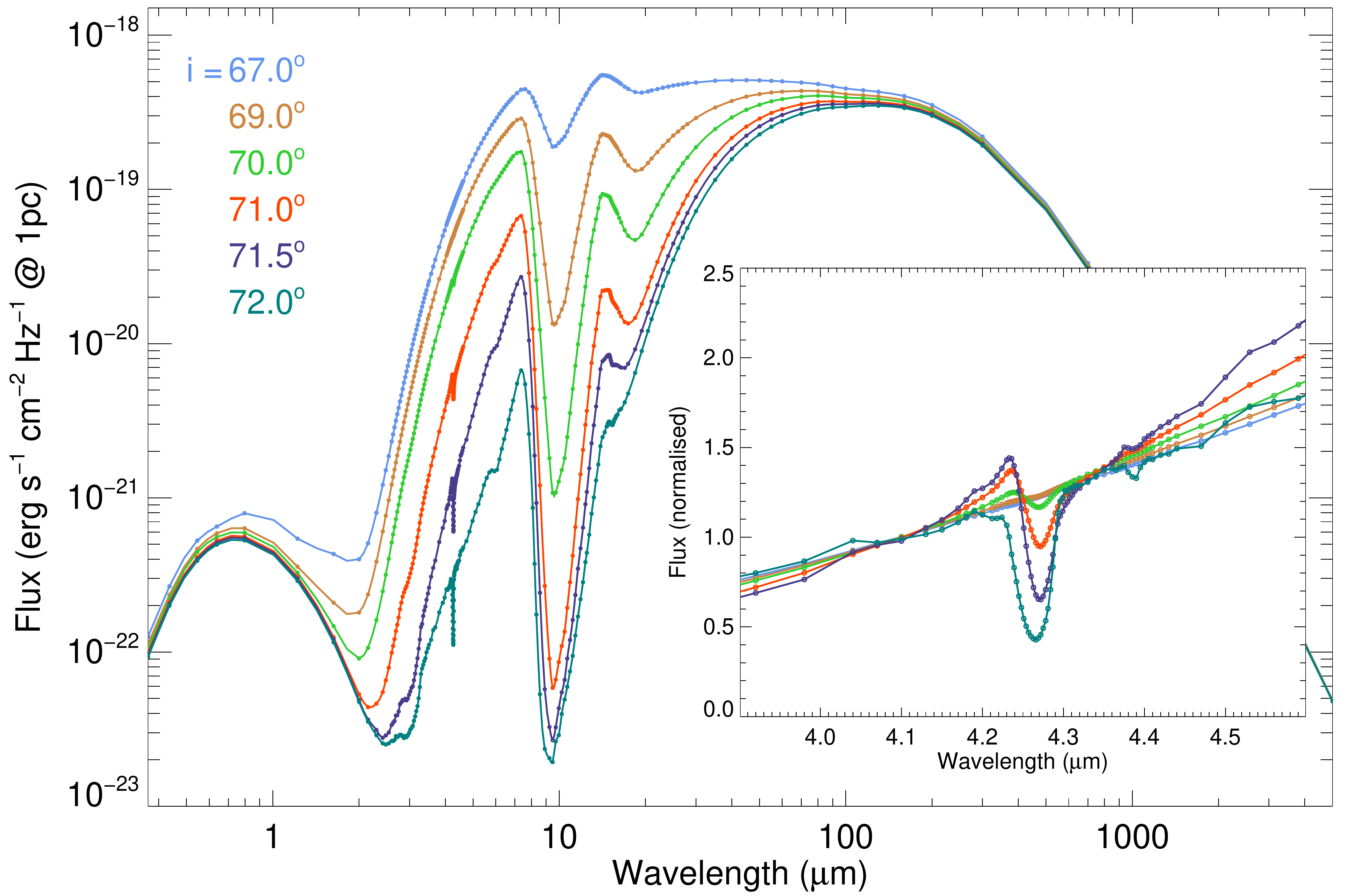}};
\begin{scope}[
x={($0.1*(image.south east)$)},
y={($0.1*(image.north west)$)}]
    \draw[very thick,green] (9.75,9.7) 
       node[below left,black]{\small Silicates + M15};
    \draw[very thick,green] (9.75,8.75) 
       node[below left,black]{\small $\tau_5$};
\end{scope}
\end{tikzpicture}
\end{minipage}
  \caption{{ Spectra of the disk radiative transfer model as observed under different disk inclinations, for the M15 ice and core composition, and different dust size distributions. Distributions corresponding to, from top left to lower right: silicates core, M15 ice mantle composition and size distribution
  MRN, $\rm a_{max}~=~0.25\mu m$ (model \#4);
  $\rm \tau_1$, $\rm a_{max}~=~1\mu m$, (model \#5);
  $\rm \tau_3$, $\rm a_{max}~=~3\mu m$, (model \#6);
  $\rm \tau_3$, $\rm a_{max}~=~3\mu m$, (model \#6) with ice threshold is set at A$\rm_V^{th}(ice)$ = 1.5 instead of 3.}
}
  \label{Fig_modeles_disk_1}
\end{figure*}

\begin{figure*}
  \centering
\begin{minipage}{\columnwidth}
\begin{tikzpicture}
\node [
    above right,
    inner sep=0] (image) at (0,0) {  \includegraphics[width=.99\columnwidth]{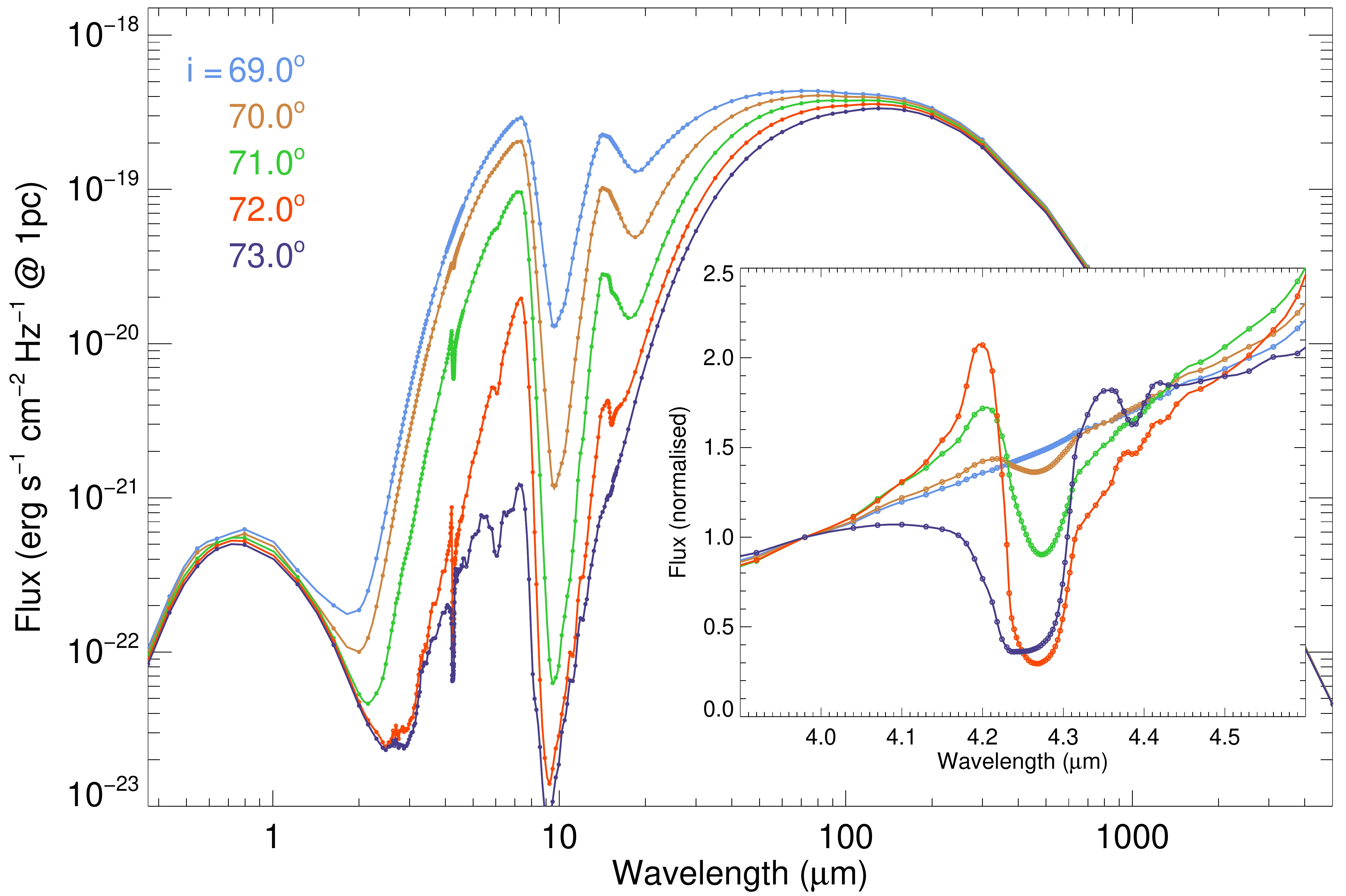}};
\begin{scope}[
x={($0.1*(image.south east)$)},
y={($0.1*(image.north west)$)}]
    \draw[very thick,green] (9.75,9.7) 
       node[below left,black]{\small Silicates + M50};
    \draw[very thick,green] (9.75,8.75) 
       node[below left,black]{\small $\tau_3$};
\end{scope}
\end{tikzpicture}
\end{minipage}
\begin{minipage}{\columnwidth}
\begin{tikzpicture}
\node [
    above right,
    inner sep=0] (image) at (0,0) {  \includegraphics[width=.99\columnwidth]{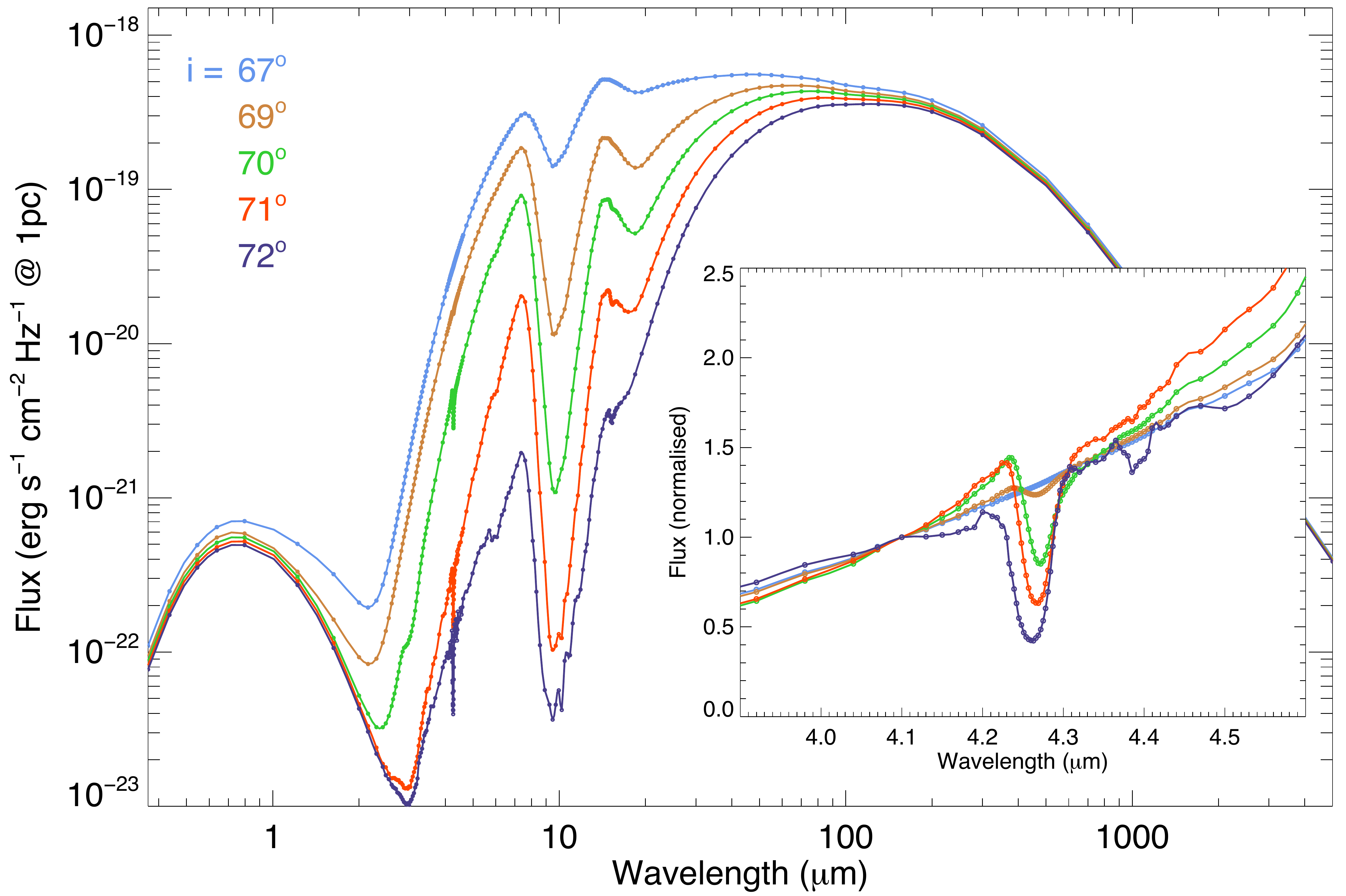}};
\begin{scope}[
x={($0.1*(image.south east)$)},
y={($0.1*(image.north west)$)}]
    \draw[very thick,green] (9.75,9.7) 
       node[below left,black]{\small Silicates + am. carbon + M15};
    \draw[very thick,green] (9.75,8.75) 
       node[below left,black]{\small $\tau_3$};
\end{scope}
\end{tikzpicture}
\end{minipage}
\begin{minipage}{\columnwidth}
\begin{tikzpicture}
\node [
    above right,
    inner sep=0] (image) at (0,0) {  \includegraphics[width=.99\columnwidth]{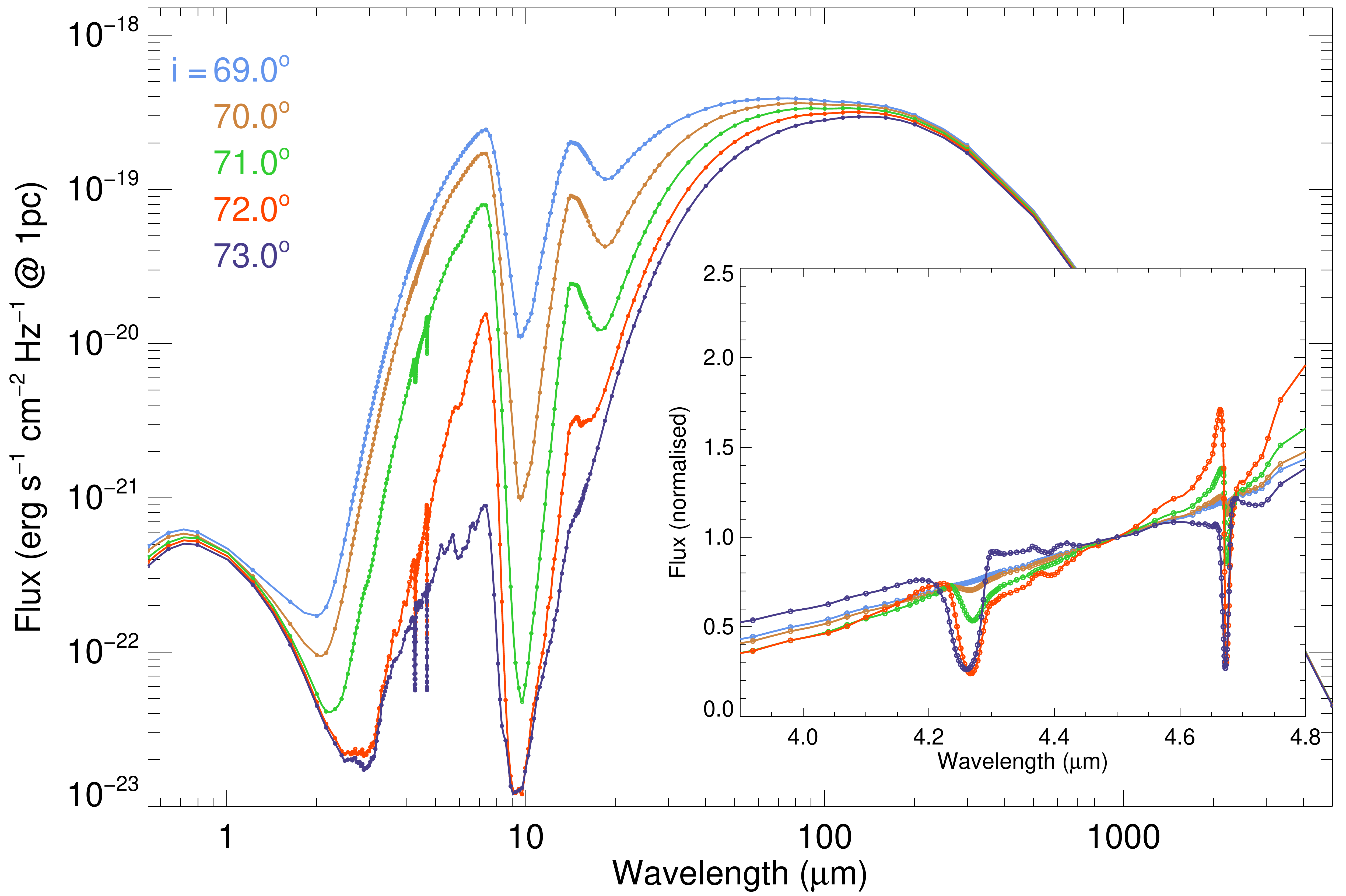}};
\begin{scope}[
x={($0.1*(image.south east)$)},
y={($0.1*(image.north west)$)}]
    \draw[very thick,green] (9.75,9.7) 
       node[below left,black]{\small Silicates + M15 + CO};
    \draw[very thick,green] (9.75,8.75) 
       node[below left,black]{\small $\tau_3$};
\end{scope}
\end{tikzpicture}
\end{minipage}
\begin{minipage}{\columnwidth}
\begin{tikzpicture}
\node [
    above right,
    inner sep=0] (image) at (0,0) {  \includegraphics[width=.99\columnwidth]{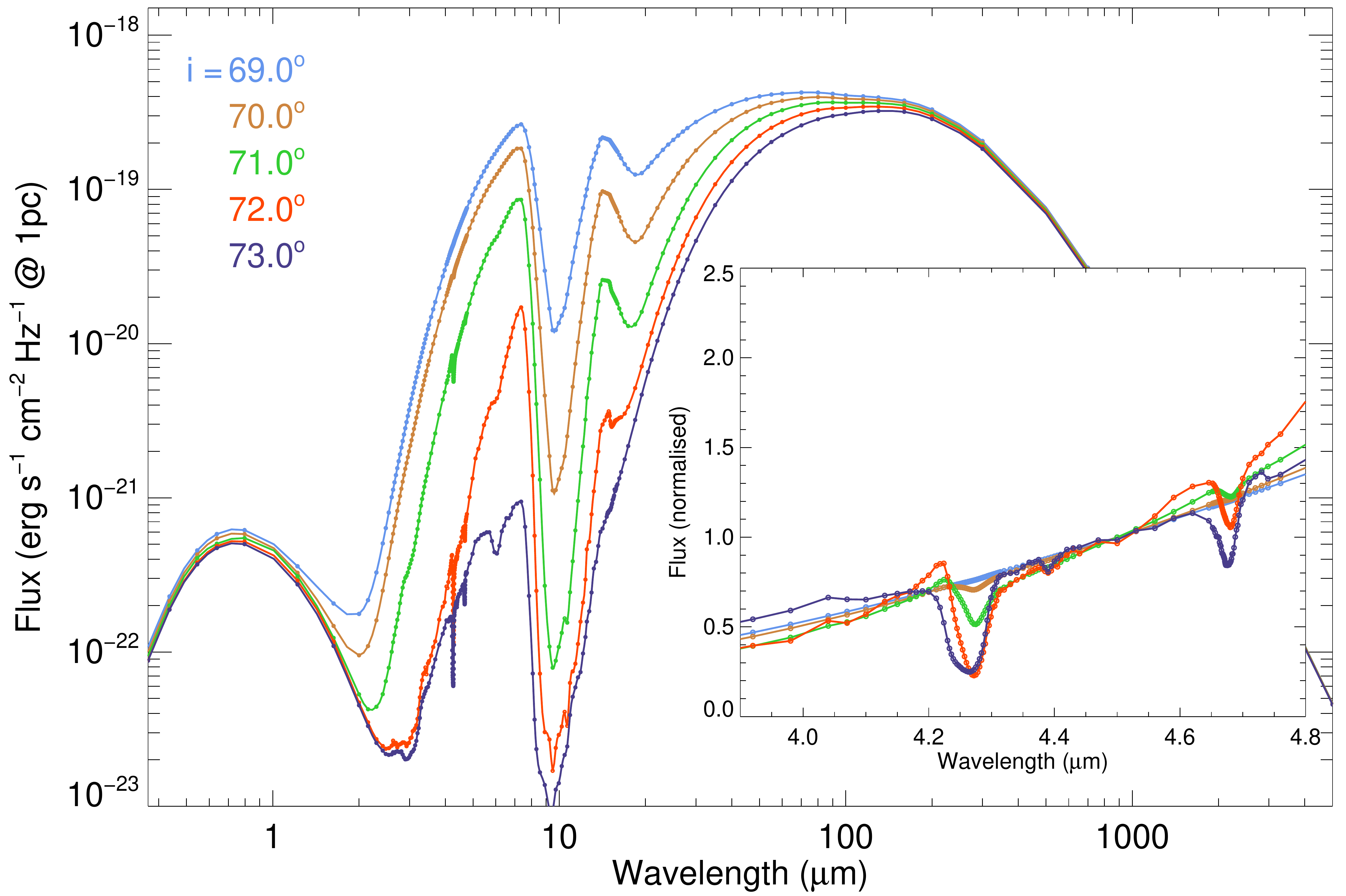}};
\begin{scope}[
x={($0.1*(image.south east)$)},
y={($0.1*(image.north west)$)}]
    \draw[very thick,green] (9.75,9.7) 
       node[below left,black]{\small Silicates + MX};
    \draw[very thick,green] (9.75,8.75) 
       node[below left,black]{\small $\tau_3$};
\end{scope}
\end{tikzpicture}
\end{minipage}
  \caption{Spectra of the disk radiative transfer model as observed under different disk inclinations, for different considered models for the ice and core composition, and dust size distribution. Distributions corresponding to, from top left to lower right: silicates core, M15 ice mantle composition and size distribution
  MRN, $\rm a_{max}~=~0.25\mu m$ (model \#4);
  $\rm \tau_1$, $\rm a_{max}~=~1\mu m$, (model \#5);
  $\rm \tau_3$, $\rm a_{max}~=~3\mu m$, (model \#6);
silicates core, M50 ice mantle composition and size distribution $\rm \tau_3$, $\rm a_{max}~=~3\mu m$, (model \#12);
silicates and amorphous carbon core, M15 ice composition and size distribution $\rm \tau_3$, $\rm a_{max}~=~3\mu m$, (model \#15);
silicates core, M15 ice mantle composition, including pure CO and size distribution $\rm \tau_3$, $\rm a_{max}~=~3\mu m$, (model \#18)
}
  \label{Fig_modeles_disk_2}
\end{figure*}

\begin{figure*}
  \centering
\includegraphics[width=2\columnwidth]{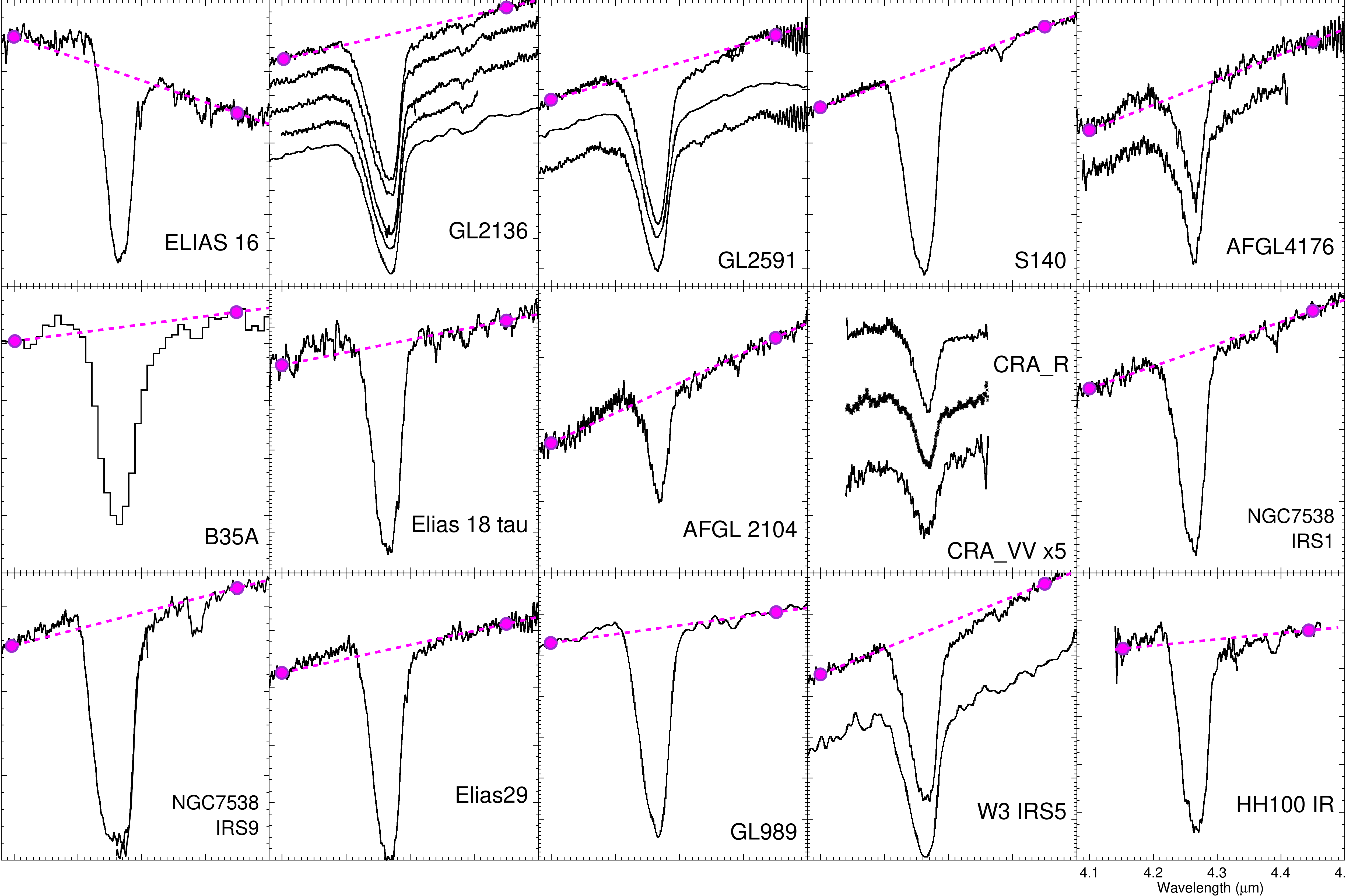}
  \caption{{ Diversity of} CO$_2$ antisymmetric stretching mode observed fluxes with the infrared space observatory (ISO), and Akari (source B35A), toward various astronomical lines of
  sights, including many massive protostellar objects. The profiles vary with some showing slightly warped asymmetric shapes (e.g. GL2591, B35A, Elias~29, GL989, HH100 IR, NGC7538).
  A linear baseline passing through continuum points on each side, far from the band center, indicated by circles, is plotted in each panel to show the various degrees of band asymmetry.}
  \label{Fig_observations}%
\end{figure*}


\begin{figure}
  \centering
\includegraphics[width=\columnwidth]{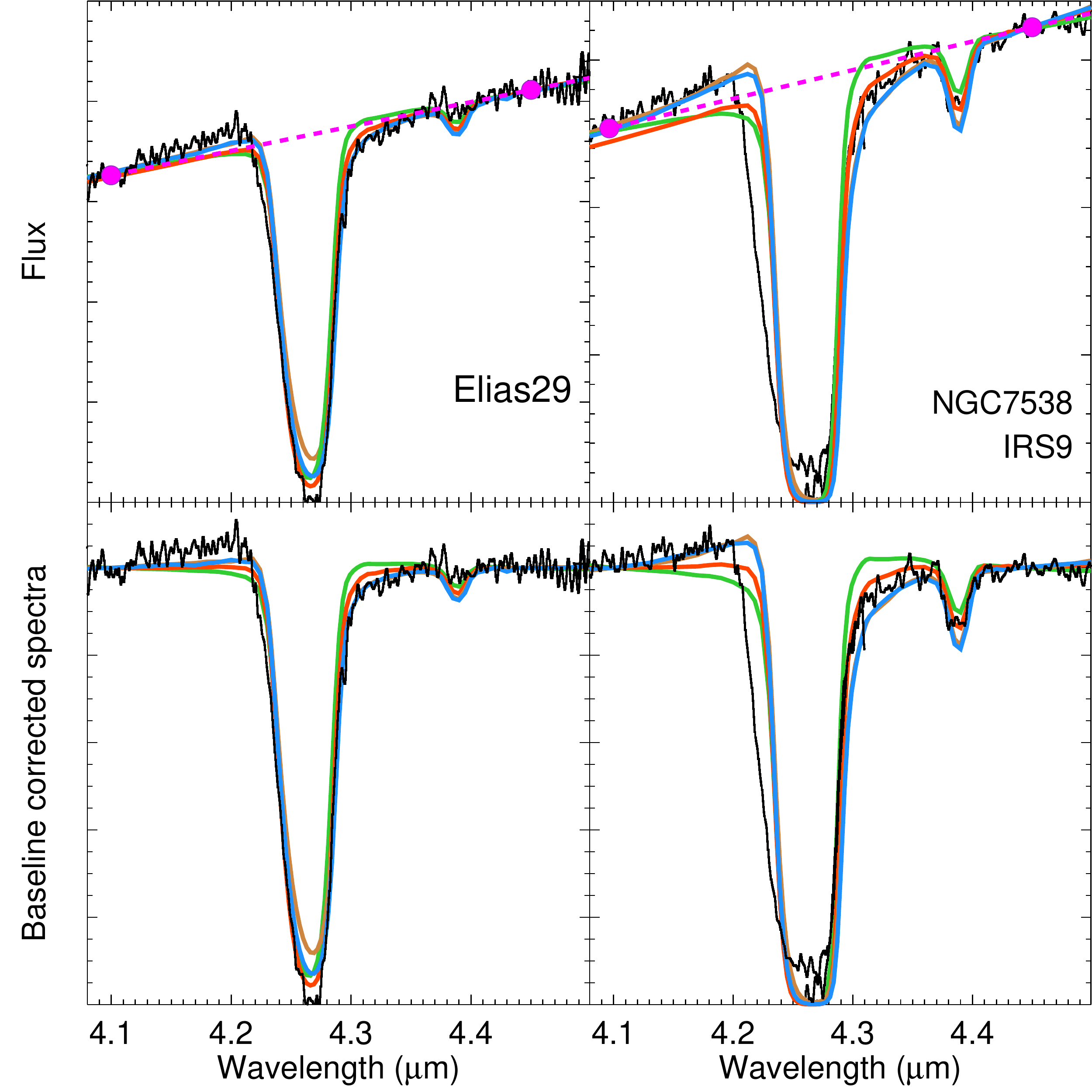}
  \caption{Example of the comparison of two sources spectra with spherical clouds models. 
  Upper panels: CO$_2$ antisymmetric stretching mode observed fluxes with the infrared space observatory (ISO) for Elias 29 and NGC7538 IRS9. A straight continuum baseline passing through points located at several times the band-FWHM from the band center (magenta) is over-plotted, as well as spherical cloud radiative transfer models outputs for M15 mixture presented in Fig.~\ref{Fig_modeles_sphere_1} with different distribution upper size limits (MRN (green), $\rm\tau_1$ (red), $\rm\tau_3$(blue), $\rm\tau_5$(brown)) scaled by an arbitrary (but common) factor for each observation. For Elias 29, the model spectra are for total visual extinction of 30. For NGC7538 IRS9, the model spectra are for total visual extinction of 60.
  Lower panels: Same comparison but in transmittance, after baseline correction applied to each spectrum, which minimises baseline effects in the comparison.
  }
  \label{Fig_observations_comparaison}
\end{figure}


\begin{figure*}
  \centering
\includegraphics[width=0.5\columnwidth]{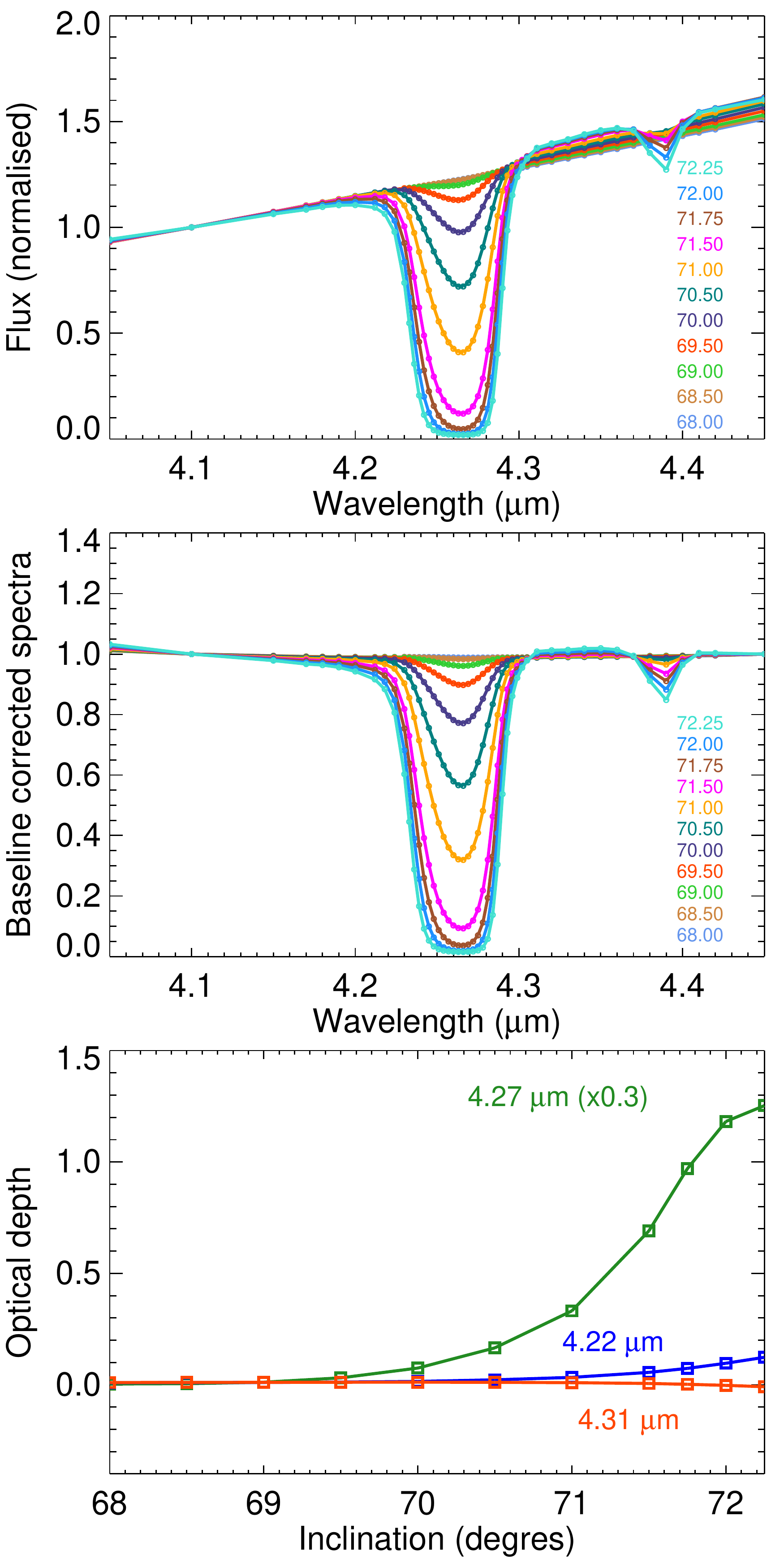}
\includegraphics[width=0.5\columnwidth]{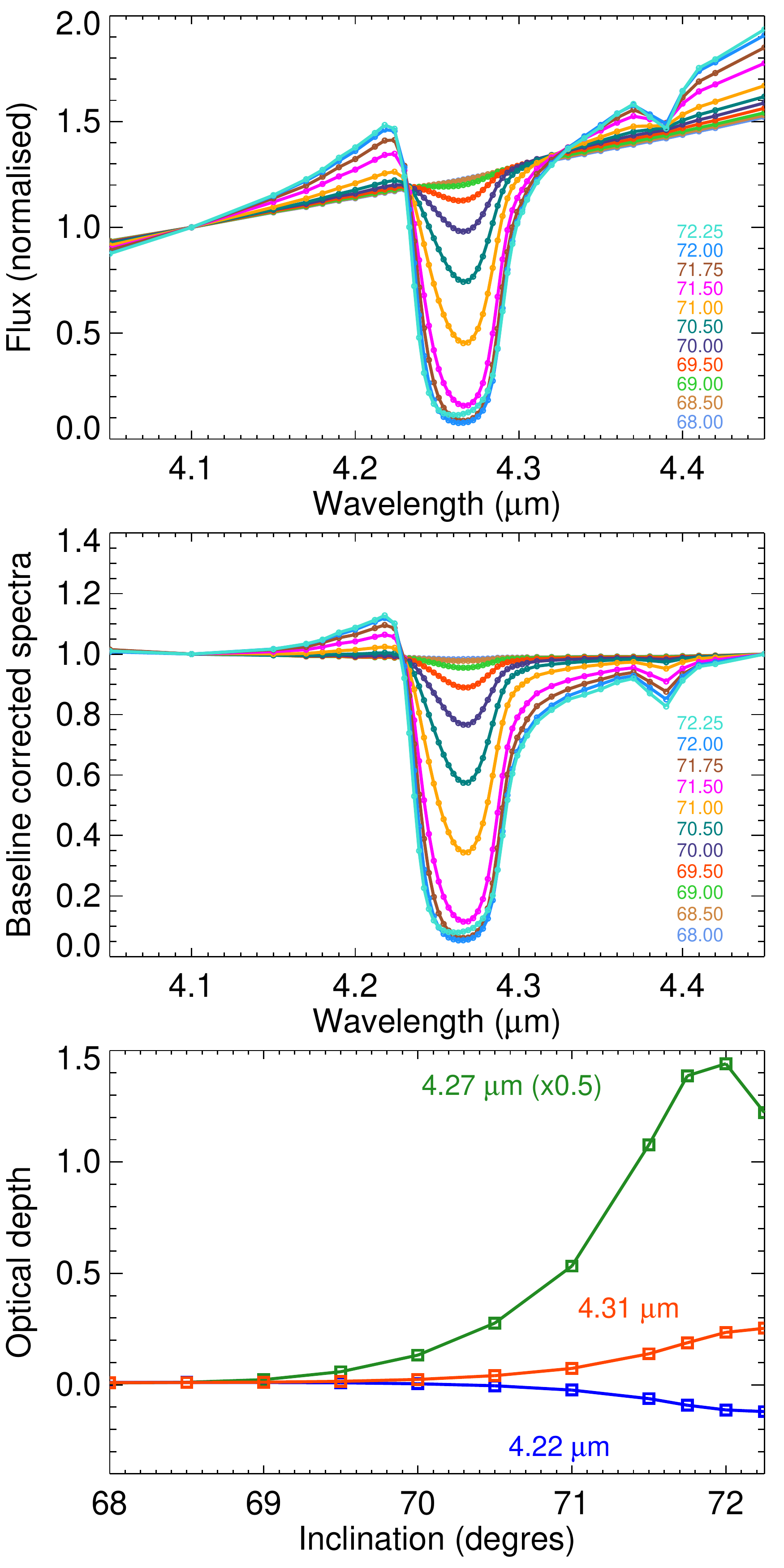}
\includegraphics[width=0.5\columnwidth]{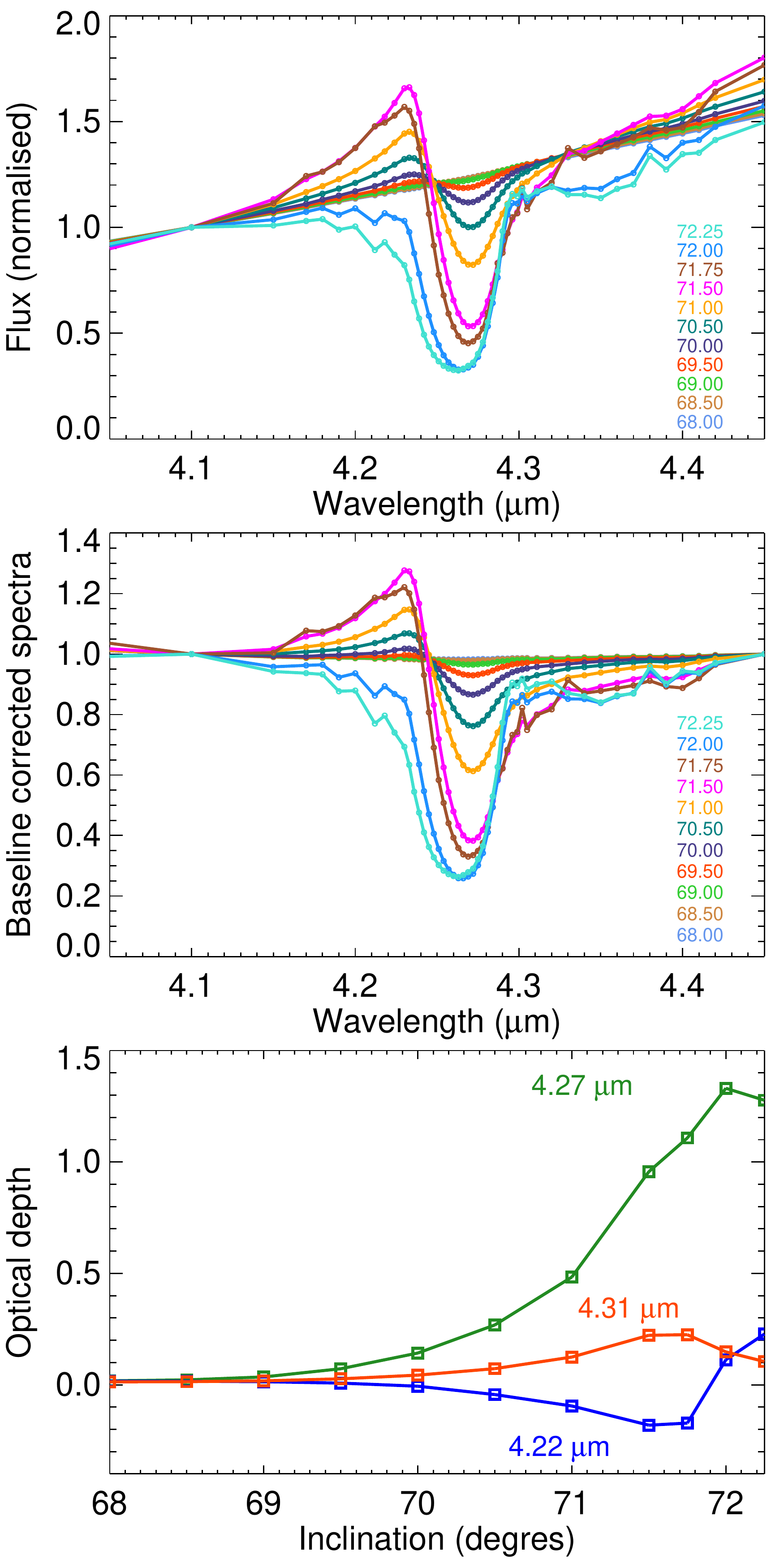}
\includegraphics[width=0.5\columnwidth]{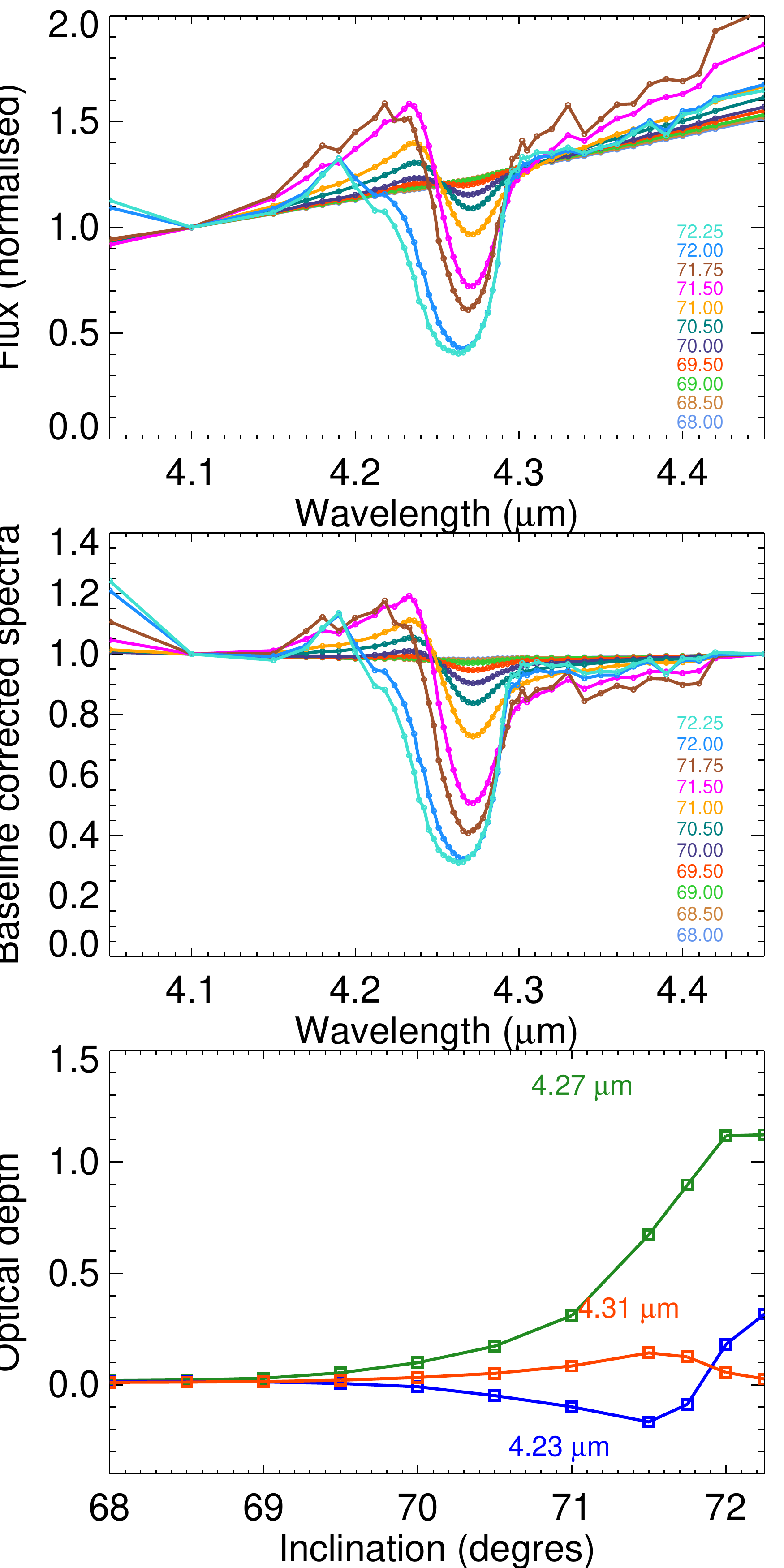}
  \caption{{ Zoom on CO$_2$ spectra of the disk radiative transfer model with M15 ice mantle composition and MRN size distribution, $\rm a_{max}~=~0.25\mu m$ (model \#4);
  $\rm \tau_1$, $\rm a_{max}~=~1\mu m$, (model \#5);
  $\rm \tau_3$, $\rm a_{max}~=~3\mu m$, (model \#6); as observed at different disk inclinations. Top panels: Flux spectra normalised at 4.1 $\mu$m. Middle panels: Linear baseline corrected spectra with baseline reference points taken at 4.1 and 4.45 $\mu$m. Lower panels: evolution of the optical depth in the core (4.27 $\mu$m), the blue wing (4.22 $\mu$m), and red wing (4.31 $\mu$m) with disk inclination.}}
  \label{Fig_modeles_disk_zoom_co2}%
\end{figure*}


\section{Discussion}

\subsection{Optical depth profiles}

Observed ice absorption close-ups are often displayed on an optical depth scale,
a display that allows for a direct view of the column density
involved and the profile of the core of an absorption band.  However,
prior evaluation of a local continuum is required to extract such an
optical depth plot. This continuum estimate can seriously modify the
original profile, especially in the profile wings. 
It tends to minimise the structures in these wings which are expected to be less intense than the core of the
absorption. Therefore, it would be desirable, for a better
understanding, to display more systematically the observed astronomical features both on a
linear flux scale and the extracted optical depth to be able to compare both the
core and wings of profiles. It would also be useful to extend the wavelength range of the display to several times that of the core of the absorption in order to understand the continuum flux behaviour.

The use of such a continuum extraction must be made in regards to the
aims of the analysis. For column density studies it is a way to
extract reasonably good first order numbers for dense clouds. 
In the case of the occurrence of intermolecular interactions, the observable effects are often significant enough on the core of the profile that it remains a valid way to proceed, at least as a first step.
 
We display some examples of observed astronomical infrared
spectra of CO$_2$ ice profiles, recorded mainly toward massive young stellar objects lines of sights, some of them presenting a slightly warped profile. These spectra are displayed in Fig.~\ref{Fig_observations}. { A linear baseline passing through continuum points on each side (indicated by circles) far from the band center, to avoid removing the wings of the absorption profile, is plotted in each panel. This allows us to show the various degrees of band asymmetry.}

These CO$_2$ ice mantle spectra were mainly extracted from the ISO SWS spectra database. The criteria to include them here was principally based on the achieved signal-to-noise ratio, in order to minimise spurious statistical effects. They thus comprise moderate resolution SWS06 and SWS01
(scanning speed 3 and 4) AOT, {with a spectral resolution between about 1000 and 3000}. We focus here on the CO$_2$
antisymmetric stretching mode due to its relatively strong integrated
intensity ($\rm 7.6 10^{-17}cm.molecule^{-1}$ \cite{Gerakines1995}) over
a restricted wavenumber range ($\sim$20cm$^{-1}$), making it an ideal
probe of grain size optical effects. The corresponding source list details are provided in Table \ref{table1}.

Upon inspection of Fig.\ref{Fig_observations} it can be seen that in many spectra
the blue wing is raised with respect to a linear continuum whereas
the red wing is lowered, as if these profiles were the sum of a
dominant and symmetric strong absorption profile and a less intense derivative profile. This is such a true profile that is erased when a
second order polynomial continuum fit is forced to lie above every
point to avoid to give rise to an {\it apparent negative absorption}.
In addition, such a forced continuum fit sometimes produces artificial
extended wing profiles. 

{ In Fig.\ref{Fig_observations_comparaison}, we zoom in on spectra of two of the sources presented in Fig.~\ref{Fig_observations}, and compare them to spherical radiative transfer results from Fig.~\ref{Fig_modeles_sphere_1}. In the upper panels, a straight continuum baseline passing through points located at several times the band-FWHM from the band center (magenta) is over-plotted, as well as spherical cloud radiative transfer model outputs for the M15 mixture, previously presented in Fig.~\ref{Fig_modeles_sphere_1}, with different distribution upper size limits (MRN,green; $\rm\tau_1$,red; $\rm\tau_3$,blue; $\rm\tau_5$,brown) scaled by an arbitrary common factor for each source observed.
The same comparison is presented below in transmittance, after linear baseline corrections applied to each model and the observational spectrum, which minimises baseline effects in the comparison. It is not a fit performing an inversion model for each of these sources, which is out of the scope of this article, but rather is designed to allow us to compare the evolution of the profiles.
Note that for the distribution with an upper bound at 5 microns in size, when compared to that at 3 microns, the profile evolution is less pronounced. Indeed, the grains getting bigger contribute as an almost flat extinction in a spherical cloud model (i.e. their cross section approaches their geometrical cross section), as can be deduced from the individual grain mass absorption contributions presented in Fig.~\ref{Fig_MAC_individuels}, model B. This can be observed in the comparison with the Elias 29 source for which the core of the CO$_2$ band in the $\rm\tau_5$ model absorbs less comparatively. The observed profiles are compatible with a size increase of the size upper bound in the 1-3 micron range (with the adopted size distributions presented in Fig.\ref{Fig_size_evolution}). 
}

The early stages of coagulation and eventual further growth is suggested by other observables such as the near infrared scattering properties and core/cloud shine effects \citep[e.g.][and references therein]{Saajasto2021,Ysard2016}.
  
In the case of the few observed edge-on disks \citep[e.g.][]{Terada2017}, with profiles departing from classical ice band profiles observed in the progenitor phases such as dense clouds, partly due to eventual composition and phase change where it is probed, and partly due to grain growth, this baseline correction will affect the interpretation.
The global effect of grain growth on the CO$_2$ ice stretching mode profile is an observable distortion with a rise in the blue
wing around 4.2~$\mu$m and an absorbing red wing. Such a slight warping of the profiles is already observed in clouds/envelopes surrounding young stellar objects and confirms the presence of grain sizes larger than those found in the diffuse ISM in the grain size distribution, following grain coagulation.

{ In Fig.~\ref{Fig_modeles_disk_zoom_co2} we display close-ups of the CO$_2$ spectra of the disk radiative transfer model with M15 ice mantle composition and MRN size distribution, with $\rm a_{max}~=~0.25\mu m$ (model \#4); $\rm \tau_1$, $\rm a_{max}~=~1\mu m$, (model \#5); $\rm \tau_3$, $\rm a_{max}~=~3\mu m$, (model \#6); as observed at different disk inclinations. The spectra are again analysed with a linear baseline drawn from 4.1 to 4.45 $\mu$m on the flux spectra (shown normalised at 4.1 $\mu$m because of large flux differences at the various inclinations).
From the baseline corrected spectra, the evolution of the optical depth in the ice absorption core (4.27 $\mu$m), in the blue wing (4.22 $\mu$m), and in the red wing (4.31 $\mu$m) are given as a function of inclination to evaluate the degree of warping of the band.}

The way local continuum baselines have generally been applied to extract the absorption profile may erase this subtle information that is sometimes only present in the wings of the profile. The profile resulting from such a baseline correction is then analysed with the sum of different ice mixture contributions using a principal component spectral deconvolution, whereas the reality is rather a more complex combination of grain size and composition effects affecting the profile when grains grow significantly in the distribution.

In the case of protoplanetary disks, as shown in our calculations, the emerging ice profiles will appear even more affected than for spherical envelopes or dense clouds.

Profiles that seem even reversed, with a pronounced blue-shifted wing absorption, in disks, for high inclination/optical depth, have been observed  \citep[e.g. IRAS04302+2247,][Fig. 3]{Aikawa2012}, as observed in some models of Fig.\ref{Fig_modeles_disk_1}-\ref{Fig_modeles_disk_2}.

\subsection{CO$_2$ and CO profiles}

CO and CO$_2$ stretching modes both fall within a relatively narrow wavelength range.
In the case of CO and CO$_2$ mixed in the same ice mantles or on the same grains in the dust size distribution, if the grain growth dominates over the influence of the intrinsic ice mixture composition, both CO$_2$ and CO profiles should display a relative common asymmetry in their respective profiles, as seems to be the case in e.g. \cite{Noble2013} (Fig 2d).
In the case of pure CO the profile asymmetry might even be more pronounced because of the narrower feature for the large grains in the distribution \citep[see, e.g.][]{Dartois2006}. 
\subsection{H$_2$O versus CO$_2$ band profile}
The spectra extracted from the disk models with an M15 ice mantle composition and the different size distributions are shown in Fig.\ref{Fig_disk_baseline_corrected_H2O_CO2}, spanning the water ice and carbon dioxide stretching modes range.
The profile distortion observed in the CO$_2$ stretching mode is also observed in the water ice band, with a red shift of its band center as well as the occurrence of a red extinction wing. The contrast is less pronounced than for the CO$_2$ ice band because of the much larger intrinsic band width for this band. 
Infrared observations in the H$_2$O, CO$_2$, and CO stretching mode region for some sources with available coverage and for which the gas phase CO does not hamper the observation of the CO ice profile are shown in Fig. \ref{Fig_observations_H2O_CO2_CO}. The profiles of the CO$_2$ and CO seem to display an increase at the lower wavelength side of each band, and for H$_2$O, CO$_2$ and CO an extinction wing contribution on the higher wavelength side of each band. This stresses the importance of recording ice profiles with a span covering not only the core, but also the surrounding continuum, over several times the band full width at half maximum, in order to accurately constrain the grain growth contribution.

{ We show for comparison spherical cloud model spectra with the more complex MX ice mixture (H$_2$O:CO$_2$:CO:NH$_3$ 100:16:8:8) for a $\tau_3$ size distribution, for cloud visual extinction of Av=15 (blue), 30 (red) and 60 (green).
We also display both observations and models after baseline correction to provide transmittance spectra for comparison.} 
The influence of grain growth on the water ice stretching mode profile is necessary to account for the red wing observed in this band toward many lines of sights, as was explored in \citep[e.g.][]{Smith1988, Smith1989, Dartois2002}, { with the additional contribution from the ammonia hydrate related band around 3.47$\mu$m (also present in the MX ice mixture models), and possibly a smaller contribution from methanol stretching mode at 3.54 $\mu$m that is not in the calculations. 
The MX model spectra} reproduces most of the observed features including the red shift of the water ice band center, the extended red wing in the absorption of CO$_2$ and CO absorption bands.
%

\begin{figure*}
  \centering
\includegraphics[width=2\columnwidth]{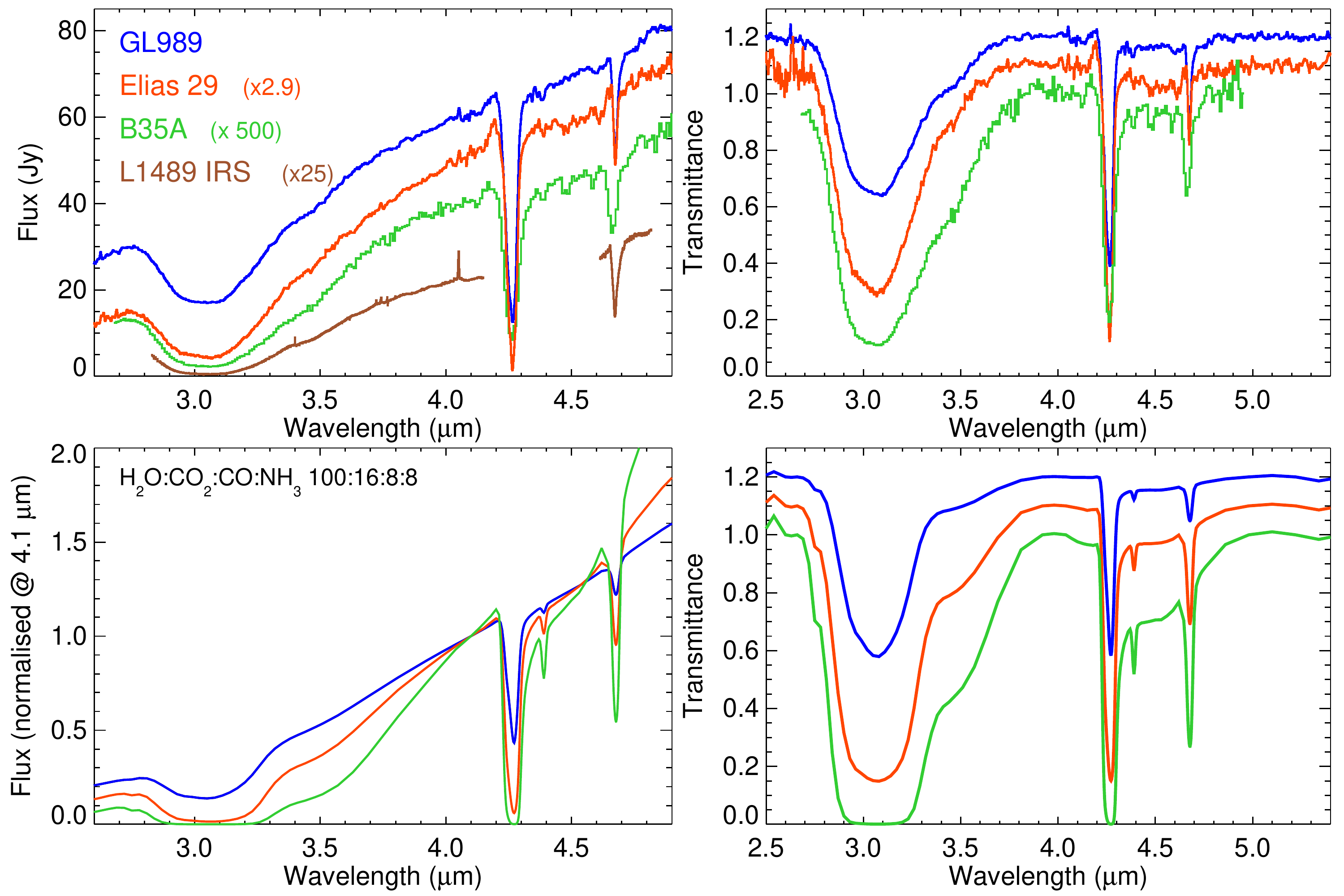}
  \caption{{ Upper left: Infrared observations in the H$_2$O, CO$_2$, CO stretching mode region for some sources with available coverage and for which gas phase CO does not hamper observation of the CO ice profile. GL989 and Elias 29 are from the ISO database, B35A is an Akari spectrum \citep{Noble2013}, L1489 IRS is ground based high resolution spectra from \citep{Boogert2002}. In the L1489 spectrum, the gas phase CO has been filtered out to show the CO ice profile. Lower left: Spherical cloud model spectra with the MX mixture for a $\tau_3$ size distribution, for cloud visual extinction of Av=15 (blue), 30 (red) and 60 (green). The right panels shown the same observations and models after baseline correction to provide transmittance spectra. The blue and red traces have been offset by 0.2 and 0.1, respectively, for clarity.}}
  \label{Fig_observations_H2O_CO2_CO}
\end{figure*}

%
\begin{figure*}
  \centering
\includegraphics[width=2\columnwidth]{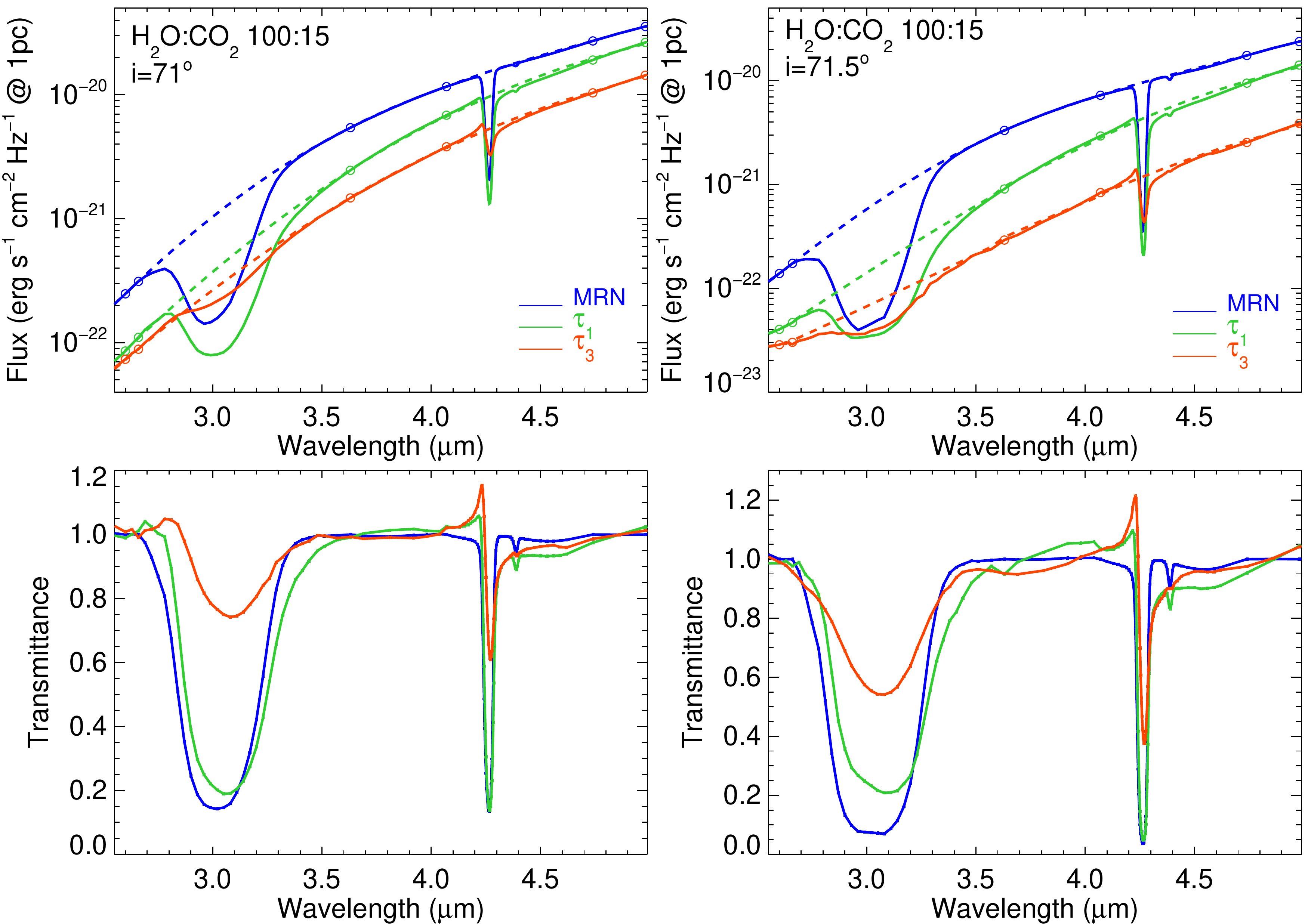}
  \caption{Close up on the disk radiative transfer models (upper panels) corresponding to silicates core, M15 ice mantle composition and size distribution MRN, $\rm a_{max}~=~0.25\mu m$ (model \#4); $\rm \tau_1$, $\rm a_{max}~=~1\mu m$, (model \#5); $\rm \tau_3$, $\rm a_{max}~=~3\mu m$, (model \#6); as observed under 70 and 71 degrees of inclination, in the water ice and carbon dioxide stretching modes range. A spline continuum passing through the indicated circle points, well apart from the core of the ice bands is overlaid (dashed line). Lower panels show the extracted resulting transmittance spectra once dividing by this estimated continuum. In the highest optical depth spectrum (i=71$^o$,$\rm \tau_3$) both the calculated spectrum and a slightly numerical noise filtered spectrum are shown. The asymmetry of the water ice band as well as the shift in the band center are evidenced and accompanies the observed higher contrast CO$_2$ profile deformation.}
  \label{Fig_disk_baseline_corrected_H2O_CO2}
\end{figure*}
%


\section{Conclusion}

We modeled the mass absorption coefficient for interstellar dust grains including ice mantles, from the classical MRN diffuse ISM dust size distribution to dust populations growing in the denser ISM phases, with increasing amounts of larger grains, up to three microns in size. These models were injected into Monte-Carlo radiative transfer calculations for a spherical dense cloud and a protoplanetary axisymmetric disk model. The influence of grain growth on the ice spectroscopic profiles is demonstrated, focusing in particular on the CO$_2$ ice stretching mode extinction.

Several ice mantle and refractory dust core association possibilities, such as constant ice/core volume ratio, constant ice thickness, and randomly mixed ice/core aggregations were modelled. While differences exist among the details of the profiles, the two dominant parameters affecting the results are on the one hand the influence of large size grains in the distributions on the resulting profiles and the chosen starting ice mixture's optical constants, reflecting inter-molecular interactions in the solid phase.

If this grain growth turns out to be significant in the observed disk regions, this study shows it will strongly impact the observed band profiles, and the retrieval of the underlying ice mantle compositions will be more complex than the usually assumed principal component analysis using a basis set of planar thin film ice mixture spectra, performed on dense clouds observations.

Sources where grain growth has significantly impacted the dust size distribution should display a set of ice bands with noticeable spectral distortion, particularly in the wings of the bands. Too highly constrained continuum extraction baselines and the resulting spectra shown in optical depth tend to minimize the significance of these distortions. We recommend strongly to take care during baseline subtraction to both use a large spectral window span around the band and not to correct for apparent negative absorption.

Detailed ice profile analyses observed in protoplanetary disks with next generation observatories, such as the JWST, and in particular for the CO$_2$ ice mantles profile, will provide by comparison to such models a strong constraint on the extent of grain growth, in the micron range, for the dust distributions.

\section*{acknowledgements}
We gratefully acknowledge H. Fraser for kindly providing us the B35A Akari spectrum data. 
%



\bibliographystyle{mnras}
\bibliography{ice_mantles_growth_modeling.bib}




\appendix
\section{Observations log}
\begin{table}
\caption{Infrared Space Observatory observations: sources, observation number (TDT), observing template (AOT), integration time}             
\label{table1}      
\centering          
\begin{tabular}{l l l l}     
\hline\hline       
Source & TDT$\;$\# & AOT & Int. time (s)\\ 
\hline                    
AFGL~2136 & 12000925 & SWS$06$ & 2994\\
          & 12800302 & SWS$06$ & 3732\\
          & 31101023 & SWS$06$ & 4391\\
          & 33000222 & SWS$01$ & 3554\\
          & 51601403 & SWS$06$ & 1404\\
AFGL~2591 & 14200503 & SWS$06$ & 2908\\
          & 02800582 & SWS$06$ & 1972\\
          & 35700734 & SWS$01$ & 3454\\
          & 35701221 & SWS$07$ & 7002\\
W3~IRS5   & 42701224 & SWS$06$ & 5668\\
          & 42701302 & SWS$01$ & 3454\\
HH100~IR  & 70400725 & SWS$06$ & 2616\\
AFGL~989  & 71602619 & SWS$01$ & 3454\\
AFGL~2104 & 14900501 & SWS$06$ & 880\\
AFGL~4176 & 11701404 & SWS$06$ & 1706\\
          & 30601344 & SWS$06$ & 4268\\
CRA~R     & 12400103 & SWS$06$ & 862\\
          & 71502103 & SWS$06$ & 862\\
CRA~W         & 12400406 & SWS$06$ & 1190\\
Elias~16      & 68600538 & SWS$06$ & 8682\\
Elias~29      & 29200615 & SWS$06$ & 5668\\
NGC7538~IRS1  & 28301235 & SWS$06$ & 4552\\
NGC7538~IRS9  & 56801802 & SWS$06$ & 3328\\
Elias~18~Tau  & 68502539 & SWS$06$ & 3140\\
S140          & 26301731 & SWS$06$ & 5270\\
GCS~3I        & 32701543 & SWS$06$ & 3226\\
\hline                  
\end{tabular}\\
SWS01 and SWS06 spectral resolution are dependent on the integration time, and are typically between R~1000 and 3000.
\end{table}

\end{document}